\documentclass[
reprint,
 amsmath,amssymb,
 aps,
]{revtex4-2}

\usepackage{graphicx}% Include figure files
\usepackage{dcolumn}% Align table columns on decimal point
\usepackage{bm}% bold math
\usepackage{hyperref} 
\usepackage{xcolor}
\usepackage[normalem]{ulem}
\numberwithin{equation}{section}

\newcommand{\dd}{\text{d}}

\begin{document}

\preprint{APS/123-QED}

\title{Extreme-mass ratio inspirals in Schwarzschild-de Sitter spacetime I: Weak-field orbits}% Force line breaks with \\

\author{John Adrian N. Villanueva}
\altaffiliation[]{javillanueva@nip.upd.edu.ph}%Lines break automatically or can be forced with \\
\author{Ian Vega}%
\email{ivega@nip.upd.edu.ph}
\affiliation{%
 National Institute of Physics, University of the Philippines, Diliman, Quezon City 1101, Philippines 
}%


%\affiliation{
%\affiliation{
 %Third institution, the second for Charlie Author
%\affiliation{%



\begin{abstract}
The inspiral of a compact object into a black hole is a key source of low-frequency gravitational waves for future space-based detectors like LISA. While models of this process have advanced considerably, they typically focus on asymptotically flat spacetimes. In this paper, we  investigate how departures from asymptotic flatness, whether driven by cosmic expansion or large-scale galactic environments, alter adiabatic orbital evolution. Using the Schwarzschild–de Sitter (SdS) metric, we parametrize these deviations via an `SdS parameter' ($\lambda$) and analyze its impact on bound orbits. We calculate how $\lambda$ shifts the separatrix between bound and plunging states and modifies the relationship between a binary’s energy, angular momentum, and orbital geometry. By applying a modified quadrupole formula in the weak-field limit, we investigate the effect of $\lambda$ on circularization, plunge times, and orbital trajectories.  We show that a positive SdS parameter accelerates eccentricity decay and shortens inspiral timescales. We also generate adiabatic waveforms from inspirals evolving under radiation reaction, exhibiting an increase in amplitude and a cumulative phase advance induced by $\lambda$. While these effects are negligible if $\lambda$ is strictly cosmological, they become observationally relevant if the parameter serves as a proxy for astrophysical environments that induce $\sim r^2$ potential corrections. Our results suggest that such environmental coupling could meaningfully bias event rate estimates and waveform templates for space-based gravitational-wave astronomy.
\end{abstract}

\maketitle

%\tableofcontents

\section{Introduction}
\label{sec:intro}

Gravitational waves have emerged as powerful probes of the universe, opening a fundamentally new window onto its most energetic and previously inaccessible phenomena. The direct detections of gravitational waves \cite{abbott2023gwtc} have firmly established their role as essential tools in modern astrophysics. Together with solar and supernova neutrinos, diffuse gamma-ray backgrounds, and electromagnetic counterparts in binary neutron star mergers \cite{branchesi2016multi,meszaros2019multi}, gravitational waves are now an indispensable pillar of multimessenger astronomy. 

%Gravitational wave  astronomy is also in line to assist the problem of the \blue{expansion} of the universe \cite{palmese2024standard}. The general rate of \blue{expansion}, known as the Hubble parameter \cite{carroll2001cosmological,peebles2003cosmological} is closely related to the \blue{cosmological constant}, $\Lambda$.  While estimates of this parameter have been obtained, most notably by the Planck collaboration \cite{aghanim2020planck}, this will still be tested more thoroughly using gravitational wave signals \cite{seto2018prospects}. Borrowing the idea of standard candles from stars to determine the gravitational redshift distance between celestial bodies, standard sirens have also been put forward for gravitational wave sources \cite{holz2005using,feeney2019prospects,tamanini2016science}. 

%\red{Relevant sources for gravitational wave astronomy include black hole binary mergers  \cite{roulet2019constraints} and neutron star binaries \cite{siellez2014simultaneous}, both of which have already been observed by the ground-based triple gravitational wave interferometer network consisting of LIGO (Hanford and Livingston) and Virgo (Italy).} 
The types and number of gravitational-wave sources available for analysis will increase significantly with the upcoming launch of the space-based LISA (Laser Interferometer Space Antenna) gravitational wave detector \cite{tamanini2017late,del2018stellar,arun2022new,colpi2024lisa}.
%Both ground-based and future space-based detectors will be pivotal in analyzing gravitational wave signals to estimate the Hubble parameter and the size of the universe. 
Low-frequency gravitational wave sources, particularly in the millihertz range, will be key targets for space-based detectors like LISA. Extreme-mass ratio inspirals (EMRIs) \cite{amaro2017laser, karnesis2024laser} in particular, where compact objects spiral into massive black holes with a mass ratio of approximately $10^{-6}$, are expected to have large number of gravitational wave cycles within the millihertz band (typically on the order of the inverse of the mass ratio) enabling us to precisely probe strong-field gravitational effects \cite{barausse2020prospects,amaro2023astrophysics}.

Models of EMRIs typically treat them as isolated two-body systems. However, external agents like matter \cite{barausse2007relativistic,macedo2013into,barausse2008influence,cardoso2020environmental} and external fields \cite{eatough2013strong, frolov2024motion} may influence the gravitational waves and orbital dynamics of real EMRIs. These external influences are often referred to as environmental effects, and have been extensively discussed in the literature  \cite{barausse2014can,toubiana2021detectable,zwick2023priorities,caneva2024first}.

%One could envision an EMRI as a binary system containing a stellar-mass compact object, and a supermassive black hole. However, the typical isolated two-body problem in classical physics provides an idealized model of a celestial binary system, neglecting various external physical factors in a real astrophysical context. To accurately model the motion of such a binary, one must account for physically realistic external agents that contribute gravitationally. These external influences are often referred to as environmental effects, and have been extensively discussed in the literature  \cite{barausse2014can,toubiana2021detectable,zwick2023priorities,caneva2024first}.

In this work, we study EMRIs in the Schwarzschild-de Sitter (SdS) spacetime, 
\begin{equation}
\dd s^2 = -f(r) \dd t^2 + f(r)^{-1} \dd r^2 + r^2 \dd \Omega^2,
\end{equation}
where $f(r) = 1-2M/r + \Lambda r^2/3$ and $\dd \Omega^2$ is the metric on a unit two-sphere. This metric is traditionally understood to be a model for a black hole of mass $M$ immersed in an expanding universe characterized by a cosmological constant $\Lambda$ \cite{kottler1918physical}. The study of EMRIs in this spacetime is of intrinsic theoretical interest, as it offers a clean and minimal setting in which to examine the impact of non-asymptotic flatness on EMRI dynamics.  Beyond this theoretical curiosity, we also view the SdS spacetime as a simplified framework for exploring environmental effects on EMRIs.

If $\Lambda$ is confined to a cosmological interpretation, its measured value $(\simeq 1.1 \times 10^{-52} \textrm{m}^{-2})$ \cite{aghanim2020planck} is sufficiently small that its impact on the evolution of extreme mass-ratio inspirals may be safely ignored \cite{kagramanova2006solar,ashtekar2015asymptotics2}. EMRIs in SdS spacetime would then be observationally indistinguishable from their counterparts in the extensively studied Schwarzschild geometry. 

Adopting a broader perspective, one may treat the $\Lambda$ term in the metric can be reinterpreted as encoding corrections to the gravitational potential of the form $\delta V \sim \Lambda r^2$. In this capacity, $\Lambda$ serves as a phenomenological proxy for $r^2$-type perturbations, facilitating a qualitative investigation into how large-scale environmental modifications affect EMRI dynamics and their resulting gravitational-wave signatures.

An example of $r^2$-type perturbations arises in studies of accretion or tidal interactions by an external body in a binary system. Suppose a tidal perturbation to the motion of a binary is contirbuted by the additional term in the potential
\begin{equation}
    \delta V=- \frac{M_{\textrm{ext}}}{|R_{\textrm{ext}}-r|}\sim - \frac{M_{\textrm{ext}}}{R_{\textrm{ext}}^3}r^2\sim K r^2
    \label{eq:tidalpot}
\end{equation}
where $R_{\textrm{ext}}$ is the radial distance of the external body of mass $M_{\textrm{ext}}$ to the central BH at the hierarchical limit, $r\ll R_{\textrm{ext}}$. We see here a $\sim K r^2$ correction in the potential, which could also be seen in tidally distorted spacetimes like in \cite{poisson2005metric}, where $K$ may be related to the components of the tidal tensor $\mathcal{E}_{ij}$. 

Closely related is the modelling of an EMRI embedded in an external uniform galactic magnetic field.  One may consider the Preston-Poisson spacetime \cite{preston2006light,konoplya2006particle,kovacs2011accretion}, which has $\sim B^2r^2$ and $\sim K r^2$ (magnetic field and tidal) corrections to the Schwarzschild metric functions:
\begin{eqnarray}
g_{t t}&=& -\left(1-\frac{2 M}{\bar{r}}\right)
\nonumber\\
&&+\frac{1}{3} K(4 M-3 \bar{r})(2 M-\bar{r})\left(3 \cos ^2 \theta-1\right) \nonumber\\
&& -\frac{1}{3} B^2\left[3(2 M-\bar{r}) \cos ^2 \theta-2 M\right](2 M-\bar{r}).\nonumber\\
\label{eq:ppmetric}
\end{eqnarray}

These examples motivate us to explore a much wider range of $\Lambda$ values than would be required by cosmology. We refer to the dimensionless quantity $\lambda$ as the SdS parameter, which parametrizes deviations from asymptotic flatness (specifically $r^2$ perturbations) potentially originating from several sources:
\begin{equation}
    \lambda := \Lambda M^2 \sim B^2M^2 \sim K M^2.
\end{equation} 
In this work, we shall take $\lambda$ to be $\lambda\sim10^{-8}-10^{-5}$. 

If this were to arise from the cosmological constant $\Lambda\approx10^{-46}$ km$^{-2}$ \cite{aghanim2020planck}, then this range of values would correspond to a central black hole mass $M$ ranging from $\approx 10^{17}-10^{20} M_{\odot}$. This is significantly heavier than typical galactic supermassive black holes ($\gtrsim 10^6 M_\odot$) and is in fact much heavier than the heaviest black hole ($\approx 10^{10} M_\odot$) observed in the universe $\cite{mehrgan201940}$. This confirms earlier expectations. An SdS parameter of \emph{purely} cosmological origin would be negligibly small, $\lambda \sim 10^{-34}-10^{-28}$, and irrelevant to EMRI dynamics. 
% In the cosmological constant scenario, we note that a non-negligible (nondimensional) $\lambda := \Lambda M^2 \sim 10^{-5}$, for $\Lambda\approx10^{-46}$ km$^{-2}$ \red{[shouldn't this be $10^{-58}$]}, would imply a central black hole mass of $M \approx 10^{20} M_{\odot}$, which is physically unrealistic in our universe. 
% But if we look at the sample environments in Eqs. (\ref{eq:rochepotential}-\ref{eq:ppmetric}), we can roughly map $\lambda$ into the environmental parameters $B$ and $\Omega$:
% \begin{equation}
%     \lambda^{1/2} \leftrightarrow (B,\Omega).  \,\,\,\red{\Lambda^{1/2} \leftrightarrow (B,\Omega).}
% \end{equation}
% \red{[Are these environmental parameters nondimensional? Dapat yata $\lambda^{1/2} = BM = \Omega M $]}. 

However, an SdS parameter of astrophysical origin, if it were to arise from a magnetic field or an external tidal perturbation, may not be as miniscule. Consider, for instance, how $\lambda$ depends on an external magnetic field, $B$. Expressed in terms of the mass of the central supermassive, $M_{BH}$, we have,
\begin{align}
    B_{\text{real}}=&10^9 \text{ Gauss} \left(\frac{\lambda}{10^{-8}}\right)^{1/2}\left(\frac{10^6 M_\odot}{M_{BH}}\right).
\end{align}
From these, we see that SdS parameter values in the range $\lambda\sim10^{-8}-10^{-5}$ correspond to magnetic field strengths $B\approx10^{-4}-10^{-2}$ $M^{-1}$ ($\sim 10^9-10^{11}$ Gauss). These field strengths have been used in exploratory theory papers such as \cite{kovacs2011accretion,konoplya2006particle}. For more realistic magnetic field strengths $B=10^1-10^{4}$ Gauss  ($\sim 10^{-19}-10^{-16}$ $M_{\odot}^{-1}$), in the inner accretion disk hosted by a supermassive black hole (SMBH) of mass $10^9M_{\odot}$ \cite{event2021first}, the SdS parameter ranges from $\lambda\sim10^{-20}-10^{-14}$, which is still tiny but already much larger than what one would get from a cosmological constant. 

Achieving a non-negligible $\lambda \sim 10^{-8}$ requires the extreme conditions of an \emph{ultramassive} black hole ($M \sim 10^{11} M_\odot$) and a robust magnetic field ($B \sim 10^4$ G). While such fields might arise from the surrounding environment, they are constrained by a theoretical ``Eddington magnetic limit," which caps the field strength at approximately $10^4$ G for a $10^9 M_\odot$ black hole \cite{beskin2009mhd}.

Finally, consider the tidal field in Eq. \eqref{eq:tidalpot}, for which $K$ may be parametrized as follows:
\begin{equation}
    K M_{BH}^2\sim \frac{M_{ext}}{R_{ext}^3} M_{BH}^2\sim \left(\frac{M_{BH}}{R_{ext}}\right)^3\left(\frac{M_{ext}}{M_{BH}}\right)\sim\lambda.
\end{equation}
A value of $\lambda = 10^{-6}$ corresponds to an external perturber ($M_{\text{ext}} \sim M_{\text{BH}}$) at $100 M_{\text{BH}}$, a configuration physically plausible in binary SMBH systems.

In this work, we investigate how the SdS parameter, interpreted here as an environmental perturbation, impacts binary inspirals. Building on the weak-field flux formulas from \cite{hoque2019quadrupolar}, we model the orbital evolution of an EMRI-like system driven by gravitational-wave radiation reaction. We assume the compact object survives tidal forces to complete its inspiral and plunge, with the background spacetime effectively described by the Schwarzschild-de Sitter metric. We analyze both conservative and dissipative effects of the SdS parameter on the binary's dynamics. 

This paper is the first in a series assessing the SdS parameter $\lambda$ and its effects on EMRI dynamics and their gravitational waves. Here, we focus primarily on the weak-field regime, leaving the extension to the strong-field regime for a later paper.

We analyze generic orbits ranging from circular (with eccentricity $e=0$) to unbound $(e\rightarrow 1)$. Although \cite{hoque2019quadrupolar} had previously examined the effect of radiation on orbital evolution in de Sitter space, they did not modify the full Keplerian dynamical quantities to include the de Sitter terms nor fully explore how the cosmological constant affects orbital dynamics. We extend their work by providing a complete treatment of the weak-field, adiabatic dissipative dynamics driven by radiation reaction for a compact binary. We also present preliminary results on strong-field energy fluxes, offering a check on the validity of the weak-field flux approximations. Additionally, we explore other conservative effects of the SdS parameter on the binary, such as the separatrix and the parameters of the last stable orbit, as well as the influence of radiation reaction on the periapsis advance, which is also modulated by the SdS parameter. Finally, we calculate adiabatic waveforms for a continuously evolving orbit due to radiation reaction in Schwarzschild-de Sitter spacetime, highlighting shifts in the waveform's amplitude and phase.

The paper is divided into five parts. Section \ref{sec:sec2} contains an elementary review of bound timelike geodesics in Schwarzschild-de Sitter spacetime, which represents the conservative trajectories followed by particles. We parametrize bound orbits using either their energy and angular momentum $(E,L)$ or their semilatus rectum and eccentricity $(p,e)$. An important result of this section is the effect of the $\lambda$ parameter on the separatrix surface, the boundary in parameter space separating unstable and stable bound orbits. In Section \ref{sec:sec3}, we build on the results of Hoque and Aggarwal to describe the influence of the cosmological constant on orbital evolution due to radiation reaction in the weak-field, adiabatic limit. We check the accuracy of these results with a preliminary calculation of the fluxes derived from black hole perturbation theory in Schwarzschild-de Sitter spacetime. Additionally, we provide sample orbital shapes to illustrate the impact of the $\lambda$ parameter on the dynamics of the binary system. Section \ref{sec:sec4} focuses on the calculation of adiabatic waveforms, demonstrating how the $\lambda$ parameter modifies the amplitude and phase of gravitational waveforms for both circular and eccentric binaries. We also explore the constraints that the adiabatic approximation places on the mass ratios and the $\lambda$ parameter, determining the conditions under which these approximations remain valid. Lastly, in Section \ref{sec:sec5} we investigate the effect of the $\lambda$ parameter on the plunge times of generic orbits and examine how this relates to the timescales for detecting gravitational wave sources.

We adopt the usual conventions by Misner, Thorne and Wheeler, which uses the mostly plus metric signature and the geometrized units of naturalizing the constants $G=c=1$. For astrophysical applications, these naturalized constants applied to conventional units translate to $1\,M_{\odot}\simeq 5\times10^{-6}$ s $\simeq1.5$ km and $1$ pc $\simeq10^{8}$ s.

\section{Orbital mechanics in Schwarzschild-de Sitter spacetime}
\label{sec:sec2}

When a nonrotating black hole is placed within an expanding universe, the resulting spacetime may be modeled as Schwarzschild-de Sitter (Eq.  \eqref{eq:metricsds}), also known as Kottler spacetime \cite{kottler1918physical}. This solution represents the simplest black hole solution to Einstein's field equations that includes a cosmological constant \cite{stuchlik1999some} and has the following form:
%The first exact solution to the Einstein field \blue{equation} representing a black hole immersed in an \blue{expanding} universe is known as the Schwarzschild-de Sitter (SdS) solution \cite{de1916einstein}:
\begin{equation}
	g_{\mu\nu}=\text{diag}(-f(r),\,f(r)^{-1},\,r^2,\,r^2\sin^2\theta),
	\label{eq:metricsds}
\end{equation}
with the metric function, $$f(r)=1-\frac{2M}{r}-\frac{1}{3}\Lambda\,r^2.$$
This metric is a special case of the McVittie solution \cite{mcvittie1956general} for positive spatial curvature and a constant Hubble parameter, $H(t)=H_0$. The SdS metric satisfies the vacuum Einstein field equation with a positive cosmological constant, $\Lambda$:
\begin{equation}
	G_{\mu\nu}-\Lambda\,g_{\mu\nu}= T_{\mu\nu}=0.
	\label{eq:efelambda}
\end{equation}
When $0 < \Lambda < \Lambda_{\textrm{max}} := (3M)^{-2}$ (or $1/\sqrt{\Lambda} > 3M$), the spacetime features two horizons (i.e., the roots of $f(r)$ in Eq. \eqref{eq:metricsds}): the event horizon, which marks the boundary of a black hole, and the cosmological horizon, which sets the scale of the observable universe. When $\Lambda > (3M)^{-2}$, the spacetime is not static for any $r$ (i.e. there no longer exists a static patch), and the SdS spacetime is no longer a good representation of a black hole immersed in an expanding universe. We shall only be interested in the former case, and we will focus on orbits between the event horizon and cosmological horizon. 

%\blue{The background curvature of the de Sitter universe also fundamentally modifies the definition of gravitational flux at infinity, a property that has been rigorously analyzed in recent literature \cite{ashtekar2014asymptotics,ashtekar2016gravitational,bishop2016gravitational, bonga2017power,ashtekar2017implications,ashtekar2015asymptotics,ashtekar2015asymptotics2,date2016gravitational,date2017cosmological,hoque2018propagation,bonga2023gravitational}. }

It is convenient to use dimensionless variables $\tilde{r} := r/M$ and $\lambda := M^2\Lambda/3$. For both horizons to exist, as required in the preceding paragraph, we thus require $\lambda < \lambda_{\textrm{max}} := 1/27$. A value of $\lambda$ approaching this critical value would imply an unrealistic scenario in which a black hole is comparable in size to the entire universe.
Based on our best estimates for the cosmological constant and the Hubble parameter \cite{scolnic2025hubble}, we instead find that $\lambda \ll 1$. 
We are thus fully justified in exploiting $\lambda$ as a small parameter in all of our calculations. As an example, consider the locations of the horizons, which are obtained by solving for the roots of the metric function:
\begin{eqnarray}
	\label{eq:horizoneq}
	f(\lambda;r)=1-\frac{2}{r}-\lambda r^2=0.
\end{eqnarray}
Up to $\mathcal{O}(\lambda^2)$, the roots are
\begin{eqnarray}
	\label{eq:sdsmetricroots}
	r_{h}&=&2+8\lambda+\mathcal{O}(\lambda^2)\\
	r_c&=&-1+\lambda^{-1/2}-\frac{3}{2}\lambda^{1/2}-4\lambda+\mathcal{O}(\lambda^{2})\\
	r_{o}&=&-1-\lambda^{-1/2}+\frac{3}{2}\lambda^{1/2}-4\lambda+\mathcal{O}(\lambda^{2})
\end{eqnarray}
with $r_{h}$ as the horizon of the SdS black hole, $r_c$ as the cosmological horizon and  $r_o$ as  an unphysical negative root. We further see that requiring  that the black hole is inside the universe,  $r_{h}<r_{c}$, implying $\lambda\lesssim 0.06$, satisfies the stricter condition $\lambda < \lambda_{\textrm{max}}:=1/27$. 

\subsection{Timelike geodesics in Schwarzschild-de Sitter spacetime}
\label{sec:sec2.1}

Test particles with mass $\mu$ move along timelike geodesics. In any static and azimuthally symmetric spacetime, geodesics will possess conserved quantities 
\begin{eqnarray}
	\label{eq:killingE}
	\xi_tu^t&=&-f(r)\frac{dt}{d\tau}=\tilde{E}/\mu\\
	\label{eq:killingL}
	\eta_\phi u^\phi&=&r^2\sin^2(\theta)\frac{d\phi}{d\tau}=\tilde{L}/\mu,
\end{eqnarray}
which are associated with the timelike and azimuthal Killing vectors, $\xi^\mu$ and $\eta^\mu$, respectively. The conserved quantities may be interpreted as the energy and angular momentum conserved during time translation and azimuthal translation of a particle. Inserting Eqs.~\eqref{eq:killingE} and \eqref{eq:killingL} into the normalization condition of the four-velocity, $u^{\mu}u_{\mu}=-1$, we obtain a radial equation of motion familiar from elementary orbital mechanics:
\begin{eqnarray}
	\label{eq:energyeq}
	\left(\frac{dr}{d\tau}\right)^2&=&\tilde{E}^2-V_{\text{eff}}(\tilde{L}, \lambda; r),\\
	V_{\text{eff}}(\tilde{L}, \lambda; r)&=&f(\lambda;r)\left(1+\frac{\tilde{L}^2}{r^2}\right). 
\end{eqnarray} 
The equation is uniquely characterized by an ``effective" potential, $V_{\text{eff}}$. In the SdS spacetime, the shape of the effective potential $V_{\text{eff}}$ depends on parameters $(\tilde{L},\lambda)$.
More complete treatments of geodesics in the SdS spacetime can be found in \cite{stuchlik1999some,hackmann2008complete,yanchyshen2025gaussian}.

\subsection{Bound orbits in Schwarzschild-de Sitter}
\label{sec:sec2.2}

Bound orbits in static and spherically symmetric spacetimes are typically described by their energy and angular momentum. Another common and often more informative parametrization of bound orbits uses orbital parameters  \cite{cutler1994gravitational}, such as the semi-latus rectum, $p$, and eccentricity, $e$. One imagines an ``instantaneous'' (or osculating) Keplerian orbit  
\begin{equation}
	\label{eq:radialeq}
	r(\psi)= \dfrac{p\, m}{1+e \cos \psi},
\end{equation}
representing the true trajectory. Here, $m=m_1+m_2$ is the total mass of the orbiter (with mass $m_1$) and the central object (with mass $m_2$) in a reduced two-body problem. The true anomaly, $\psi$, is the angle swept by the orbiter around the central object, with $\delta$ as the argument at the periapsis. The coordinate azimuthal angle $\phi$ thus equals $\phi=\psi+\delta$. For this paper we will be considering a compact object-black hole (CO-BH) system with reduced mass $\mu=m_1m_2/m$. This parametrization is particularly useful when studying processes that induce slow changes in the energy and angular momentum, such as radiation reaction. In this context, the adiabatic evolution can be pictured as a continuous sequence of Keplerian orbits of different sizes and shapes. 

For a CO-BH system with the compact object much smaller than the central black hole (EMRI), the reduced mass will be approximately equal to the mass of the compact object solely, $\mu=\mu_{CO}$, and the total mass will be approximately equal to mass of the black hole, $m=M_{BH}=M$. We shall rescale $r$ in Eq.~(\eqref{eq:radialeq}) by the total mass, i.e. $r\rightarrow r/M$, henceforth.

The transformation from $(\tilde{E},\tilde{L})$ to $(p,e)$ is obtained starting with the orbit equation
\begin{equation}
	\label{eq:radialangulareq}
	\left(\frac{1}{r^2}\frac{dr}{d\phi}\right)^2=\frac{1}{\tilde{L}^2}\left(\tilde{E}^2-V_{\text{eff}}\right).
\end{equation}
A bound orbit will have two turning points, $r_a$ and $r_p$, respectively corresponding to the maximum (apoapsis) and minimum distance (periapsis) of the small compact object from the black hole.  The radial equation of motion, Eq.~\eqref{eq:energyeq}, evaluated at the two turning points $r=r_{a(p)}$ gives us two algebraic equations that can be solved for $(\tilde{E},\tilde{L})$, giving expressions in terms of $(r_a,r_p)$. In turn, $r_a$ and $r_p$ can be expressed in terms of $p$ and $e$ using Eq.~(\eqref{eq:radialeq}):
\begin{eqnarray}
	\label{eq:apoapsis}
	r_a&=&\max[r(\psi)]=r(\psi=\pi)=p(1-e)^{-1}\\
	\label{eq:periapsis}
	r_p&=&\min[r(\psi)]=r(\psi=0)=p(1+e)^{-1}.
\end{eqnarray}
For the Schwarzschild metric, one can work out explicit expressions for  $(\tilde{E}(r_a,r_p),\tilde{L}(r_a,r_p))$ in terms of $(p,e)$. However, for the more complicated SdS metric, this procedure does not yield explicit expressions. Recognizing that the SdS parameter is small, we shall content ourselves with working out approximate expressions valid only to first order in $\lambda$. In terms of the scaled radial variable $r\rightarrow r/M$ and scaled SdS parameter $\lambda\rightarrow(1/3)M^2\Lambda$, the conserved quantities are given by:
\begin{eqnarray}
	\label{eq:sdsL}
	L^2&=&\left(\frac{p^2}{p-3-e^2}\right)\left(1-\lambda\frac{p^3 }{(1-e^2)^2}+\mathcal{O}(\lambda^2)\right),\\
	\label{eq:sdsE}
	E^2&=&\left(\frac{(p-2)^2-4e^2}{p(p-3-e^2)}\right)\nonumber\\
    &&\times\left(1-\lambda\frac{p^3(p-2+(p-6)e^2)}{((p-2)^2-4e^2)(e^2-1)^2}+\mathcal{O}(\lambda^2)\right)\nonumber\\
\end{eqnarray}
where we have similarly non-dimensionalized the energy and angular momentum according to $E^2\rightarrow(\tilde{E}/\mu)^2$ and $L\rightarrow(\tilde{L}/M\mu)$. 

We note that the above construction of the parameter transformation forbids us from including circular orbits $e=0$ .   Even though the resulting expressions $E(p,e)$ and $L(p,e)$ are regular in the limit $e\rightarrow0$,  two distinct bound turning points (Eqs.  \eqref{eq:apoapsis} and \eqref{eq:periapsis} which become degenerate at $e\rightarrow 0$) are required to keep the transformation $(E(r_a,r_p),L(r_a,r_p))\rightarrow(E(p,e),L(p,e))$ defined, thus restricting the eccentricity to be within $0<e<1$.

The effect of the SdS parameter on the energy and angular momentum can be understood more clearly if we translate the equations in terms of the classical solutions to the Keplerian two-body problem. Expressing the energy equation in terms of the Newtonian binding energy $\varepsilon=(E^2-1)/2$, we get the $\lambda$ corrections for weak-field slow-moving orbits:
\begin{eqnarray}
	\label{eq:largepL}
	L^2&=&p-\lambda\frac{ p^4}{\left(1-e^2\right)^2} + O(\lambda^2 p^3)\\
	\label{eq:largepE}
	\varepsilon&=&-\frac{1-e^2}{2 p}-\lambda\frac{\left(e^2+1\right)   p^2}{\left(1-e^2\right)^2} + O(\lambda^2 p).
\end{eqnarray}

The relations above can be used to put limits on the possible orbital parameters of bound orbits in SdS. We see from these Eqs. \ref{eq:largepL} and \ref{eq:largepE} that the SdS parameter makes the binding energy of an orbit more negative, making more eccentric orbits tighter, while decreasing the angular momentum required to reach a certain eccentricity value. 
We see therefore, that bound orbits in a $\lambda>0$ spacetime require more energy to separate.
Looking at the angular momentum in Eq. \eqref{eq:largepL}, and requiring it to be always positive along with $p$, we could also limit the possible values of the eccentricity of an orbit. A particular SdS orbit can therefore only have eccentricity values, $0 < e < 1-\sqrt{\lambda} p^{3/2}$. 
These effects becomes more pronounced at higher eccentricities due to the $(1-e^2)$ singular factor. 

We can also get the Newtonian limit of geodesics in the SdS spacetime. 
In isotropic coordinates $\lbrace T,R,\theta,\phi\rbrace$, the areal radius $r$ is related to the isotropic radius $R$ by the function
\begin{equation}
    r=R\left[\left(1+\frac{M}{2R}\right)^2+\Lambda\frac{ R^2}{4}\left(1+\frac{2M}{R}\right)\right]+\mathcal{O}(\Lambda^2)
\end{equation}
and the SdS line element up to leading order in $\Lambda$ is 
\begin{equation}
	ds^2=-A(R)^2dT^2+B(R)^{2}\left(dR^2+R^2d\Omega^2\right),
	\label{eq:isocoords}
\end{equation}
with the metric functions
\begin{eqnarray}	
	\label{eq:tempiso}
	A(R)^2&=&\left(\frac{1-\frac{M}{2R}}{1+\frac{M}{2R}}\right)^2
	-\frac{\Lambda}{3}\Bigg(R^2\left(1+\frac{M}{2R}\right)^4\nonumber\\
    &&-\frac{MR}{2}\left(\frac{1+\frac{2M}{R}}{\left(1+\frac{M}{2R}\right)^4}\right)\Bigg)+\mathcal{O}(\Lambda^2)\\
	\label{eq:spatialiso}
	B(R)^2&=&\left[\left(1+\frac{M}{2R}\right)^2+\frac{\Lambda R^2}{12}\left(1+\frac{2M}{R}\right)\right]^2+\mathcal{O}(\Lambda^2)\nonumber\\
\end{eqnarray}
and $d\Omega^2$ as the line element on the unit 2-sphere. Note that we have brought back the dimensional parameters $(M,\Lambda)$, which means that the coordinate $R$ has units of length. 

We consider the region simultaneously far from the black hole event horizon and the cosmological horizon, i.e. $2M \ll R \ll \Lambda^{-1}$, which we assume to exist because $2M \ll \Lambda^{-1}$. In this region, the metric components are that of Minkowski with corrections:  
\begin{eqnarray}
	\label{eq:timeisometric}
	g_{00}&=&-1+\frac{2M}{R}+\frac{\Lambda}{3}R^2\left(1+\frac{5M}{2R}\right)
    \nonumber\\
    &&+\mathcal{O}(\Lambda^2)+\mathcal{O}\left(\left(\frac{2M}{R}\right)^2\right)\\
	\label{eq:spatisometric}
	g_{jk}&=&\delta_{jk}\left(1+\frac{2M}{R}-\frac{\Lambda}{3}R^2\left(\frac{1}{2}+\frac{3M}{2R}\right)\right)
    \nonumber\\
    &&+\mathcal{O}(\Lambda^2)+\mathcal{O}\left(\left(\frac{2M}{R}\right)^2\right).
\end{eqnarray}
The spatial acceleration of a free-falling particle can then be calculated to be 
\begin{eqnarray}
	\label{eq:spataccelsds}
	\frac{d v^{j}}{d T}&=&\partial_jU+\frac{\Lambda}{3}\Bigg[R+\frac{1}{2}U+\frac{7}{4}R^2\partial_jU
    \nonumber\\
    &&-\frac{v^2}{2}\left(R+8RU+4R^2\partial_jU\right)\Bigg]\nonumber\\&&+\mathcal{O}(\Lambda^2)+\mathcal{O}\left(\frac{2M}{R}\right)^2
\end{eqnarray}
which shows the Newtonian spatial acceleration with a potential, $U=2M/R$, along with the leading order correction due to the SdS parameter. 
\begin{figure*}
    \begin{minipage}{0.48\linewidth}
        \centering
        \includegraphics[width=\linewidth]{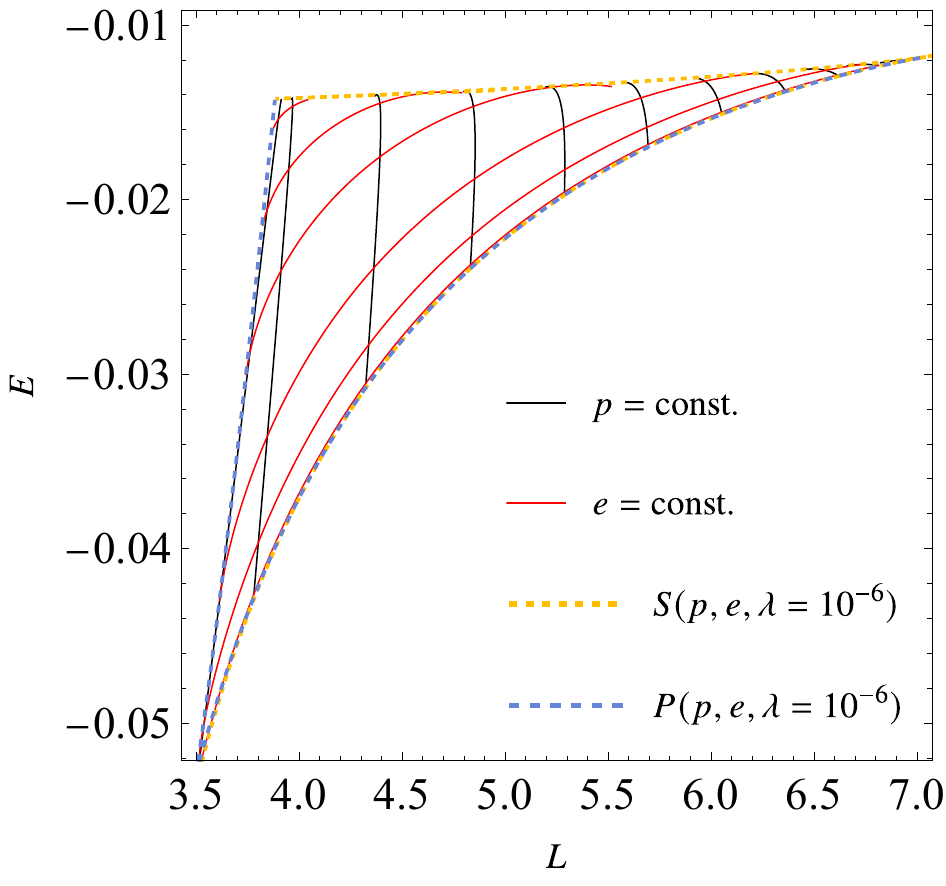}
        \text{(a)}
    \end{minipage}
    \begin{minipage}{0.48\linewidth}
        \centering
       \includegraphics[width=0.95\linewidth]{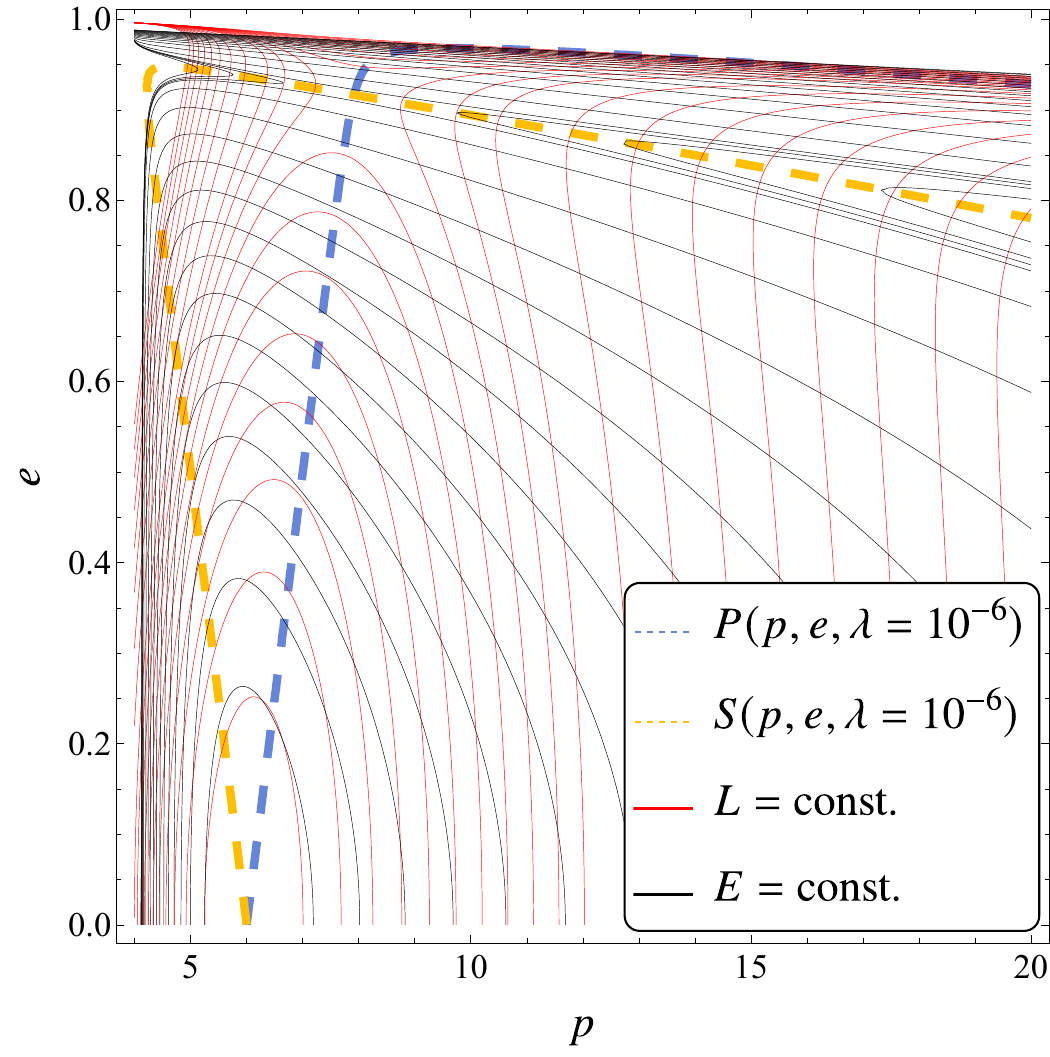}
        \text{(b)}
    \end{minipage}
	\caption{(a) Constant $p$ and $e$ curves  in SdS $(E,L)$ space with black lines as constant $p$ curves increasing to the right and red lines as constant $e$ curves increasing upwards. (b) Constant $E$ (black) and $L$ curves (red) in $(p,e)$ space in SdS spacetime. The dashed curves are the separatrices (Eqs. \eqref{eq:separatrix} and \eqref{eq:scattersep}) of bound orbits or the curves where the determinant of the Jacobian of transformation from $(p,e)\rightarrow(E,L)$ becomes zero.}
    \label{fig:constant-El-in-pe}
\end{figure*}

Finally, we show how constant $(p,e)$ curves live in the $(E,L)$ space and vice versa both in Schwarzschild and SdS in Fig. \ref{fig:constant-El-in-pe}a and \ref{fig:constant-El-in-pe}b . We show in Fig. \ref{fig:constant-El-in-pe}a the constant $p$ and $e$ curves in the $(E,L)$ space under the constraint of the separatrices that will be covered in the next subsection. The $e=0$ curve serves as a lower boundary for the possible $(E,L)$ for a bound orbit. In Fig. \ref{fig:constant-El-in-pe}b we show the constant $E$ and $L$ curves in $(p,e)$ space. We see here that as we approach the $e=0$ limit, the $(E,L)$ curves become parallel to each other restricting the mapping of the $(p,e)$ to $(E,L)$ transformation. We show this further analytically by looking at the points where the Jacobian becomes degenerate or when the determinant of the transformation becomes zero. The transformation $(E,L)\rightarrow(p,e)$ is facilitated by the matrix $J_{ij}$ wherein $J_{ij}=\partial_{i}\rho_{j}$ with $i$ referring to the elements of $(p,e)$ and $\rho_j$ referring to the elements of $(E,L)$. The zeroes of the determinant $|J_{ij}|=0$ are located at $e=0$ as well as in regions that will be shown to be the separatrices $p_{\text{inner}}$ and $p_{\text{outer}}$ in the $(p,e)$ space which will be discussed in Section \ref{sec:sec2.3}. This shows that the transformation breaks down at these regions.

\subsection{Separatrices for bound orbits}
\label{sec:sec2.3}

Compact binaries serve as sources of continuous gravitational wave signals during their inspiral phase, terminating at the merger and ringdown. It is therefore useful to reduce to reduce the $(p,e)$ parameter space, and further the $(L,E)$ parameter space, to those regions corresponding to bound orbits, separating them from parameters that lead to plunging or scattering trajectories. 

Although the position of the last stable orbit (LSO) in Schwarzschild spacetime is already textbook knowledge found in many core general relativity books \cite{hartle2003gravity,shapiro2008black,carroll2019spacetime}, the position in the full orbital parameter space (not only for $e=0$) has only recently been determined. For generic orbits in Schwarzschild, a paper by Poisson et al. \cite{cutler1994gravitational} derived the fundamental $p=6+2e$ curve for marginally stable orbits. This curve however will be modified as a spacetime obtains other parameters. For rotating black hole spacetimes like the Kerr, numerical approximations of this separatrix are sufficient \cite{amaro2013role} but a more recent work obtained an analytic expression \cite{suzuki1998innermost,levin2009homoclinic,stein2020location}. For charged black holes, the innermost stable circular orbit \cite{pugliese2011motion} and sufficient conditions for the position of marginally stable orbits has been derived \cite{misra2010rational,hackmann2013charged} but a generalized expression for the last stable orbit is yet to be found. In spacetimes with a cosmological constant, a paper by \cite{stuchlik1983motion,stuchlik1999some,kunst2015isofrequency} and more recently with an added contribution of a spin from \cite{zhang2019motion} presented the condition for marginally stable generic orbits.

As discussed briefly in the last subsection (Section \ref{sec:sec2.2}), the zeroes of the Jacobian of the transformation $(E,L)\rightarrow (p,e)$ coincide with what we will be calling the \textit{separatrix curves} for bound and unbound orbits in the $(p,e)$ space. We will show two kinds of separatrices which will be a feature of the SdS orbital dynamics. The usual separatrix in the Schwarzschild spacetime is a plunging separatrix in which orbital parameters outside of the region bounded by the curve eventually lead to plunging orbits. We show in this section a scattering separatrix wherein orbital parameters outside the region bounded by this curve will lead to a scattering orbit up to infinity. We show the conventional way to derive these curves in this section which is to analyze the family of points that merges the turning points with the unstable maxima in the effective potential. Alternatively, the separatrix polynomials can also be derived by finding the zeroes of the Jacobian of the transformation $(E,L)\rightarrow (p,e)$ which means that at the separatrix itself, the orbital parameters will not be valid to describe the dynamics of the orbit anymore.

\subsubsection{Plunging separatrix}

\begin{figure*}[htbp]
	\centering
	\begin{minipage}{0.45\textwidth}
		\includegraphics[width=\linewidth]{"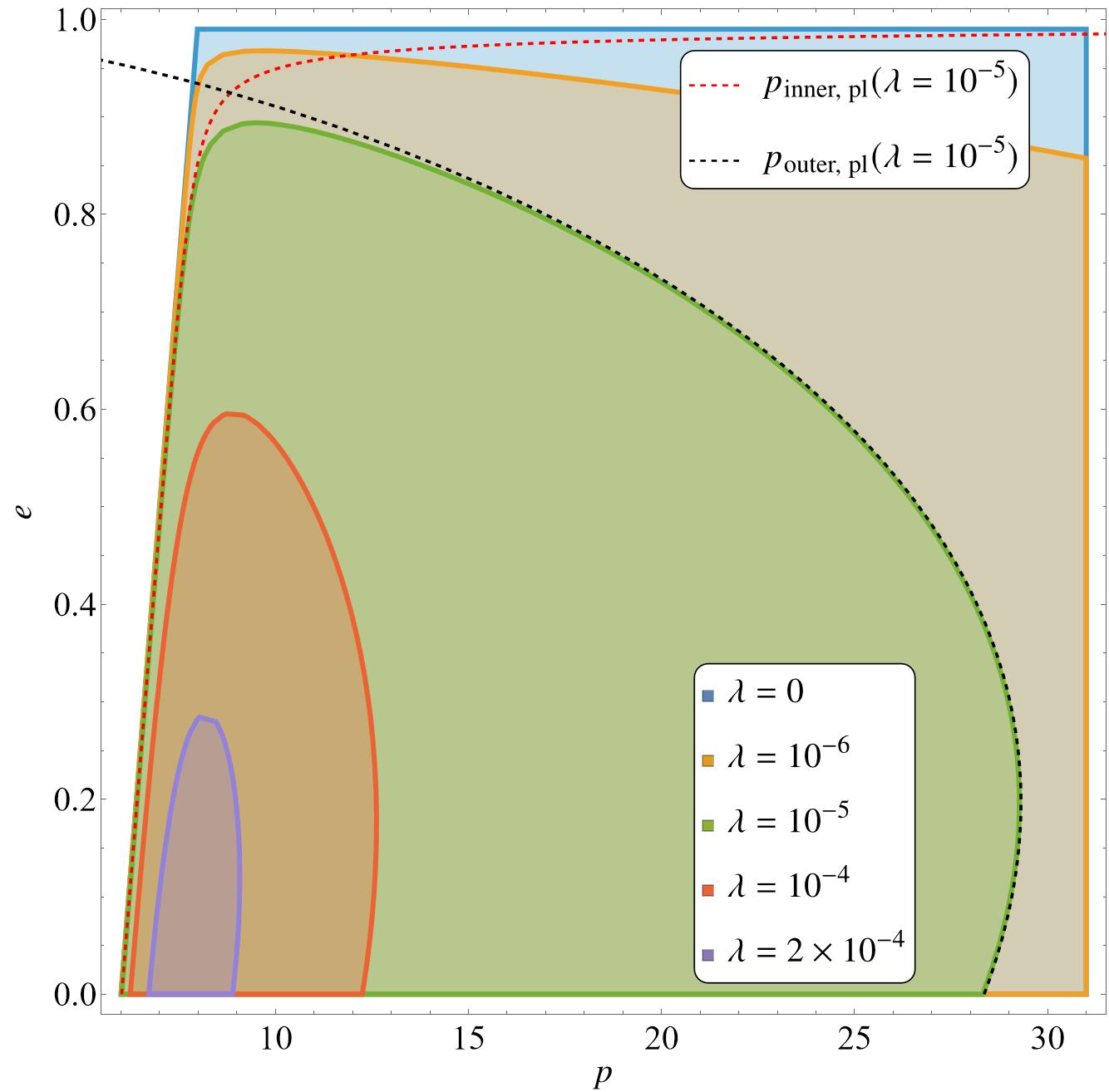"}
        \text{(a)}
	\end{minipage}
	\hspace{0.3cm}
	\begin{minipage}{0.45\textwidth} 
		\includegraphics[width=\linewidth]{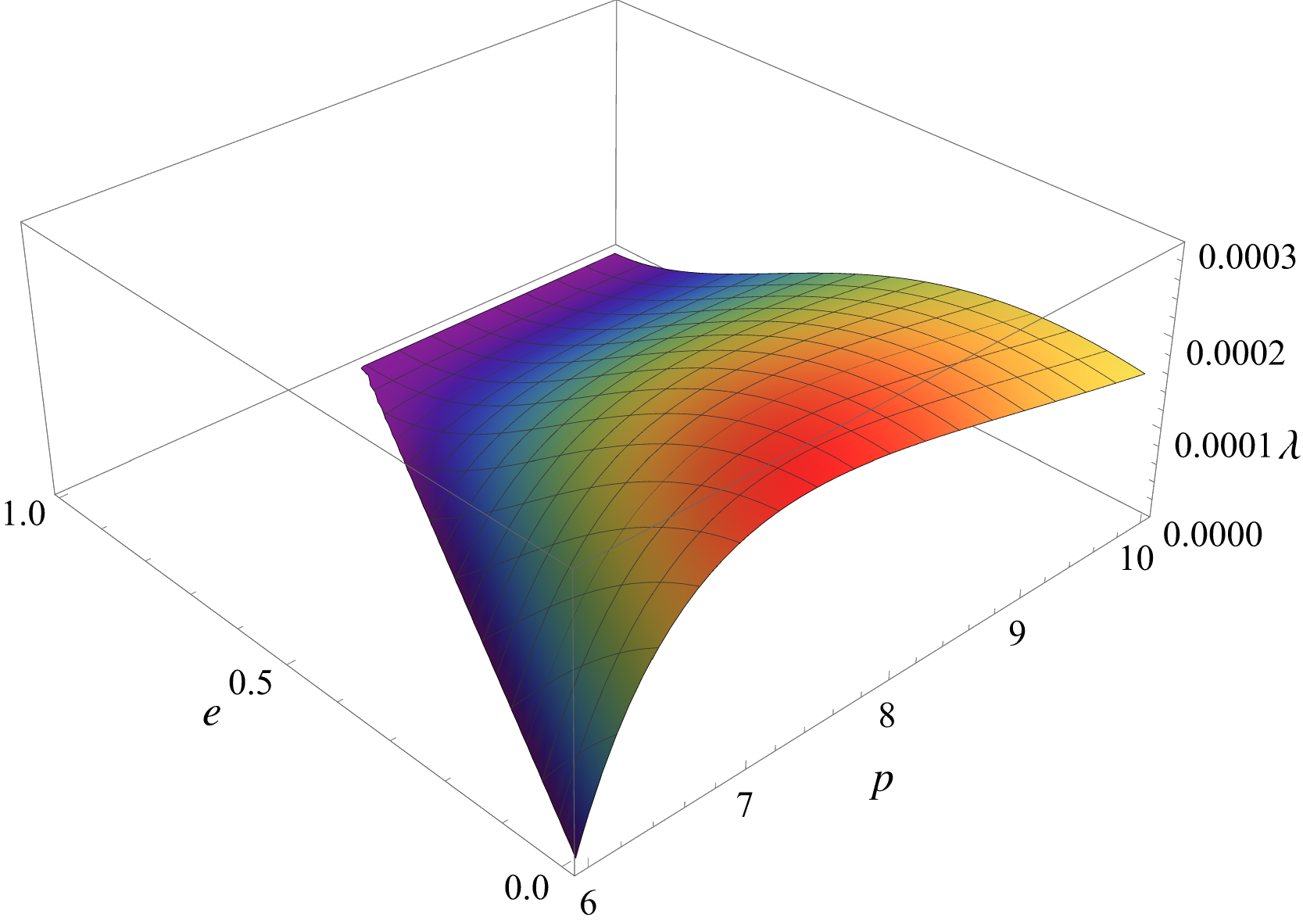}
        \text{(b)}
	\end{minipage}
	\caption{The separatrix equation $P(p,e,\lambda)$  (Eq. \eqref{eq:separatrix}) in $(p,e)$ space. In (a) the contours have varying $\lambda$ with the filled region as the allowable parameter space for stable bound orbits.  Introducing a $\lambda>0$ makes most eccentric orbits unbound. We see in the red and violet curve that a sufficiently high SdS parameter would also make orbits with high separation (larger orbits) also unbound. In (b) we show a continuous variation of $\lambda$ onto the phase space $(p,e)$ wherein the marginally bound orbits are the points lying on the surface.}
	\label{fig:separatrix}
\end{figure*}

In this part, we show a representation of the plunging separatrix curve, or the location of the last stable orbit (LSO), of marginally bound orbits in $(p,e)$ space to show at which point in the parameter space is the region of orbital plunge. The locus of all parameter values that will create a merged plunge and bound region in the effective potential will be the plunging separatrix in the $(p,e,\lambda)$ parameter space. The maxima of the potential must be located at the periapsis, and is expressed mathematically as \cite{stein2020location}:
\begin{eqnarray}
	\label{eq:potentialderivative}
	\dfrac{d\,V_{\text{eff}}}{dr}\Big|_{r=r_p}=0.
\end{eqnarray}
This method of deriving the separatrix means that we are considering that at the periapsis, $r=r_p=p(1+e)^{-1}$, the bound orbiter will be marginally stable and thus will either plunge or scatter away depending on the evolution of the orbit. Together with the energy and angular momentum for bound orbits (Eqs. \eqref{eq:sdsL} and \eqref{eq:sdsE}), we derive an equation for the separatrix:
\begin{align}
	\label{eq:separatrix}
	P(p,e,\lambda)=(&e-1)^2 (e+1)^3  (-p+6+2e) \nonumber \\ &+\lambda\; p^3 \left(e^3-5 e^2-5 e+4 p-15\right)
    \nonumber\\
    &\quad+\mathcal{O}(\lambda^2)=0
\end{align}
which reduces to the usual Schwarzschild separatrix $p=6+2e$ at $\lambda=0$.  
We see in  Fig. \ref{fig:separatrix} the evolution of the separatrix from Schwarzschild to SdS both with a fixed $\lambda$ in a 2D contour plot, and a continuous effect of $\lambda$ in a 3D contour plot. The main effect of the SdS parameter is to eliminate most eccentric orbits and larger orbital sizes in the stable bound parameter space. This can be already seen in the energy and angular momentum, (Eqs. \eqref{eq:sdsL} and \eqref{eq:sdsE} ), wherein more eccentric and larger orbits are affected greater by the $\lambda$ term, with the SdS parameter dominating making the orbits unbound. The maximum $\lambda$ for bound orbits to occur is $\lambda=4/16875\sim 2.37\times 10^{-4}$ at $(p=7.5,e=0)$ which produces a single stable bound orbit $\left(p=7.5,e=0,\lambda=4/16875\right)$.

\begin{figure*}[htbp]
	\centering
	\begin{minipage}{0.45\textwidth}
		\includegraphics[width=\linewidth]{"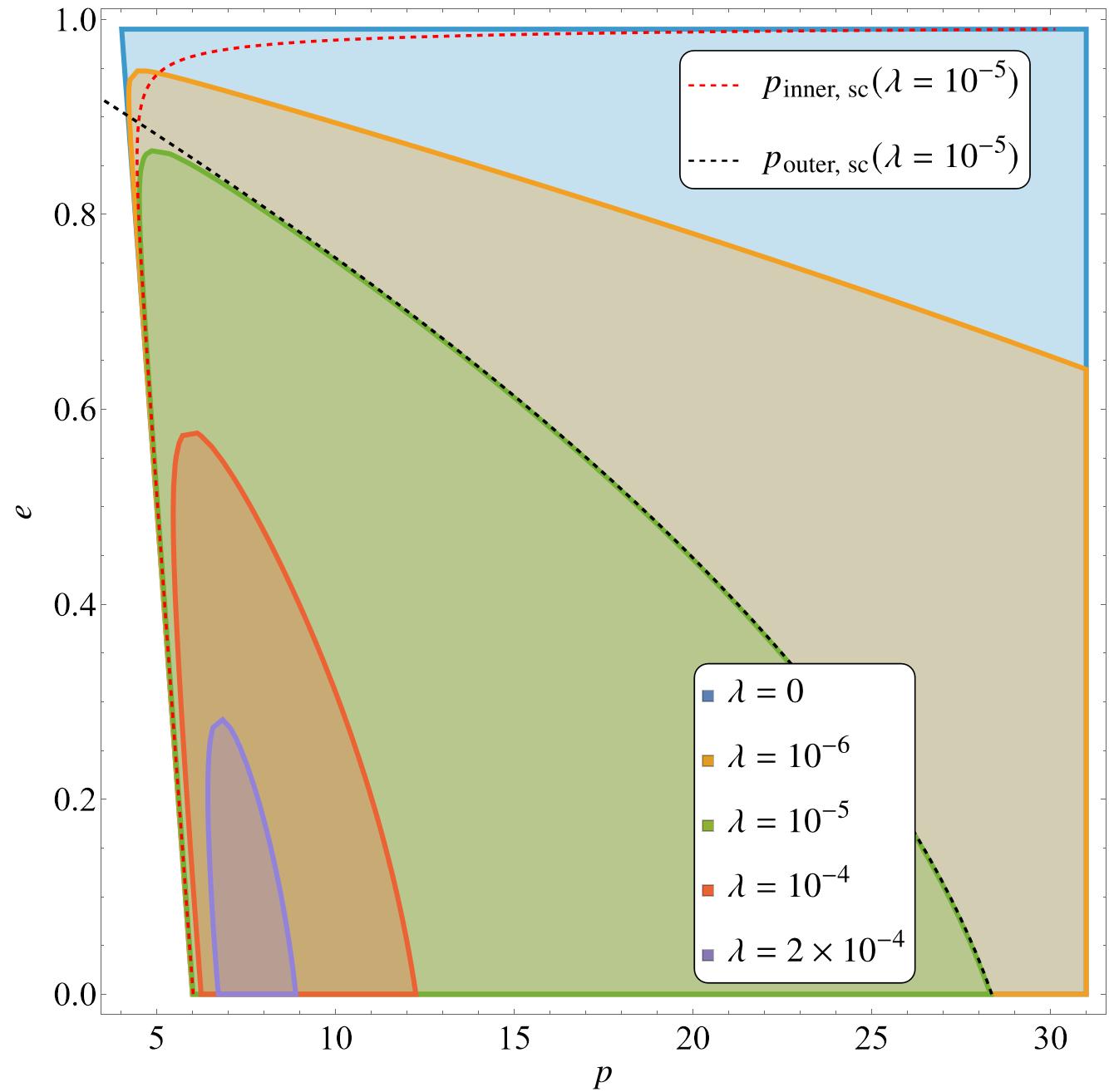"}
	\end{minipage}
	\hspace{0.3cm}
	\begin{minipage}{0.45\textwidth}
		\includegraphics[width=\linewidth]{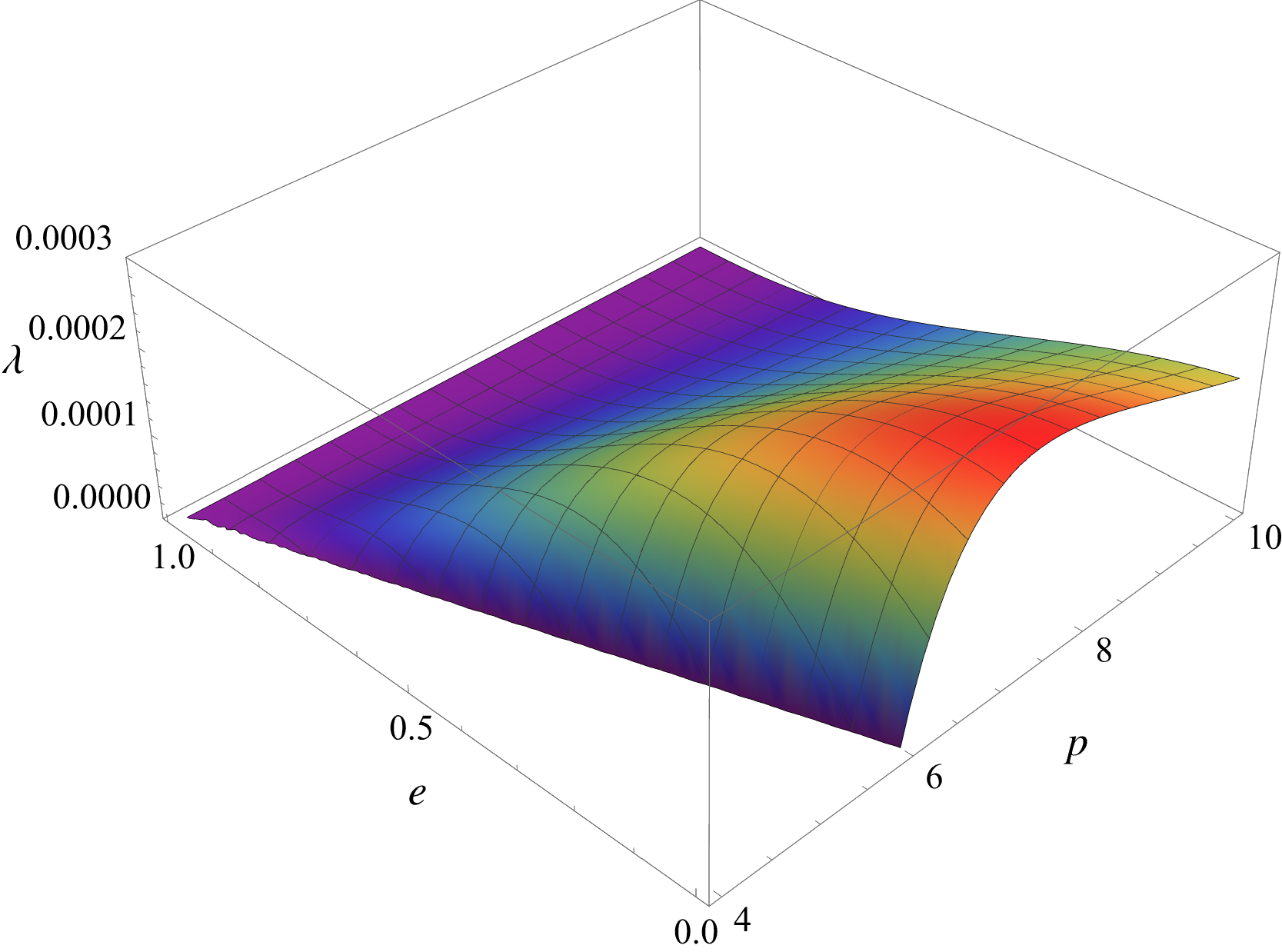}
	\end{minipage}
		\caption{Similar contour plot and 3D plot to Fig. \ref{fig:separatrix} but with the scattering separatrix equation $S(p,e,\lambda)$ (Eq. \eqref{eq:scattersep}).}
	\label{fig:outerseparatrix}
\end{figure*}

We can solve for the real roots in $p$ of Eq. \eqref{eq:separatrix} perturbatively using the method of dominant balance \cite{bender2013advanced}
and see that there are two relevant curves,
\begin{eqnarray}
	\label{eq:sdsseparatrix}
	p_{\mathrm{inner,pl}}&=&6+2e+\lambda\frac{8 (3-e)^2 (3+e)^3 }{\left(1-e^2\right)^2}
    \nonumber\\
    &&\quad+ O(\lambda^2),
\end{eqnarray}
\begin{eqnarray}
	\label{eq:outersep}
	p_{\mathrm{outer,pl}}&=&-\frac{1}{12} (e-3)^2 (e+1)+\left(\frac{(e-1)^2 (e+1)^3}{4 \lambda}\right)^{1/3}
    \nonumber\\
    &&\quad+\mathcal{O}(\lambda^{1/3}).
\end{eqnarray}
The two curves (Eqs. \eqref{eq:sdsseparatrix} and \eqref{eq:outersep}) form approximately the boundary of the region of bound orbital parameters, one as an ``inner" separatrix, and one as an ``outer" separatrix. The main observation here is the coupling of the SdS parameter with eccentricity, producing a tighter constraint for allowable orbits  of higher eccentricity.

Both separatrices, Eqs. \eqref{eq:sdsseparatrix} and \eqref{eq:outersep}, are also placed in Fig. \ref{fig:constant-El-in-pe}a and \ref{fig:constant-El-in-pe}b to see how they limit the constant curves of $(E,L)$ and $(p,e)$. This boundary extends to very large $p$ for small enough $\lambda$ but for a nontrivial $\lambda$, we see here a distinct feature of the SdS spacetime. The separatrices may imply that the presence of the SdS parameter limits the existence of very large orbits, making them unstable since the dynamics due to the $\lambda$ dominates the opposing central gravitational attraction of the black hole. We further check this claim in the next part by examining the nature of marginally bound orbits associated with an outer unstable extrema.

\subsubsection{Scattering separatrix}

The earlier analysis only applies on marginally bound orbits in relation to the inner unstable extrema, wherein we have analyzed the region of bound orbits merging with the plunge region. This amounts to solving for the extremas with the radial distance evaluated at the periapsis. We can now examine the locus of parameters that merges the apoapsis to the outer unstable extrema by examining:
\begin{eqnarray}
	\label{eq:outerpotentialderivative}
	\dfrac{d\,V_{\text{eff}}}{dr}\Big|_{r=r_a}=0
\end{eqnarray}
which produces a distinct separatrix polynomial obtained earlier:
\begin{align}
	S(p,e,\lambda)=(&e-1)^3 (e+1)^2 (2 e+p-6)\nonumber\\
	&- \lambda\, p^3\left(e^3+5 e^2-5 e-4 p+15\right)
    \nonumber\\
    &\quad+O\left(\lambda^2\right)=0.
	\label{eq:scattersep}
\end{align}
\begin{figure*}
	\centering
	\includegraphics[width=\linewidth]{"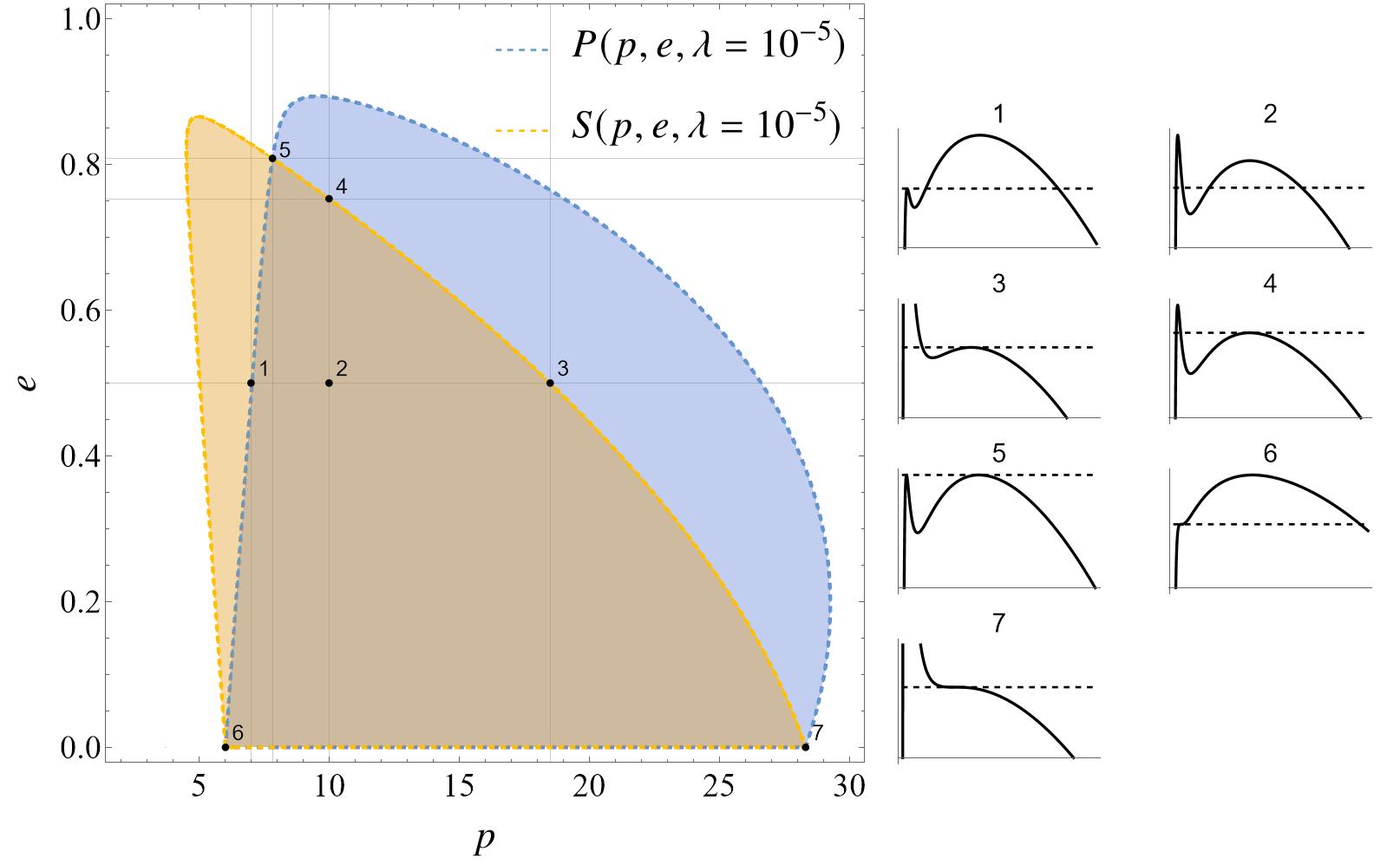"}
	\caption{Potential shapes, $V(r)$ vs $r$, (black curves in the right with dashed line as the orbital energy, $E$) of selected orbital parameters (black dots) in $(p,e)$ space. The intersection of orange and blue region contains the space of all bound orbits. The orange boundary (orange dashed line) marks the point at which orbits are marginally bound and may scatter away to the cosmological horizon while the blue boundary (blue dashed line) marks the points at which orbits are also marginally bound but may plunge to the black hole. The complement of the colored regions are unbound and may either plunge or scatter away.}
	\label{fig:separatrix2}
\end{figure*}
The contours of $\lambda$ satisfying $S(p,e,\lambda)=0$ is distinct from the contours of the separatrix polynomial, $P(p,e,\lambda)=0$, (Eq. \eqref{eq:separatrix}), and can be seen in Fig. \ref{fig:outerseparatrix}. We call $P(p,e,\lambda)$ as the plunge separatrix since orbits outside the region it bounds will plunge to the black hole while we call $S(p,e,\lambda)$ as the scattering separatrix since orbits  outside the region it bounds will scatter to the cosmological horizon. The scattering separatrix also has maximum parameters $\lambda=4/16875\sim 2.37\times 10^{-4}$ at $(p=7.5,e=0)$ equivalent to the parameters derived from the limits of the plunging separatrix. When solved for $p$ using a dominant balance method, we also obtain two real physical solutions in $p$ distinct from Eqs. \eqref{eq:sdsseparatrix} and \eqref{eq:outersep},
\begin{eqnarray}
	\label{eq:sdsseparatrix2}
	p_{\mathrm{inner,sc}}&=&2 (3-e)-\frac{8 (e+3)^2 (e-3)^3 \lambda }{\left(e^2-1\right)^2}
    \nonumber\\
    &&\quad+\mathcal{O}(\lambda^{2}),
\end{eqnarray}
\begin{eqnarray}
	\label{eq:outersep2}
	p_{\mathrm{outer,sc}}&=&\frac{1}{12} (3+e)^2 (1+e)+\left(\frac{(1-e)^3 (1+e)^2}{4 \lambda}\right)^{1/3}
    \nonumber\\
    &&
    \quad+\mathcal{O}(\lambda^{1/3}).
\end{eqnarray}
This solution marks the boundary at which orbits will stay bounded and not scatter away as opposed to Eq. \eqref{eq:outersep} which is the curve that marks the boundary at which orbits will stay bounded and not plunge to the black hole. 
We can now establish a hierarchy of lengthscales where our perturbative solutions will be valid, by examining the colored regions in Fig. \ref{fig:separatrix2}. We impose the hierarchy of the following appropriate lengthscales as:
\begin{equation}
	\label{eq:hierarchy}
	p_{h}<p_{\text{inner,pl}}<p<p_{\text{outer,sc}}\ll p_{c}
\end{equation}
with $p_{h}=2+8\lambda$, $p_c=\lambda^{-1/2}-1$ as the position of the black hole and cosmological horizon in $(p,e)$ space. Stable bound orbits occur in the region, $p_{\text{inner,pl}}<p_{\text{stable}}<p_{\text{outer,sc}}$. An interesting feature in the region of bound orbits is the intersection between the plunging and scattering separatrix (Eqs. \eqref{eq:separatrix} and\eqref{eq:scattersep}) that can be seen in Fig. \ref{fig:separatrix2}. The point of intersection (Point 5 in Fig. \ref{fig:separatrix2}) are marginally bound orbits and can either plunge or scatter away, depending on what direction $p$ or $e$ is changed by a small perturbation. These points are located in the parameter space at i.) $(p=\frac{1}{2} (e^2+15),\,e=1 - 8 \lambda^{1/3})$, at ii.) the innermost stable circular orbit (ISCO) and iii.) at the outermost stable circular orbit (OSCO).

The shifted ISCO can now be obtained by examining the circular limit, $e\rightarrow0$, of the separatrix equation (Eq. \eqref{eq:sdsseparatrix}). This transforms the separatrix to $(p-6) (1836 \lambda-1)+1944 \lambda=0$, leading to the shifted ISCO,
\begin{equation}
	p_{\text{ISCO}}(\lambda)=r_{\text{ISCO}}(\lambda)=6+1944\,\lambda + \mathcal{O}(\lambda^2).
\end{equation}
This means that a compact object orbiting circularly a $M=10^{6}M_{\odot}$ SMBH under a $\lambda=10^{-6}$ SdS spacetime are unstable and might plunge at an additional distance of $p<\mathcal{O}(10^3)$ km from the black hole. 
%For a realistic $\Lambda\approx10^{-46}$ the additional distance of plunge is extremely negligible with $\sim10^{-25}$ km for circular orbits, and  $\sim10^{-23}$ km for highly eccentric orbits of $e=0.99$. 
We also derive an outermost stable circular orbit (OSCO), $p_{\text{OSCO}}$ from Eq. \eqref{eq:outersep} by setting $e\rightarrow0$, leading to:
\begin{equation}
	p_{\mathrm{OSCO}}(\lambda)=r_{\text{OSCO}}(\lambda)=\left(\frac{1}{4 \lambda}\right)^{1/3}-\frac{3}{4}+\mathcal{O}(\lambda^{1/3}).	
	\label{eq:oscosds}
\end{equation}
As an example, for a $M=10^{6}M_{\odot}$ SMBH under a $\lambda=10^{-6}$ SdS spacetime, circular orbits are restriced to the region $p<10^{6}$ km.
%and for $\Lambda\approx10^{-46}$km$^{-2}$ the value gets pushed to $p<10^{15}$ km.

Going back to the orbital energy and angular momentum at the separatrix, we examine the effect of the SdS parameter on the minimum and maximum values these quantities can hold. We evaluate the energy and the angular momentum at the approximate plunging separatrix equation (Eq. \eqref{eq:sdsseparatrix}):
\begin{eqnarray}
	E^2|_{\mathrm{pl}}&=&\frac{8}{9-e^2}
    \nonumber\\
    &&\quad-\lambda\frac{32 (e+3) \left(3 e^2-2 e+3\right) }{(3-e) \left(e^2-1\right)^2}\\
	L^2|_{\mathrm{pl}}&=&\frac{4(e+3)^2}{(3-e) (e+1)}
    \nonumber\\
    &&\quad+\lambda\frac{32 (e+3)^4 \left(e^3-5 e^2-e-3\right) }{(3-e) (1-e)^2 (e+1)^4}
\end{eqnarray}
which evaluates at $E^2>\dfrac{8}{9}-24\,\lambda$ and $L^2>12-2592\,\lambda$ for circular orbits. For eccentric orbits, neglecting high eccentricities of order $\mathcal{O}((1-e)^{-1})$, we have $E^2<16-768\,\lambda$ and $L^2<4-48\,\lambda$. As expected from earlier results, the energy and angular momentum at the separatrix is decreased by a factor due to the SdS parameter. This would mean that lower energy and angular momentum would be required for orbits to plunge the black hole due to $\lambda$, taking it longer to inspiral if we naively assume equal rates of energy and angular momentum loss for particles in a Schwarzschild and SdS spacetime.

\section{Adiabatic orbital evolution}
\label{sec:sec3}

As the compact object inspirals toward the black hole, it emits gravitational radiation, which carries away energy and angular momentum from its motion. In the leading post-Newtonian (pN) approximation, the quadrupolar formula for the secularly averaged energy and angular momentum flux measured at the cosmological horizon has been studied extensively by Bonga among others \cite{bonga2017power,hoque2019quadrupolar,hoque2018propagation}.
For sources inside asymptotically flat spacetimes, the gravitational radiation is defined at the ``far-away wave zone" using a $1/r$ expansion and then performing a temporal average over the orbital period. For asymptotically - de Sitter ($\Lambda>0$) spacetimes, the technique used by Bonga and Ashtekar \cite{ashtekar2015asymptotics,ashtekar2017implications} was to define the radiative modes in the future null infinity ($\mathcal{I}^+$)  using a late-time expansion to replace the expansion in $1/r$.  
In particular, the radiation field was defined to be the following:
\begin{eqnarray}
    \mathcal{Q}_{i j}&:=&\Bigg(\mathcal{L}_T^3 Q_{i j}^{(\rho)}+3 H \mathcal{L}_T^2 Q_{i j}^{(\rho)}+2 H^2 \mathcal{L}_T Q_{i j}^{(\rho)}
    \nonumber\\
    &&\quad+H \mathcal{L}_T^2 Q_{i j}^{(p)}+3 H^2 \mathcal{L}_T Q_{i j}^{(p)}+2 H^3 Q_{i j}^{(p)}\Bigg)\Bigg|_{t\rightarrow t_{\mathrm{ret}}}\nonumber\\
\end{eqnarray}    
with $H=\left(\frac{\Lambda}{3}\right)^{1/2}$, $\mathcal{L}_T$ is the Lie derivative along the de Sitter time translational Killing vector and is related to the coordinate time $t$ by $\mathcal{L}_T=\partial_t-2\left(\frac{\Lambda}{3}\right)^{1/2}$. 
The energy of the gravitational radiation at  $\mathcal{I}^+$ is defined to be the conserved current associated with a modified time-translation vector field.
The averaged fluxes of energy and angular momentum radiated through $\mathcal{I}^+$ of a compact source were derived \cite{2019quadrupolar}  to be the integral of the mass ($\rho$) and pressure ($p$) quadrupole moments $\mathcal{Q}_{ij} = \partial_t^3Q_{ij}^{(\rho)}-(3\Lambda)^{1/2}\partial_t^2Q_{ij}^{(\rho)}+\frac{2}{3}\Lambda\partial_tQ_{ij}^{(\rho)}+\left(\frac{\Lambda}{3}\right)^{1/2}\partial_t^2Q_{ij}^{(p)}-\frac{\Lambda}{3}\partial_tQ_{ij}^{(p)}$,
\begin{equation}
    \begin{aligned}
    \langle\mathcal{P}\rangle=&\frac{G}{5}\Big\langle\mathcal{Q}_{ij}\mathcal{Q}_{ij}-\frac{1}{3}\mathcal{Q}^2\Big\rangle
    \\
    \langle\dot{L}_k\rangle=&\frac{2G}{5}\epsilon_{ikm}\Bigg\langle\mathcal{Q}_{ij}\Bigg(\mathcal{L}_T^2 Q_{ij}^{(\rho)}
    \\
    &
    +\left(\frac{\Lambda}{3}\right)^{1/2} \left(\mathcal{L}_T Q_{ij}^{(\rho)}+\mathcal{L}_T Q_{ij}^{(p)}\Bigg)+\frac{\Lambda}{3}Q_{ij}^{(p)}\right)\Bigg\rangle
    \end{aligned}
\end{equation}
evaluated at the retarded time $t_{ret}=-\left(\frac{\Lambda}{3}\right)^{-1/2}\ln\left(\left(\frac{\Lambda}{3}\right)^{1/2}(r-\eta)\right)$ with $\eta$ as the de Sitter conformal time. We reexpress the fluxes in terms of the orbital parameters $(p,e)$ noting that we also neglect the pressure quadrupole moment in the leading order,

\begin{eqnarray}
	\label{eq:Edot}
	\dfrac{dE}{dt}&=&-\frac{32}{5}\frac{1}{p^5}\left(1+\frac{73}{24}e^2+\frac{37}{96}e^4\right)(1-e^2)^{3/2}\nonumber\\&&-\lambda\frac{8}{3p^2}\left(4-(1-e^2)^{1/2}\right)(1-e^2)^{3/2}+\mathcal{O}(\lambda^2)\nonumber\\\\
	\label{eq:Ldot}
	\dfrac{dL}{dt}&=&-\frac{32}{5}\frac{1}{p^{7/2}}\left(1+\frac{7}{8}e^2\right)(1-e^2)^{3/2}\nonumber\\&&-\lambda\frac{8}{p^{1/2}}(1-e^2)+\mathcal{O}(\lambda^2).
\end{eqnarray}
It is important to note that these expressions are valid only within the range specified by Eq. \eqref{eq:hierarchy}. Under the conditions described by Eq. \eqref{eq:hierarchy}, for $p=p_{\text{stable}}$, the flux contribution from the SdS parameter remains subdominant compared to the Peters-Mathews fluxes. However, this contribution becomes comparable in magnitude only for $p>p_{\text{stable}}$, corresponding to unbound orbits.

The expressions above (Eqs. \eqref{eq:Edot} and \eqref{eq:Ldot}) indicate that a nonzero energy and angular momentum flux implies the energy and angular momentum of a bound orbit must change over time, although implicitly through the orbital parameters $(L,E)\rightarrow(L(p(t),e(t)),E(p(t),e(t)))$. 
%\begin{equation}
%(L,E)\rightarrow(L(p(t),e(t)),E(p(t),e(t))).  
%\end{equation}
The flux of energy and angular momentum will cause the orbital trajectory to evolve continuously, resulting in a dynamic orbit characterized by time-dependent orbital parameters. Given that we have defined the bound orbits as ellipses (Eq. \eqref{eq:radialeq}), long-term secular changes  \cite{maggiore2008gravitational} in the orbital parameters will manifest as an evolving ellipse, or an osculating orbit.

We now outline the derivation for the orbital evolution equations. To begin, we observe that the time derivatives of energy and angular momentum can be directly related to time derivatives of the orbital elements, 
\begin{equation}
	\label{eq:evoeqderivation1}
	\dfrac{dE}{dt}=\dfrac{\partial E}{\partial p}\dfrac{dp}{dt}+\dfrac{\partial E}{\partial e}\dfrac{de}{dt}\,,\quad
	\dfrac{dL}{dt}=\dfrac{\partial L}{\partial p}\dfrac{dp}{dt}+\dfrac{\partial L}{\partial e}\dfrac{de}{dt}.
\end{equation}
We utilize the Jacobian $J_{ij}=\partial_{i}\rho_{j}$ where $i$ corresponds to the elements of $(p,e)$ and $\rho_j$ represents the elements of $(L,E)$. The evolution of the orbital parameters can then be expressed as:
\begin{eqnarray}
	\label{eq:evoeqderivation2}
	\left(\begin{array}{c}
		\dfrac{dp}{dt} \\\\ 
		\dfrac{de}{dt}
	\end{array}\right)=(J_{ij})^{-1} \left(\begin{array}{c}
		\dfrac{dL}{dt} \\\\ 
		\dfrac{dE}{dt}
	\end{array}\right).
\end{eqnarray}
In this derivation, circular orbits, $e=0$ are excluded, as they render $J_{ij}^{-1}$ singular, as discussed in Section \ref{sec:sec2.2}. The evolution of circular orbits will instead be addressed separately in the next subsection, Section \ref{sec:sec3.1}.

Using the fluxes (Eqs.\eqref{eq:Edot} and \eqref{eq:Ldot}) and the derivatives of the corrected energy and angular momentum (Eqs.  \eqref{eq:largepL} and \eqref{eq:largepE}), the orbital evolution equations given by:
\begin{eqnarray}
	\label{eq:pdotsdsflux}
	\dot{p}(p,e)&=&2 \sqrt{p} \;\dot{E}_N\nonumber\\
	&&-\lambda\Bigg[2 \sqrt{p} \;\dot{E}_\lambda+\epsilon\frac{3 p^{7/2} }{\left(e^2-1\right)^2}\dot{L}_N
    \nonumber\\
    &&\quad-\epsilon\frac{4 p^5 }{\left(e^2-1\right)^3}\dot{E}_N\Bigg]+\mathcal{O}(\lambda^2)
    \\
	\dot{e}(p,e)&=&\frac{p }{e}\;\dot{L}_N+\frac{\left(e^2-1\right) }{\sqrt{p}e }\;\dot{E}_N\nonumber\\
	&&-\frac{\lambda}{e}\Bigg[p\;\dot{L}_\lambda+\frac{\left(e^2-1\right) }{\sqrt{p}}\;\dot{E}_\lambda+\epsilon\frac{7 p^{5/2} }{2 \left(e^2-1\right)}\dot{L}_N
    \nonumber\\
   &&\quad-\epsilon\frac{4 \left(e^2+1\right) p^4 }{ \left(e^2-1\right)^3}\dot{E}_N\Bigg]+\mathcal{O}(\lambda^2).
   \label{eq:edotsdsflux}
   \nonumber\\
\end{eqnarray}
The orbital evolution equations involve de Sitter terms of order $\epsilon(=1)$, a fiducial parameter, to highlight the effect of the corrected conserved energy, $E=E_{N}+\epsilon \lambda E_{\lambda}$, and angular momentum, $L=L_{N}+\epsilon \lambda L_{\lambda}$ (Eqs. \eqref{eq:sdsL} and \eqref{eq:sdsE}). These orbital evolution equations represent the correct form of the orbital dynamics up to leading $\mathcal{O}(\lambda)$ and leading $p$ - dependence. Notably, the earlier work by Hoque and Aggarwal \cite{hoque2019quadrupolar} did not account for these linear $\lambda$ terms, which should also contribute to the evolution of the orbital parameters.

We further observe that the orbital evolution equations are only valid within the eccentricity range $0<e<1$, evidenced by the $e^{-1}$ divergence for $e=0$ orbits. In standard literature, the $e^{-1}$ divergence is typically removed by using the Peters-Mathews fluxes or by incorporating modified fluxes due to the SdS parameter, as discussed in \cite{hoque2019quadrupolar}. It will be evident later in this section that the new terms due to the modified energy and angular momentum retain this divergence in eccentricity.

The gravitational fluxes (equations \eqref{eq:Edot} and \eqref{eq:Ldot}) are valid only in the weak-field regime as noted earlier. Therefore, we now present the form of the orbital evolution equations for bound orbits far from the black hole, using a large $p$ expansion for the equations under the relation, Eq. \eqref{eq:hierarchy}. The resulting secular evolution equations $\dot{p}$ and $\dot{e}$ are given by:
\begin{eqnarray}
	\label{eq:pdotsds}
	\dot{p}(p,e)&=&-\frac{64}{5}p^{-3}(1-e^2)^{3/2}\left(1+\frac{7}{8}e^2\right)\nonumber\\
	&&-\lambda\Bigg[16(1-e^2)-\epsilon\frac{4 \left(26 e^4-283 e^2-168\right)  }{15 \left(1-e^2\right)^{3/2}}
    \nonumber\\
    &&\quad+\mathcal{O}(\lambda p)\Bigg]\\
	\label{eq:edotsds}
	\dot{e}(p,e)&=&-\frac{304}{15}p^{-4}e(1-e^2)^{3/2}\left(1+\frac{121}{304}e^2\right)\nonumber\\
	&&-\lambda\Bigg[\frac{32}{3}p^{-1}e^{-1} \left(1-e^2\right)^{3/2} \left(1-(1-e^2)^{1/2}\right)\nonumber\\
	&&-\epsilon p^{-1}e^{-1}\frac{2 \left(73 e^6-784 e^4-965 e^2-24\right)}{15  \left(1-e^2\right)^{3/2} }
    \nonumber\\
    &&\quad
    +\mathcal{O}(\lambda p^{-2})\Bigg].
\end{eqnarray}

Equations \eqref{eq:pdotsds} and \eqref{eq:edotsds} reduce to the post-Newtonian (2.5 pN) orbital evolution equation when $\lambda=0$ \cite{peters1964gravitational}, but with the $p$ dependence replacing the semi-major axis $a$. In the following subsections, we will demonstrate how initially circular and eccentric orbits evolve using these equations.

\subsection{Evolution of circular orbits}
\label{sec:sec3.1}

Earlier work \cite{peters1964gravitational,hoque2019quadrupolar} has shown that the circular limit of the orbital equations (\eqref{eq:pdotsds} and \eqref{eq:edotsds}) is regular at the $e\rightarrow0$ limit, implying that circular orbits will remain circular due to the vanishing of $\dot{e}$. However, the circular limit of the eccentricity evolution equation (Eq.  \eqref{eq:edotsds}) is ill-defined, as the terms contributed by the SdS parameter do not eliminate the $e^{-1}$ divergence present in Eq. \eqref{eq:edotsdsflux}. Therefore, we must develop an independent method to analyze how circular orbits will evolve under the release of gravitational fluxes.

One approach to understanding how circular orbits evolve is to shift the analysis to $(E,L)$ space (visualized in Fig. \ref{fig:constant-El-in-pe}a) and examine how the fluxes $(\dot{E},\dot{L})$ influence the space of circular orbits, represented by the $e=0$ curve. The tangents, $dE/dL$, lying in the $e=0$ curve correspond to the rates of energy and angular momentum change that will make an initially circular orbit circular. If the tangents stay close and trace the $e=0$ curve, this would indicate that the fluxes of energy and angular momentum necessarily maintain the circularity of the orbit. However, if the tangents deviate away from the $e=0$ curve, it would suggest that circular orbits will lose its circularity due to the fluxes. For the Peters-Mathews fluxes, this can be easily demonstrated by first taking the $\lambda=0$ limit, followed by the $e=0$ limit of the fluxes (Eqs. \eqref{eq:Edot} and \eqref{eq:Ldot}), 
\begin{equation}
	\frac{dE}{dL}\Bigg|_{\mathrm{flux},\,\lambda=0}=\frac{dE}{dt}\left(\frac{dL}{dt}\right)^{-1}=\frac{p^{-5}}{p^{-7/2}}=p^{-3/2}.
	\label{eq:circfluxpm}
\end{equation}
In the conservative case, the energy and angular momentum in terms of $(p,e)$ (Eqs.  \eqref{eq:largepL} and \eqref{eq:largepE}), yield the ratio of energy and angular momentum as
\begin{equation}
	\frac{dE}{dL}\Bigg|_{\mathrm{circ},\,\lambda=0}=\frac{dE}{dp}\left(\frac{dL}{dp}\right)^{-1}=\frac{ p^{-2}}{p^{-1/2}}=p^{-3/2}.
	\label{eq:circratepm}
\end{equation}
The flux ratio (Eq. \eqref{eq:circfluxpm}) match the rates at which a circular orbit remains circular (Eq. \eqref{eq:circratepm}), indicating that under the Peters-Mathews fluxes, circular orbits will indeed remain circular until plunge.

Repeating the analysis with the inclusion of the SdS parameter $\lambda$, we see that the flux ratio take the form:
\begin{eqnarray}
	\frac{dE}{dL}\Bigg|_{\mathrm{flux},\,\lambda\neq0,\,p=\tilde{p}}=\frac{\sqrt{\lambda }}{\tilde{p}^{3/2}}+\mathcal{O}(\lambda\tilde{p}^{3/2}),
	\label{eq:circfluxsds}
\end{eqnarray}
for the intermediate scale $\tilde{p}=p \lambda^{1/3}\ll1$, which satisfies the condition for stable orbital parameters from Eq.  \eqref{eq:hierarchy}. On the other hand, the tangents to the SdS $e=0$ curve follow the form:
\begin{eqnarray}
	\frac{dE}{dL}\Bigg|_{\mathrm{circ},\,\lambda\neq0,\,p=\tilde{p}}=\frac{\sqrt{\lambda }}{\tilde{p}^{3/2}}-\frac{1}{2} \sqrt{\lambda } \tilde{p}^{3/2}+\mathcal{O}(\lambda\tilde{p}^{5/2}).
	\label{eq:circratesds}
\end{eqnarray} Comparing Eqs. \eqref{eq:circfluxsds} and \eqref{eq:circratesds}, 
we see an additional term in the expression for the tangents, unlike the match between Eqs. \eqref{eq:circfluxpm} and \eqref{eq:circratepm}. This indicates that the SdS fluxes drives the orbits away from the $e=0$ curve, causing initially circular orbits to break their circularity. As $\lambda\rightarrow 0$, the SdS flux ratio reduce (Eq. \eqref{eq:circfluxsds}) to Peters-Mathews flux ratio (Eq. \eqref{eq:circfluxpm}) and the Schwarzschild $e=0$ tangents (Eq. \eqref{eq:circratepm}) become,
\begin{eqnarray}
	\frac{dE}{dL}\Bigg|_{\mathrm{flux},\,\lambda\rightarrow 0}=\frac{dE}{dL}\Bigg|_{\mathrm{circ},\,\lambda=0}=p^{-3/2}.
	\label{eq:circfluxsdspmlimit}
\end{eqnarray}
Thus for $\lambda\rightarrow 0$, a Newtonian orbit will remain circular. This result was also demonstrated by \cite{hoque2019quadrupolar} using the orbital evolution equations rather than the rates of energy and angular momentum change. However, a key point of this paper is that the energy and angular momentum of a bound orbit will be influenced by the SdS parameter, which means that a Newtonian description of a bound orbit will be inaccurate. 

The tangents to the $e=0$ curve (Eq.  \eqref{eq:circratesds}) indicate that inside the region where $p\ll \lambda^{-1/3}$ or at $\tilde{p}\ll1$, the $\lambda$ term is small  and gravitational fluxes will indeed drive circular orbits to remain circular up to the leading order $\mathcal{O}(\tilde{p}^{-3/2})$. 
Closer to the outer separatrix, $p_{\text{OSCO}}-p\ll1$, far from the intermediate scale $\tilde{p}$, the flux ratio becomes:
\begin{eqnarray}
	\frac{dE}{dL}\Bigg|_{\mathrm{flux},\,p\rightarrow p_{\mathrm{OSCO}}}&=&\frac{1}{p_{\text{OSCO}}^{3/2}}+\frac{3 }{2 p_{\text{OSCO}}^{5/2}}\left(p_{\text{OSCO}}-p\right)
    \nonumber\\
&&\quad+\mathcal{O}\left(\left(p_{\text{OSCO}}-p\right){}^2\right)
	\label{eq:circfluxosco}
\end{eqnarray}
which reduces, in the leading order in $p$, to the Peters-Mathews fluxes for $\lambda\rightarrow0$ or $p_{OSCO}\rightarrow\infty$ (where the OSCO extends to infinity in the Schwarzschild spacetime). The tangents to the SdS $e=0$ curve near $p_{\text{OSCO}}$ are given by (factoring out the flux \eqref{eq:circfluxosco} for easy comparison):
\begin{eqnarray}
	\frac{dE}{dL}\Bigg|_{\text{circ},\,p\rightarrow p_{\text{OSCO}}}&=&\frac{dE}{dL}\Bigg|_{\text{flux}}\left(1-\lambda \frac{ p_{\text{OSCO}}^3}{2}\right)	+
    \nonumber\\
    &&+\mathcal{O}\left(\lambda\left(p_{\text{OSCO}}-p\right){}^2\right).
	\label{eq:circrateosco}
\end{eqnarray}
Comparing the flux ratio and the $e=0$ tangents near the OSCO (Eqs. \eqref{eq:circfluxosco} and \eqref{eq:circrateosco}), we see that an additional term proportional to $\lambda$ and $p_{\text{OSCO}}$. This means that in these regions, gravitational fluxes will not preserve the circularity of the orbit.  Thus in the region where $\left(p_{\text{OSCO}}-p\right)\ll 1$, near the outer separatrix, the SdS gravitational fluxes will not be sufficient to maintain the circularity of the orbital evolution. 

\begin{figure}
	\centering
	\includegraphics[width=\linewidth]{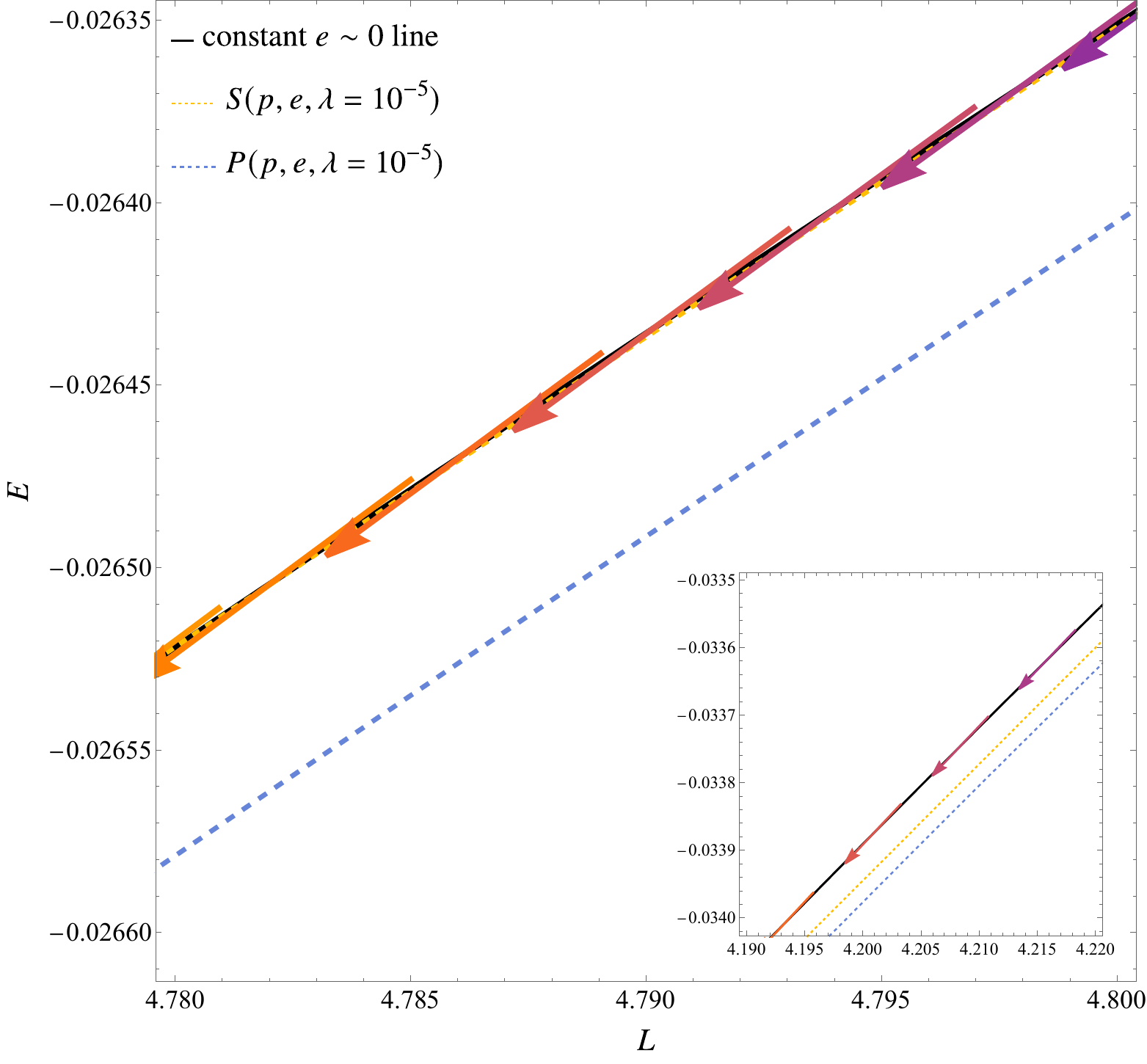}
	\caption{Gravitational flux (vectors) tangent to the $e=0$ curve (black) in the $(E,L)$ space. We see here as discussed in Section \ref{sec:sec3.1} that far from the OSCO (inset, low $(E,L)$), circularity is maintained (vectors lie along the black curve) but as for orbits with separation near the OSCO, the orbits will cross the scattering separatrix (yellow) and become unbound  with the vectors crossing the yellow dashed line.}
	\label{fig:circ-dynamics}
\end{figure}

As illustrated in Fig. \ref{fig:circ-dynamics}, at higher orbital separations, closer to the OSCO, gravitational fluxes will push the orbits outside the scattering separatrix (Eq. \eqref{eq:scattersep}) making the orbits unbound. In the inset of Fig. \ref{fig:circ-dynamics}, we see that for orbits far from the OSCO, (with low energy and angular momentum), gravitational fluxes will drive the orbits along the $e=0$ curve, maintaining their circularity.

\subsection{Evolution of eccentric orbits}
\label{sec:sec3.2}

We now analyze Eqs. \eqref{eq:pdotsds} and \eqref{eq:edotsds} to understand the evolution of eccentric orbits. By examining the signs of the terms, we observe that the de Sitter contributions generally increase the magnitude of orbital parameter evolution. At first glance, the de Sitter terms seems to enhance the rate at which orbits shrink, leading to a faster inspiral and circularization. However, the second de Sitter term introduces a positive contribution, which counteracts this effect by decreasing the rate of inspiral and circularization. The competing influence highlights the complexity of the de Sitter modifications in the orbital evolution.

\begin{figure}
	\centering
	\begin{minipage}{0.46\textwidth}
		\includegraphics[width=\linewidth]{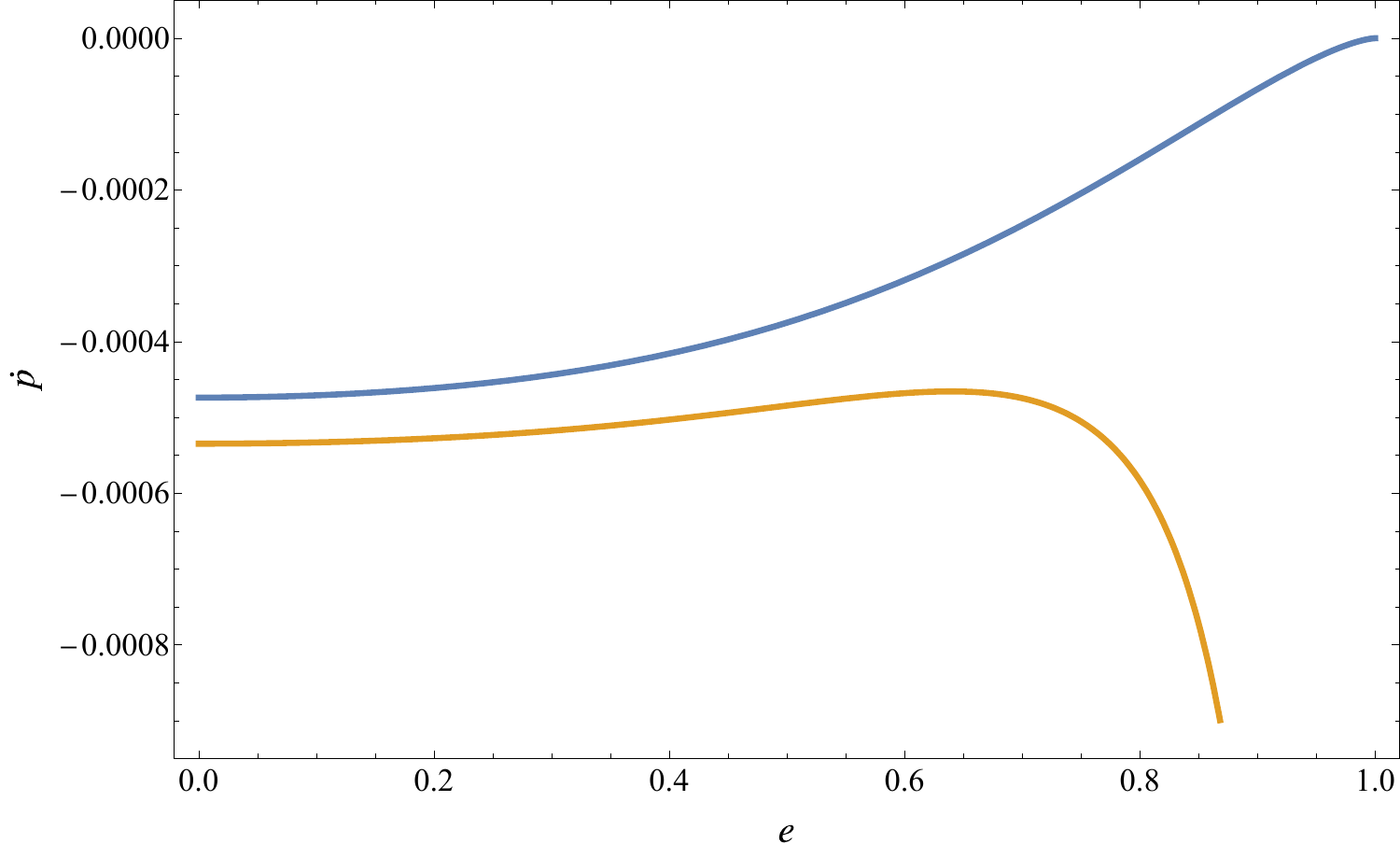}
	\end{minipage}
	\hspace{1cm}
	\begin{minipage}{0.46\textwidth}
		\includegraphics[width=\linewidth]{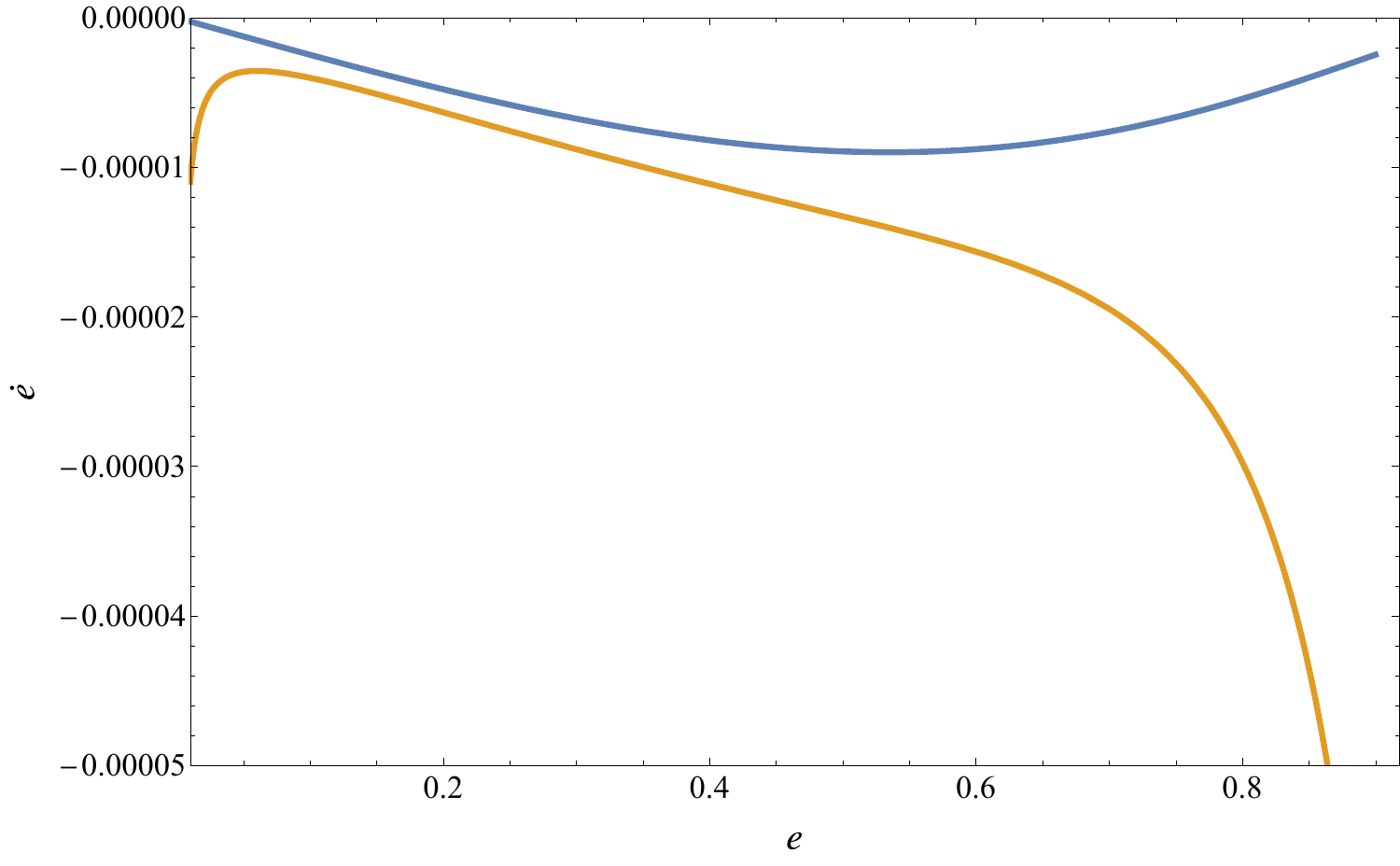}
	\end{minipage}
	\caption{Dependence of orbital evolution, $\dot{p}$ and $\dot{e}$, on eccentricty, $e$. Fixing the semi-latus at $p=30$ and SdS parameter $\lambda=10^{-6}$ for the SdS (yellow) case and $\lambda=0$ for the Schwarzschild case (blue), we show the rate of orbital parameter evolution with respect to eccentricity. We show in (a) the evolution of $p$ as steadily decreasing reaching a constant maximum at the circular limit. We show in (b) the evolution of $e$ is large for highly eccentric orbits and $e \rightarrow 0$ orbits. The SdS parameter amplifies the decay of the orbital parameters continuously increasing at the low $e$ region where the evolution in $e$ is dominated by the de Sitter term. }
	\label{fig:p-e-ecc-dependence}
\end{figure}

A particularly interesting feature emerges in the dependence of the de Sitter terms on the orbital parameters. Specifically, in the eccentricity evolution equation, the second de Sitter term exhibits an $e^{-1}$ scaling, which diverges for very nearly circular orbits ($e\rightarrow 0$). This is a novel effect introduced by the presence brought of the SdS parameter, affecting the late-time behavior of eccentricity evolution. In the regime where the second de Sitter term dominates, the eccentricity changes more rapidly than expected. Fig. \ref{fig:p-e-ecc-dependence} illustrates the dependence of the orbital evolution equations on eccentricity, further demonstrating these effects.
The divergence in the low-eccentricity regime can be observed in Eq. \eqref{eq:edotsdsflux} and can be attributed to a specific contribution of the SdS parameter to the eccentricity evolution, this contribution is given by:
\begin{eqnarray}
	\label{eq:edotlowe}
	\frac{de}{dt}\Big|_{e\rightarrow 0}&=&\epsilon\frac{\frac{\partial L_{\lambda}}{\partial p}}{\frac{\partial L_{N}}{\partial p}}\dot{E}\left(\frac{\partial E_{N}}{\partial e}\right)^{-1}-\epsilon\frac{\frac{\partial E_{\lambda}}{\partial p}}{\frac{\partial L_{N}}{\partial p}}\dot{L}\left(\frac{\partial E_{N}}{\partial e}\right)^{-1}
    \nonumber\\
    &&\sim e^{-1}.
\end{eqnarray}
The derivative of the Keplerian energy with respect to $e$ follows the scaling relation $\frac{\partial E_{N}}{\partial e}\sim e$.  For small eccentricities $e$, this term becomes significantly smaller than the factors $\frac{\partial L_{\lambda}}{\partial L_{N}}\dot{E}\sim 1$ and $\frac{\partial E_{\lambda}}{\partial L_{N}}\dot{L}\sim 1$. Furthermore, this effect persists because the relation $\frac{\partial L_{\lambda}}{\partial p}\dot{E}<\frac{\partial E_{\lambda}}{\partial p}\dot{L}$ or $\frac{\partial L_{\lambda}}{\partial L}<\frac{\partial E_{\lambda}}{\partial E}$ holds. As a result, the evolution of eccentricity is significantly accelerated in the low - eccentricity parameter space.

As discussed earlier in this section, initially eccentric orbits tend to decrease in eccentricity throughout their evolution. The decrease in eccentricity accelerates as the orbits approach circularity. For orbits far from the OSCO, circular orbits will remain circular, while eccentric orbits eventually circularize. From the previous subsection, it was shown that orbits close to the outer separatrix, near the outermost stable circular orbit (OSCO, Eq.\eqref{eq:oscosds}), the SdS gravitational fluxes fail to maintain the circularity of an initially circular SdS orbit. Such orbits will become unbound over time. Near the OSCO, where $p_{OSCO}-p\ll 1$, the eccentricity evolution equation (Eq. \eqref{eq:edotsds}) vanishes when:
%\begin{eqnarray}
%	\frac{de}{dt}\Big|_{p\rightarrow p_{\text{OSCO}},\,e\ll1}&=& -\frac{304 e}{15 p_{\text{OSCO}}^4}-\frac{1216 e \left(p_{\text{OSCO}}-p\right)}{15 p_{\text{OSCO}}^5}
%    \nonumber\\
%    &&+\lambda\frac{  \left(-640 e^2 p_{\text{OSCO}}^3+37407 e^2+17064\right)}{120 e p_{\text{OSCO}}^4}\nonumber\\
%	&&+\mathcal{O}(\left(p_{\text{OSCO}}-p\right)^2)+\mathcal{O}(\lambda\left(p_{\text{OSCO}}-p\right)).\nonumber\\
%\end{eqnarray} 
\begin{eqnarray}
	p\Big|_{\dot{e}=0}&=&\frac{5 p_{\text{OSCO}}}{4}+\lambda  \Bigg(-\frac{2133 p_{\text{OSCO}}}{1216 e^2}
    \nonumber\\
    &&+\frac{p_{\text{OSCO}} (640 p_{\text{OSCO}}^3-37407)}{9728}\Bigg)\nonumber\\
	&&+O\left(\lambda \;e^2\right).
\end{eqnarray}
The first term is a constant which exceeds $p_{\text{OSCO}}$ by a fixed factor. Hence, for bound orbits with $p< p_{\text{OSCO}}$, the eccentricity evolution is always decreasing, reaching zero only if the initial semi-latus rectum is at $p=5 p_{\text{OSCO}}/4$, which lies outside the region of stable orbits. This implies that, within the bound region, eccentricity will always decrease.  We summarize the eccentricity evolution of bound orbits as:
\begin{itemize}
	\item \textit{Far from the OSCO:}
	\begin{itemize}
		\item Circular orbits up to $\mathcal{O}(\tilde{p}^{3/2})$ to will maintain circularity .
		\item Eccentric orbits will gradually circularize.
	\end{itemize}
	\item \textit{Near the OSCO:}
	\begin{itemize}
		\item Circular orbits will be pushed outside OSCO, becoming unbound.
		\item Eccentric orbits near circularity will also be pushed outside OSCO, becoming unbound.
	\end{itemize}
\end{itemize}

\begin{figure*}
	\centering
	\includegraphics[width=0.5\linewidth]{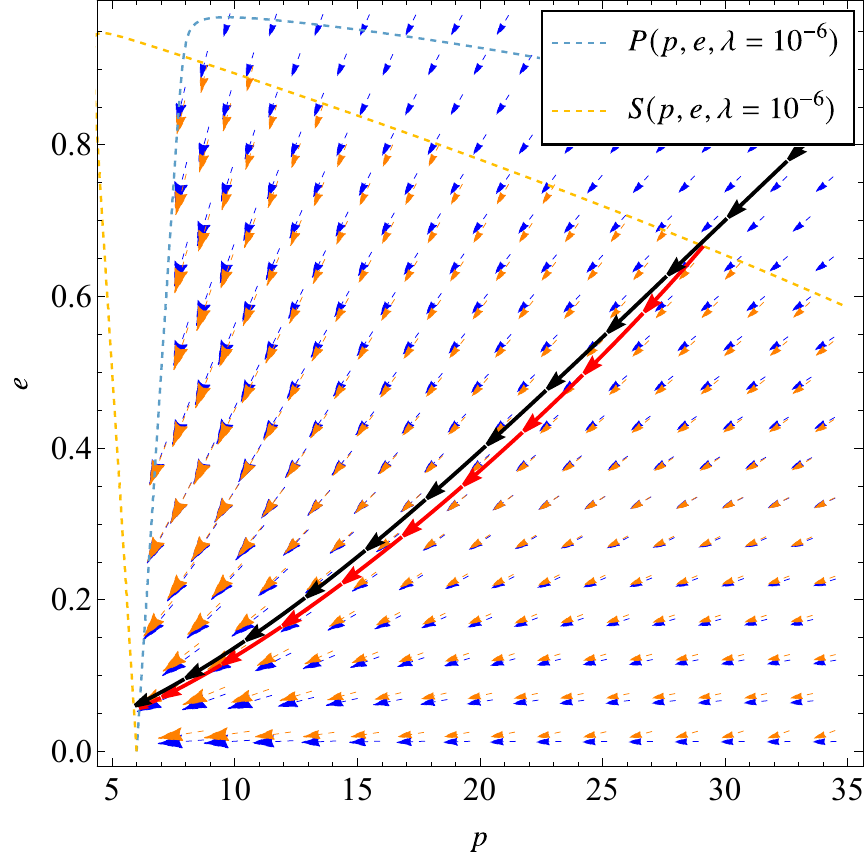}
	\caption{The flow of $\dot{p}$ and $\dot{e}$ in the phase plane $(p,e)$ and the vector slope $de/dp$ in Schwarzschild and SdS spacetime. In (a), we show the $\dot{p}$ and $\dot{e}$ as vectors following Eqs. \eqref{eq:pdotsds} and \eqref{eq:edotsds}. Blue vectors denote orbital evolution in Schwarzschild spacetime $(\lambda=0)$ while orange vectors are in Schwarzschild $-$ de Sitter $(\lambda=10^{-6})$.  Black vectors $(\lambda=0)$ and red vectors $(\lambda=10^{-6})$ are fixed trajectories that meets at the same parameters, $(p_0=30,e_0=0.7)$. We see here that the black trajectory has a lower slope indicating slower eccentricity decay with respect to $p$. Sizes of the vectors indicate relative magnitudes.}
	\label{fig:phase-space-stream}
\end{figure*}

The orbital evolution is obtained by solving Eqs. \eqref{eq:pdotsds} and \eqref{eq:edotsds}, treating them as a system of coupled, nonlinear, ordinary differential equations where the orbital parameters, $p(t)$ and $e(t)$, evolve in time. To gain insight into the system before solving explicitly, we analyze Eqs. \eqref{eq:pdotsds} and \eqref{eq:edotsds} by interpreting the right-hand side as a set of vector fields in the $(p,e)$ phase space. This allows us to construct a phase portrait of the differential equations, as shown in Fig. \ref{fig:phase-space-stream}. Similar to the Schwarzschild case, gravitational radiation reaction in the SdS spacetime generally leads to $(p,e)$ decay. The key quantity in determining the direction of orbital evolution in phase space is the slope $de/dp$. When the slopes are larger, the vectors are steeper, indicating that eccentricity decay dominates the evolution. Conversely, smaller slope correspond to flatter vectors indicating that the inspiral dominates the evolution.

\subsection{Inspiral, circularization and orbital trajectory}
\label{sec:sec3.3}

\begin{figure}
	\centering
	\begin{minipage}{0.45\textwidth}
		\includegraphics[width=\linewidth]{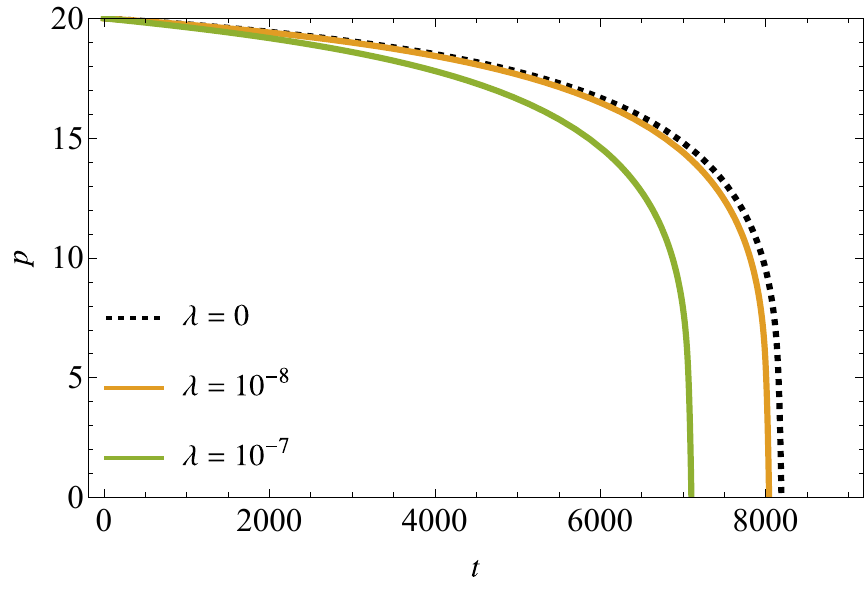}
	\end{minipage}
	\hspace{1cm}
	\begin{minipage}{0.45\textwidth}
		\includegraphics[width=\linewidth]{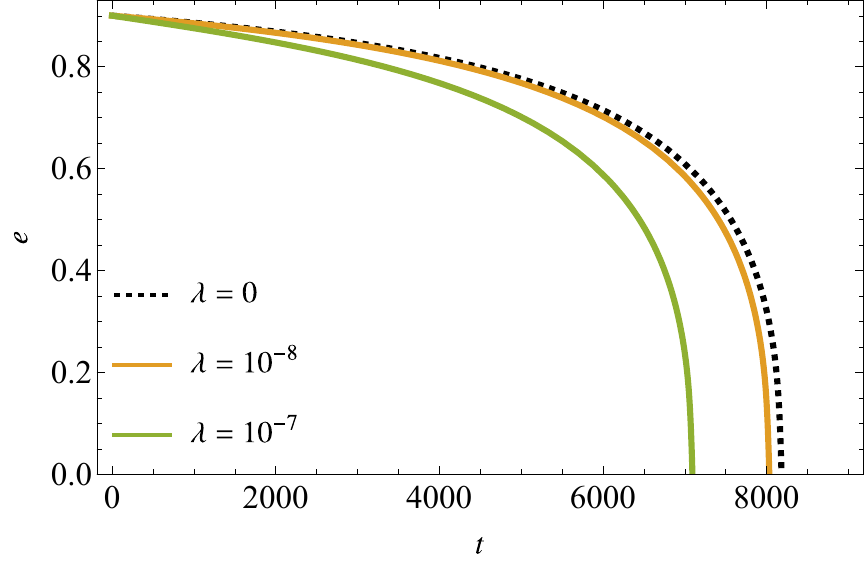}
	\end{minipage}
	\caption{Numerical curves of $p(t)$ and $e(t)$ with initial parameters $(p_0=20,\;e_0=0.9)$ for different $\lambda$ values. We show here the general effect of the SdS parameter which is to reduce plunge and circularization times as $\lambda$ increases.}
	\label{fig:p-e-solution}
\end{figure}

\begin{figure}[ht!]
	\centering
	\includegraphics[width=\linewidth]{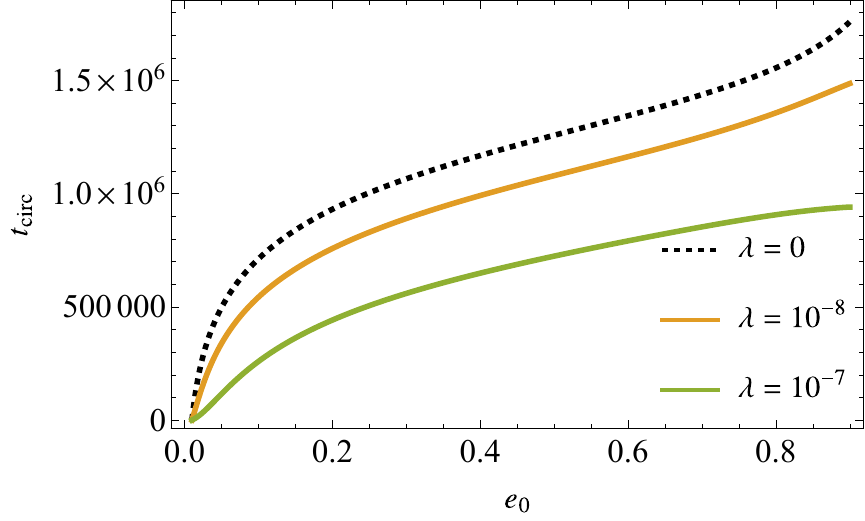}
	\caption{Numerical curves for $t_{circ}$ for a fixed $p=50$ (with the assumption that it stays constant) and its dependence on $e$ for varying $\lambda$. We see here that circularization time are reduced by the SdS parameter. As we increase the value of $\lambda$, highly eccentric orbits circularizes much faster.}
	\label{fig:circ-time-vs-e}
\end{figure}

To analyze the time evolution of the orbital parameters $(p,e)$, we numerically solve the system of ODE (Eqs. \eqref{eq:pdotsds} and \eqref{eq:edotsds}) for given initial configurations. The resulting solutions, illustrated in Fig. \ref{fig:p-e-solution}, demonstrate that the presence of the SdS parameter is to increase the rate of orbital decay. As a consequence, the SdS parameter accelerates the inspiral of the CO, leading to an earlier plunge into the central BH. For a fixed $p$, we can numerically integrate the equation for $\dot{e}$ (Eq. \ref{eq:edotsds}) to estimate the circularization time and its dependence on $e$ and $\lambda$. Specifically, we compute $t_{circ}=t(e,p_{\text{fixed}})=\int(de/dt)^{-1}de$. Fig. \ref{fig:circ-time-vs-e} shows the results for a fixed orbit size of $p_{\text{fixed}}=50$. We observe that at a given value of $\lambda$, the circularization time is longest at high eccentricities. However, at larger values of $\lambda$ the circularization time shortens at higher eccentricities.

We now reconstruct hybrid orbital trajectories in the following manner. The method of osculating orbits \cite{pound2008osculating,poisson2014gravity} supposes the true physical orbit evolves in time through its orbital parameters, such that at every instant, the orbit is a geodesic. The quasi-Keplerian equation of the orbit is of the form:
\begin{equation}
    r(\psi)=\frac{pM}{1+e\cos{\psi}}\rightarrow r(t)=\frac{p(t)M}{1+e(t)\cos{\psi(t)}}.
\end{equation}
where the orbital parameters $\{p,e,\psi\}$ are now functions of time. 

The angular frequency, $\omega$, of the CO orbiting the central BH is derived directly from the geodesics of the SdS spacetime. Using the definition
$\omega=(d\phi/d\tau)(dt/d\tau)^{-1}$, and considering the orbital plane at $\theta=\pi/2$, with the angular momentum $L$ and energy $E$ expressed in terms of $(p,e)$ and $r(\psi)$ for a bound orbit, the angular frequency in the large-$p$ limit is given by:
\begin{equation}
	\omega=\left(\frac{L}{r^2}\right)\left(\frac{E}{f(r)}\right)^{-1}=n(1+e \cos (\psi ))^2
	\label{eq:phifrequency}
\end{equation}
where $n$ is the corrected mean motion or the angular frequency of a circular orbit modified by the SdS parameter,
\begin{equation}
	n=\frac{1}{p^{3/2}} -\lambda\left(\frac{p^{3/2} }{2 }+\mathcal{O}(\lambda p^{1/2})\right).
	\label{eq:meanmotion}
\end{equation}
To eliminate oscillations in the angular frequency due to dependence on the true anomaly $\psi$, we perform an averaging procedure over a full orbital cycle. 
\begin{eqnarray}
	\frac{1}{P}\int_{t_0}^{t_0+P}\omega dt&=&\frac{1}{P}\int_{t_0}^{t_0+P}n(1+e \cos (\psi ))^2 dt
	\label{eq:nderivation1}
\end{eqnarray}
Assuming that the argument $\delta$ of the periapsis remains constant over one complete orbit, 
\begin{equation}
    \omega=\frac{d\phi}{dt}=\frac{d\psi}{dt}+\frac{d\delta}{dt}\sim\frac{d\psi}{dt},
\end{equation}
and we set $dt=\omega^{-1}d\psi$. Then we integrate:
\begin{equation}
	\int_{0}^{2\pi}(1+e \cos (\psi ))^{-2}d\psi=\int_{t_0}^{t_0+P}n \,dt=\mathcal{M}=2\pi
	\label{eq:meananomaly}
\end{equation}
over $\psi=(0,2\pi)$ corresponding to the angle swept after a period $P$ telling us that over a complete radial cycle, the orbit with angular frequency equal to the mean motion $n$, completes a full cycle in $\psi$. Since the right hand side corresponds to the mean anomaly $\mathcal{M}$, which is equal to $2\pi$, the mean motion $n$ serves as the averaged angular frequency over a single cycle from $\psi=0$ to $\psi=2\pi$ for a generic eccentric orbit. We consider the mean motion as the characteristic angular frequency of the orbit, which will directly govern the phase evolution of the emitted gravitational wave. 

The osculating equations of the orbital parameters is appended by the equation for $\psi$, 
\begin{eqnarray}
    \frac{d \psi}{dt}=n(p)(1+e\cos{\psi})^2
\end{eqnarray}
with $n(p)$ (Eq. \eqref{eq:meanmotion}) as the orbital mean motion as a function of $p$.  Within a single cycle (an orbital period timescale), we impose that only $\psi$ changes significantly, and that the changes in $p$ and $e$ happen only as a secular drift (on the adiabatic timescale, or the radiation reaction timescale). Within the adiabatic timescale, such that the true anomaly remains the single orbital parameter that changes within an orbital period, our hybrid orbital trajectory will be valid.
Considering the radial motion along an orbital $x-y$ plane, we have the positions:
\begin{equation}
    x=r(t)\cos{\phi(t)},\quad y=r(t)\sin{\phi(t)}
\end{equation}
with the coordinate azimuthal angle $\phi=\psi+\delta$, related to the angle at the periapsis, which also becomes an evolving parameter $\delta\rightarrow\delta(t)$. We only include the leading order (1pN) rate of change in $\delta$,
\begin{equation}
    \Delta\delta_{1pN}=\frac{6\pi}{p}.
\end{equation}
%\blue{Since $\delta$ depends on the evolving orbital size, we can estimate how much it changes due to the contribution of the SdS parameter to the gravitational flux. Using the dimensionalized inspiral rate $\dot{p}$ from \eqref{eq:dimpdot}, we find that after one complete orbit, the periapsis shifts by a coordinate angle:
%\begin{equation}
%	\phi=2\pi\left(1+\frac{3}{p}\frac{G M}{c^2}\right)\sim2\pi\left(1+\left(\frac{1\;  \mathrm{ km}}{p}\right)\left(\frac{M}{M_{\odot}}\right)\right).
%	\label{eq:dimapsis}
%\end{equation}
%For an EMRI orbit over a 1-year evolution, the expected change in periapsis phase shift due of to the SdS gravitational flux is of the order $\Delta \delta=\delta_{Sch}-\delta_{SdS}\sim10^{-30}$ which is extremely negligible. Averaging over an orbital period (single radial cycle) $P=2\pi/n(p)$, the rate of periapsis precession becomes the following expression:
%\begin{equation}
%\frac{d\delta}{dt}=\frac{\Delta\delta_{1pN}}{P}=\frac{3n(p)}{p}
%\end{equation}
%which becomes implicitly an osculating orbital element through $p$ and $e$.}

\begin{figure*}
	\centering
	\begin{tabular}{ccc}
	   \includegraphics[width=0.43\linewidth]{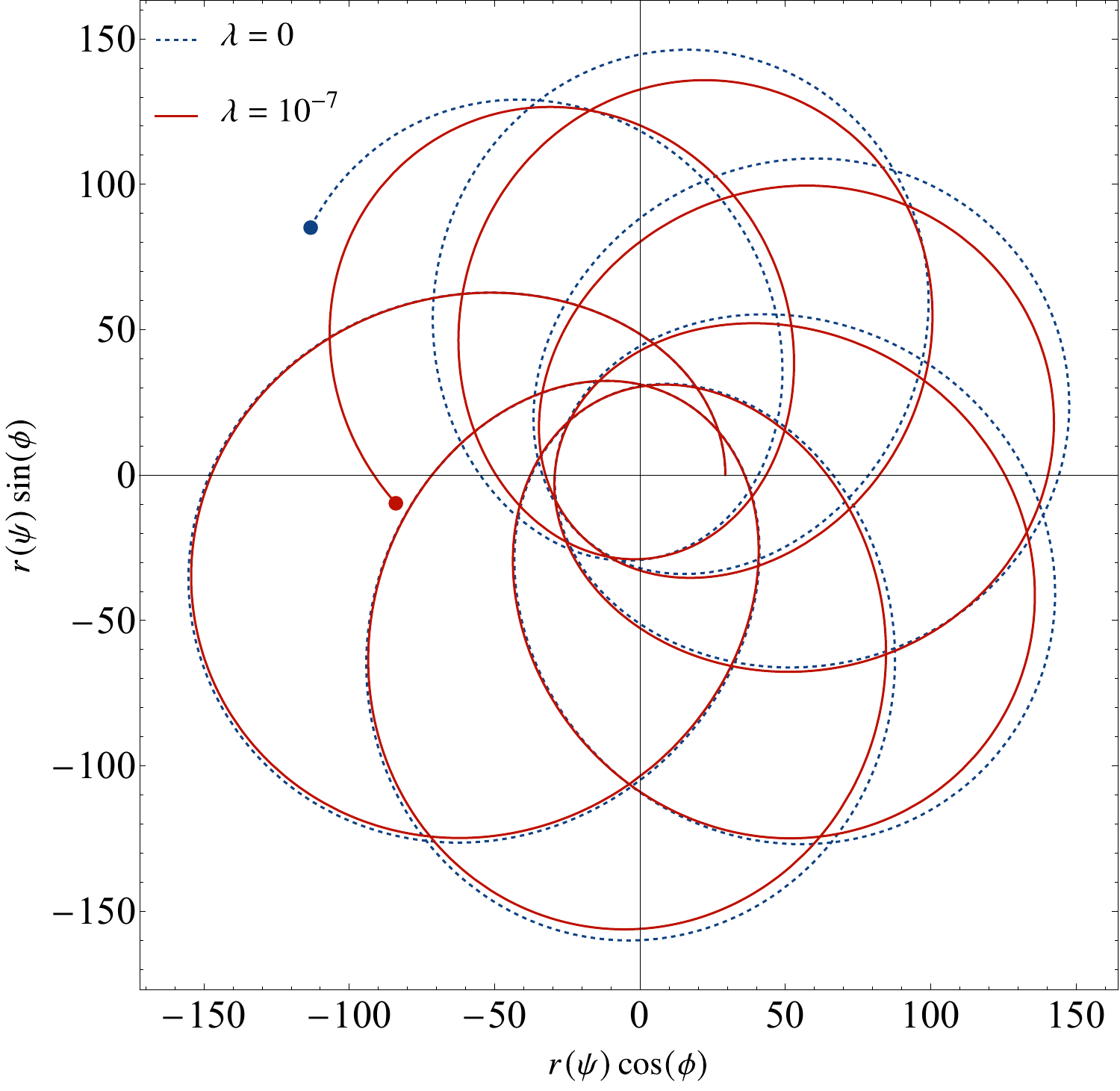}& \includegraphics[width=0.44\linewidth]{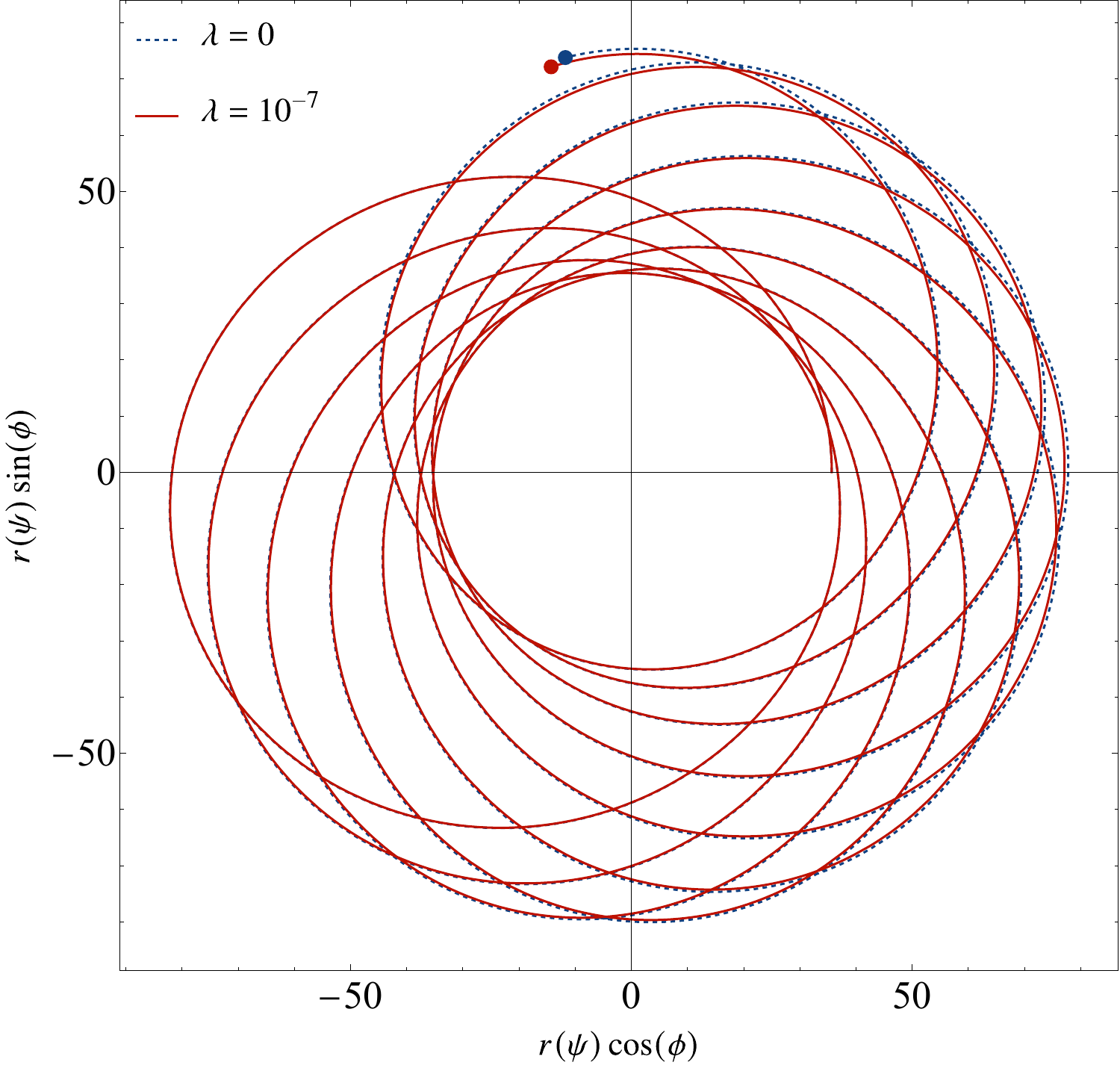}\\ 
		\\
		(a.1) $p_0=50$, $e_0=0.7$ at $t=3.1\times10^{4}$&(a.2) $p_0=50$, $e_0=0.4$ at $t= 2.6\times10^4$
		\\
		\\
		\includegraphics[width=0.43\linewidth]{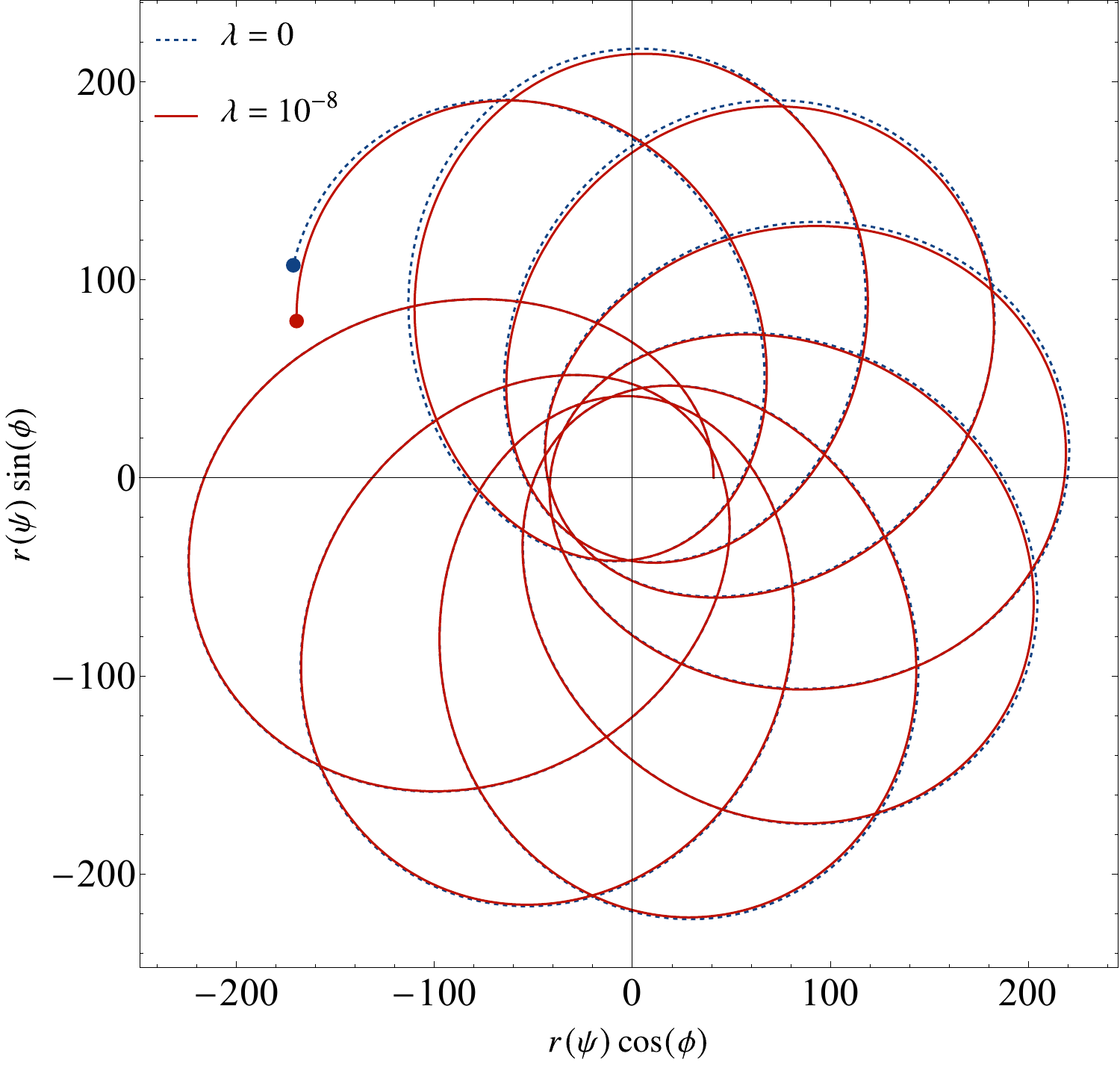}& \includegraphics[width=0.44\linewidth]{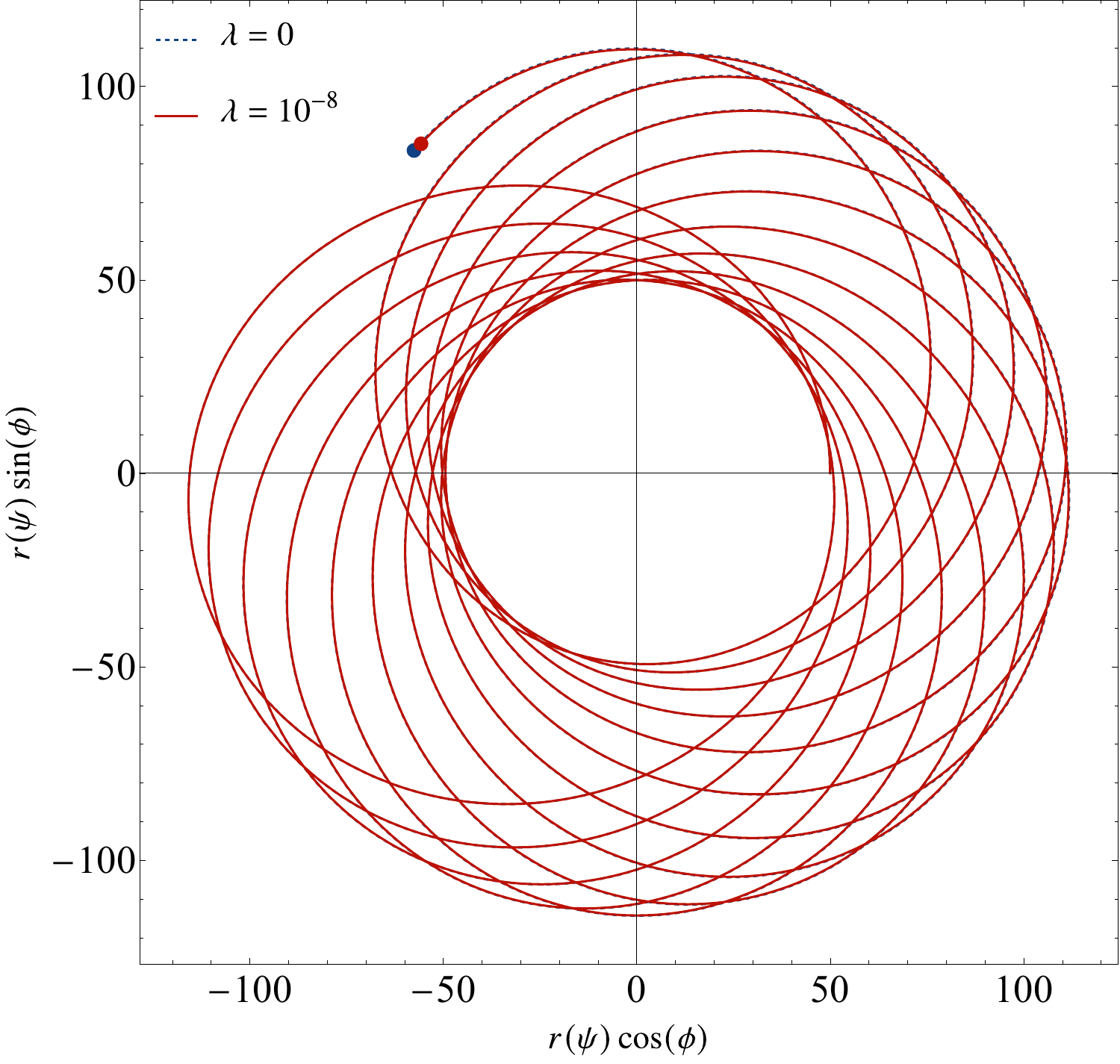}\\ 
		\\
		(b.1) $p_0=70$, $e_0=0.7$ at $t=7.3\times10^{4}$&(b.2) $p_0=70$, $e_0=0.4$ at $t=6.3\times10^{4}$
	\end{tabular} 
	\caption{Sample eccentric orbit shapes for various initial $p$ for Schwarzschild orbits (blue, dashed) and for SdS orbits (red, solid) with (a) $\lambda=10^{-7}$ and (b) $\lambda=10^{-8}$. The corresponding colored dots are the position at the given time. We show here sample  orbital trajectories achievable by initial $(p_0,\,e_0)$ indicated below each figure. We also show the time stamps below each figure for the initial parameters to achieve the corresponding orbital shape. We see here that the deviation caused by the SdS parameter is more relevant for orbits of higher eccentricity. We also notice a lag in the phase to complete a radial cycle caused by $\lambda$.}
	\label{fig:sample-ecc-orbits}
\end{figure*}

\begin{figure}
	\centering
	\includegraphics[width=\linewidth]{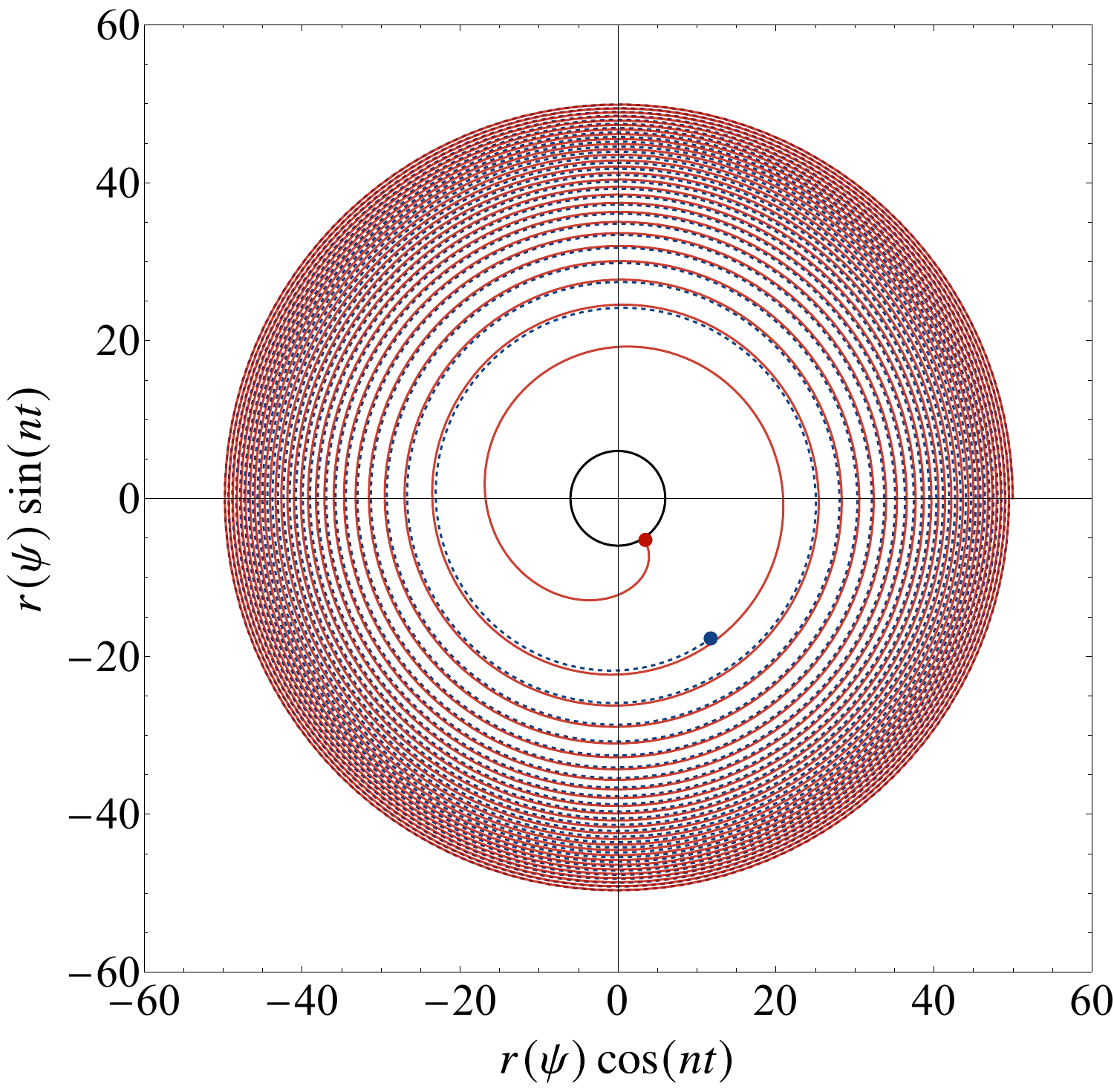}
	\\
	\caption{Sample circular orbit in Schwarzschild (blue-dashed) and  SdS (red-solid) at $\lambda=10^{-7}$ spacetime with initial separation $p_0=50$ at times $t=(0,1.18\times10^5)$. At short timescales, the orbital deviations are negligible, but at long timescales, SdS orbits deviate noticeably from the Schwarzschild orbit, with the SdS parameter facilitating faster evolution, reaching the separatrix faster (black) or plunging the black hole earlier.}
	\label{fig:sample-circ-orbits}
\end{figure}

Along with the radiation reaction equations, \eqref{eq:pdotsds} and \eqref{eq:edotsds}, which could be interpreted as the (secularly averaged) $p$ and $e$ change after a single radial cycle, the complete adiabatic system are the system of equations:
\begin{eqnarray}
    \frac{d \psi}{dt}&=&n(p)(1+e\cos{\psi})^2\\
    \frac{d\delta}{dt}&=&\frac{3n(p)}{p}\\
    \frac{dp}{dt}&=&-\frac{64}{5}p^{-3}(1-e^2)^{3/2}\left(1+\frac{7}{8}e^2\right)\nonumber\\
	&&-\lambda\Bigg[16(1-e^2)-\frac{4 \left(26 e^4-283 e^2-168\right)  }{15 \left(1-e^2\right)^{3/2}}
    \nonumber\\
    &&\quad+\mathcal{O}(\lambda p)\Bigg]\\
	\frac{de}{dt}&=&-\frac{304}{15}p^{-4}e(1-e^2)^{3/2}\left(1+\frac{121}{304}e^2\right)\nonumber\\
	&&-\lambda\Bigg[\frac{32}{3}p^{-1}e^{-1} \left(1-e^2\right)^{3/2} \left(1-(1-e^2)^{1/2}\right)\nonumber\\
	&&-p^{-1}e^{-1}\frac{2 \left(73 e^6-784 e^4-965 e^2-24\right)}{15  \left(1-e^2\right)^{3/2} }
    \nonumber\\
    &&\quad
    +\mathcal{O}(\lambda p^{-2})\Bigg].
\end{eqnarray}
Upon solving for the evolution of the osculating elements, we can reconstruct a sample orbit, with the fast evolution of $\psi$, as well as the secular drift of $\{p,e,\delta\}$.  We note here that the equations for $\psi$ and $\delta$ brings about conservative changes to the orbit while the equations for $p$ and $e$ comes from the dissipation of energy and angular momentum of the orbit. We present orbital solutions for a selected set of initial conditions, $(p_0=70,\lambda=10^{-8})$ and $(p_0=50,\lambda=10^{-7})$ with eccentricities  $e_0=0.7$ , and $e_0=0.4$ in Fig. \ref{fig:sample-ecc-orbits}. Additionally, Fig. \ref{fig:sample-circ-orbits} illustrates the evolution of circular orbits for $p_0=50$ and $\lambda=10^{-7}$. These initial conditions and timestamps were chosen to visually highlight noticeable orbital deviations induced by the SdS parameter. Beyond accelerating the decrease in binary separation $r(\psi)$, the SdS parameter also introduces a phase shift, contributing to enhanced precession of the periapsis. 

%\blue{Lastly, we also estimate the conservative corrections to the orbital period $P$ due to the SdS parameter,
%\begin{eqnarray}
%	P=\frac{2\pi}{n}=2\pi\sqrt{\frac{p^3}{G M}}\left(1+\frac{\Lambda}{6}p^3\frac{c^2}{G M}\right)
%	\label{eq:nperiod}
%\end{eqnarray}
%where we have brought back the physical constants for increased readability. Considering an EMRI of mass $\mu=10M_{\odot}$ orbiting a black hole of mass $M=10^{6}M_{\odot}$ in a de Sitter universe with a cosmological constant $\Lambda\sim10^{-46}$ km$^-2$, we obtain the following estimate for the orbital period:
%\begin{eqnarray}
%	P&\sim&2\pi\times 10^{4}\mathrm{yrs}\left(\frac{p}{1\;\mathrm{pc}}\right)^{3/2}\left(\frac{10^{6}M_{\odot}}{M}\right)^{1/2}
 %   \nonumber\\
    %&&\times\Bigg(1+10^{-24}\left(\frac{p}{1\;\mathrm{pc}}\right)^{3}\left(\frac{10^{6}M_{\odot}}{M}\right)
    %\nonumber\\
    %&&\times\left(\frac{\Lambda}{10^{-46}\mathrm{  km}^{-2}}\right)\Bigg).\nonumber\\
	%\label{eq:dimperiod}
%\end{eqnarray}
%At an orbital separation of 1 pc, the correction from the cosmological constant is negligible. However at much larger separations ($\sim\mathcal{O}(10^8)$ pc), the effect on the orbital period may become noticeable. Nonetheless, when accounting for gravitational radiation reaction, this estimate is further reduced for larger $\lambda$ value, and also since the orbital period depends on the evolving orbital size.}

\section{Gravitational waveforms}
\label{sec:sec4}

In this section, we examine the gravitational waveforms generated by an orbit under adiabatic evolution, where the dominant governing equation is the balance of energy and angular momentum flux. Adiabatic evolution happens when changes in the background metric due to variations in the geodesics are negligible, allowing us to ignore perturbations to the spacetime. The energy and angular momentum flux influences only the orbital parameters, while the trajectory remains constrained to the background geodesics. 

\subsubsection*{Adiabatic approximation}
\label{sec:sec4.0.1}

Adiabatic evolution is defined by the requirement that any relevant orbital parameter $q=(p,e)$ evolves much more slowly than the radial orbital period $P=2\pi/\Omega_r$ (where $\Omega_r$ is the radial angular frequency), such that $\Delta q=|\dot{q}|P\ll q$ \cite{cutler1994gravitational}. By analyzing the adiabatic evolution condition, we can effectively divide the full orbital parameter space into two distinct regions, (1) parameter regions far from the separatrix, where the weak-field formulation of the fluxes is sufficient to describe the orbital evolution and waveforms of the inspiral and (2) parameter regions near the separatrix, where a strong-field formulation from black hole perturbation theory must be applied. 
%\begin{itemize}
%	\item \textit{Regions of slow evolution}, where the weak-field gravitational fluxes will be sufficient to describe the evolution of the orbit.
%	\item \textit{Regions of fast evolution}, where a strong-field formalism for energy fluxes (such as black hole perturbation theory) is required, along with corrections to the full metric to account for changes in the background geodesics (self-consistent evolution with self-force effects). 
%\end{itemize}
Adiabatic evolution may still hold in both regions, but as we will see later, additional conditions must be met, imposing further constraints on the orbital parameters.

\subsection{Orbits near the separatrix}
\label{sec:sec4.1}

Near the separatrix, orbits are expected to be affected by the black hole so strongly that describing their orbital evolution using a weak-field formulation is inadequate. This happens as the CO continuously decreases its distance to the black hole, increasing the rate of gravitational energy release. A full computation of the gravitational fluxes in the strong-field regime is beyond the scope of this paper and will be presented in a companion work, where black hole perturbation theory is applied to Schwarzschild - de Sitter spacetime. However, the adiabatic approximation can still be examined in this region to determine the additional conditions necessary for its validity.

We derive the radial period $P$ near the separatrix, $p\sim p_{\text{sep}}$ (Eq.\eqref{eq:sdsseparatrix}) to examine the time scale at which adiabatic evolution holds. Assuming that changes in angular frequency and orbital period remain within the adiabatic limit, we approximate the orbital period at the separatrix up to linear in $\lambda$ and  $\Delta p\sim p-p_{\text{sep}}$:
\begin{eqnarray}
	P|_{\Delta p\sim p-p_{\text{sep}}\ll1}&\sim&12   \sqrt{6}\pi
    +3   \sqrt{6}\pi\,\Delta p
    \nonumber\\
    &&+\lambda\left(1296  \sqrt{6}\pi +972   \sqrt{6}\pi \,\Delta p\right). 
	\label{eq:period}
\end{eqnarray}
We assume here that orbits near the separatrix have already circularized, and we introduce $\Delta p$ to describe deviations from the separatrix. The validity of the adiabatic approximation is then tested by examining the evolution of $p$ under the prescribed weak-field fluxes:
\begin{eqnarray}
	\frac{|\dot{p}|P}{p}&\sim&\frac{16}{45}   \sqrt{\frac{2}{3}}\pi q-\frac{4}{27} \sqrt{\frac{2}{3}} \pi  \Delta p\; q
    \nonumber\\
    &&+\lambda  \;q\left(\frac{56 \sqrt{6} \pi }{5} \Delta p\; +\frac{672 \sqrt{6} \pi }{5}\right)
	\label{eq:adiabatic}
\end{eqnarray}
where we brought back the mass ratio scaling $q:=\mu/M$ from Eq. \eqref{eq:pdotsds} and \eqref{eq:edotsds}. This leads to a constraint on the mass ratio:
\begin{eqnarray}
	\frac{\mu}{M}&\sim&\frac{45 \sqrt{\frac{3}{2}}}{16 \pi }+\frac{75 \sqrt{\frac{3}{2}} \; }{64 \pi }\Delta p
    \nonumber\\
    &&-\lambda  \left(\frac{25515 \sqrt{\frac{3}{2}}}{8 \pi }+\frac{93555 \sqrt{\frac{3}{2}} }{32 \pi }\Delta p\right).
	\label{eq:nearmassratio}
\end{eqnarray}
We now assess whether the mass ratio constraint can remain within the physical range $q\in(0,1)$. In the absence of the SdS parameter $(\lambda=0)$, the constraint leads to unphysical scenarios, producing values outside the range of $q$ for $\Delta p$, i.e., for $p>p_{sep}$. This invalidates the use of weak-field fluxes for adiabatic evolution in the strong-field region. Furthermore, inclusion of a SdS parameter does not help loosen the validity of this approximation; for the constraint to yield physically meaningful mass ratios, an unrealistically large SdS parameter $(\lambda\sim10^{-3})$ well beyond the bound for stable orbits $(\lambda\sim10^{-4})$. Although the breakdown of our initial assumptions (such as the applicability of pN fluxes near the separatrix) suggests that this analysis is not strictly valid in the strong-field regime, this calculation serves as a preliminary attempt to explore the influence of the SdS parameter on the limits of adiabatic evolution. 

We now present preliminary results comparing the weak-field energy flux formula with the strong-field energy flux at infinity, obtained from the perturbation of spherically symmetric spacetimes due to an orbiting point mass. This comparison provides insight into the validity range of the weak-field formula. The strong-field energy flux is obtained using methods of Schwarzschild perturbation theory \cite{martel2004gravitational,martel2005gravitational,hopper2010gravitational}, generalized to spherically symmetric spacetimes to accommodate the Schwarzschild-de Sitter spacetime. We briefly outline the method, with the complete procedure  detailed in a companion paper. The goal is to numerically calculate the energy flux valid near the separatrix of the Schwarzschild-de Sitter spacetime:
\begin{equation}
	\langle \dot{E}^{\infty}_{\ell m} \rangle=\frac{1}{64}\frac{(l+2)!}{(l-2)!}\sum_n\omega^2_{mn}|C^{+}_{\ell mn}|^2.
	\label{eq:EdotSF}
\end{equation}
Here $\omega_{mn}$ is the angular frequency of the orbiting particle in terms of the $(m,n)$ modes of its radial and azimuthal oscillations.
The normalization coefficient $C^{+}_{\ell mn}$ is obtained by numerically integrating the inhomogeneous Regge-Wheeler-Zerilli equation in the frequency domain with the tortoise coordinate defined by $\frac{dr}{dr_{*}}=f^{-1}$,
\begin{equation}
	\left(\frac{d^2}{dr_{*}^2}-V_\ell+\omega_{mn}^2\right)R_{\ell mn}(r)=Z_{\ell mn}(r)
\end{equation}
where $R_{\ell mn}$ satisfies the general solution of the form:
\begin{equation}
	R_{\ell m n}(r)=c_{\ell m n}^{+}(r) \hat{R}_{\ell m n}^{+}(r)+c_{\ell m n}^{-}(r) \hat{R}_{\ell m n}^{-}(r)
\end{equation}
with the coefficients defined as:
\begin{eqnarray}
	\begin{aligned}
		& c_{\ell m n}^{+}(r) \equiv \frac{1}{W_{\ell m n}} \int_{r_{\min }}^r d r^{\prime} \frac{\hat{R}_{\ell m n}^{-}\left(r^{\prime}\right) Z_{\ell m n}\left(r^{\prime}\right)}{f\left(r^{\prime}\right)},\\
		&C_{\ell m n}^{+} \equiv c_{\ell m n}^{+}\left(r_{\max }\right)\\
        & c_{\ell m n}^{-}(r) \equiv \frac{1}{W_{\ell m n}} \int_{r}^{r_{\max}} d r^{\prime} \frac{\hat{R}_{\ell m n}^{+}\left(r^{\prime}\right) Z_{\ell m n}\left(r^{\prime}\right)}{f\left(r^{\prime}\right)}\\
		&
		C_{\ell m n}^{-} \equiv c_{\ell m n}^{-}\left(r_{\min}\right).
	\end{aligned}
\end{eqnarray}
The source term $Z_{\ell mn}(r)$ is derived from the stress-energy tensor of a point particle, of mass $\mu$, in a generic orbital trajectory, $x_p(\tau)$:
\begin{equation}
	T^{\mu \nu}\left(x^\alpha\right)=\mu \int \frac{d \tau}{\sqrt{-g}} u^\mu(\tau) u^\nu(\tau) \delta^4\left[x-x_p(\tau)\right].
\end{equation}
The homogeneous solutions $\hat{R}_{\ell m n}^{\pm}(r)$ are obtained by solving the homogeneous Regge-Wheeler-Zerilli equation with the following asymptotic ingoing and outgoing wave boundary conditions:
\begin{equation}
	\hat{R}_{\ell m n}^{\pm}(r_{*}\rightarrow\pm\infty)=\exp(\pm i\omega_{mn}r_{*}).
\end{equation}
Although this method allows for a generic orbital configuration, we restrict the preliminary analysis to the circular limit and consider only the dominant $(l,m)$ energy flux mode. This simplification is sufficient to provide a rough verification of the weak-field formula. Specifically, we solve for $E^{\infty}_{22}$ using the Zerilli potential, modified to incorporate the SdS parameter (with the full derivation included in the companion paper):
\begin{align}
	V_{\text{even}}=& \frac{f}{r^2 \tilde{\Lambda}^2}\left[2 \tilde{\lambda}^2\left(\tilde{\lambda}+1+\frac{3 M}{r}\right)+\frac{18 M^2}{r^2}\left(\tilde{\lambda}+\frac{M}{r}\right)\right]
    \nonumber\\
    &-\frac{f}{r^2 \tilde{\Lambda}^2}(6 M^2 \Lambda )
\end{align}
with constants (not related to $\lambda$ or $\Lambda$)  $\tilde{\Lambda}=\tilde{\lambda}+3M/r$ and $\tilde{\lambda}=(l+2)(l-1)/2$. 

We note here however, that \eqref{eq:EdotSF} is the expected radiation rate if the spacetime is asymptotically flat. This is a problem since we know that SdS is asymptotically de - Sitter, and the energy flux in terms of the fields should be modified by the SdS parameter. Luckily, many prescriptions on computing the energy flux have been put forward by analyzing de Sitter in Bondi frame, and computing the mass loss in terms of the news tensor, notably by \cite{ashtekar2015asymptotics,hoque2018propagation,szabados2019review}
 and more recently, de Sitter quadrupolar linearized waves by \cite{hoque2024sitter,bonga2023gravitational,compere2025so}. It was shown that the even parity, $\ell=2$ fields (at the Bondi frame of the asymptotically de Sitter null infinity) carry energy far from the source, $r\rightarrow\infty$, (a la' Teukolsky waves \cite{teukolsky1982linearized}) as
 \begin{eqnarray}
     \dot{E}^{\ell=2,m=\pm2}&=&-\frac{3}{8\pi G}\Big|\partial_{u}^3A_m-\frac{4}{3}\Lambda\;\partial_{u} A_m\Big|^2
     \nonumber\\
     &&-\Lambda\;|\partial_u^2A_m|^2+\mathcal{O}(\Lambda^2)
     \label{eq:EdotSFBonga}
 \end{eqnarray}
 where $A_m(u)$'s
 are even-parity solutions to the linearized Einstein field equations that reduces to the asymptotically flat limit at $\Lambda\rightarrow0$. Throughout our energy flux comparisons, we only use the asymptotically flat limit of the energy flux formula, since all succeeding terms from the $\Lambda$ correction have subdominant contributions compared to the leading term.
 
The energy flux computed from the solutions of the Regge-Wheeler-Zerilli equations is expected to be independent to a far-field (large-$p$) approximation, as these equations are derived from perturbations of the SdS spacetime due to an orbiting test mass. Hence, the strong-field energy flux formula should remain valid even at large orbital separations. This allows us to assess the accuracy of the weak-field energy flux formula (Eq. \eqref{eq:Edot}) by evaluating its deviation from the strong-field energy flux (Eq. \eqref{eq:EdotSF}). 

\begin{figure*}
	\centering
	\includegraphics[width=0.8\linewidth]{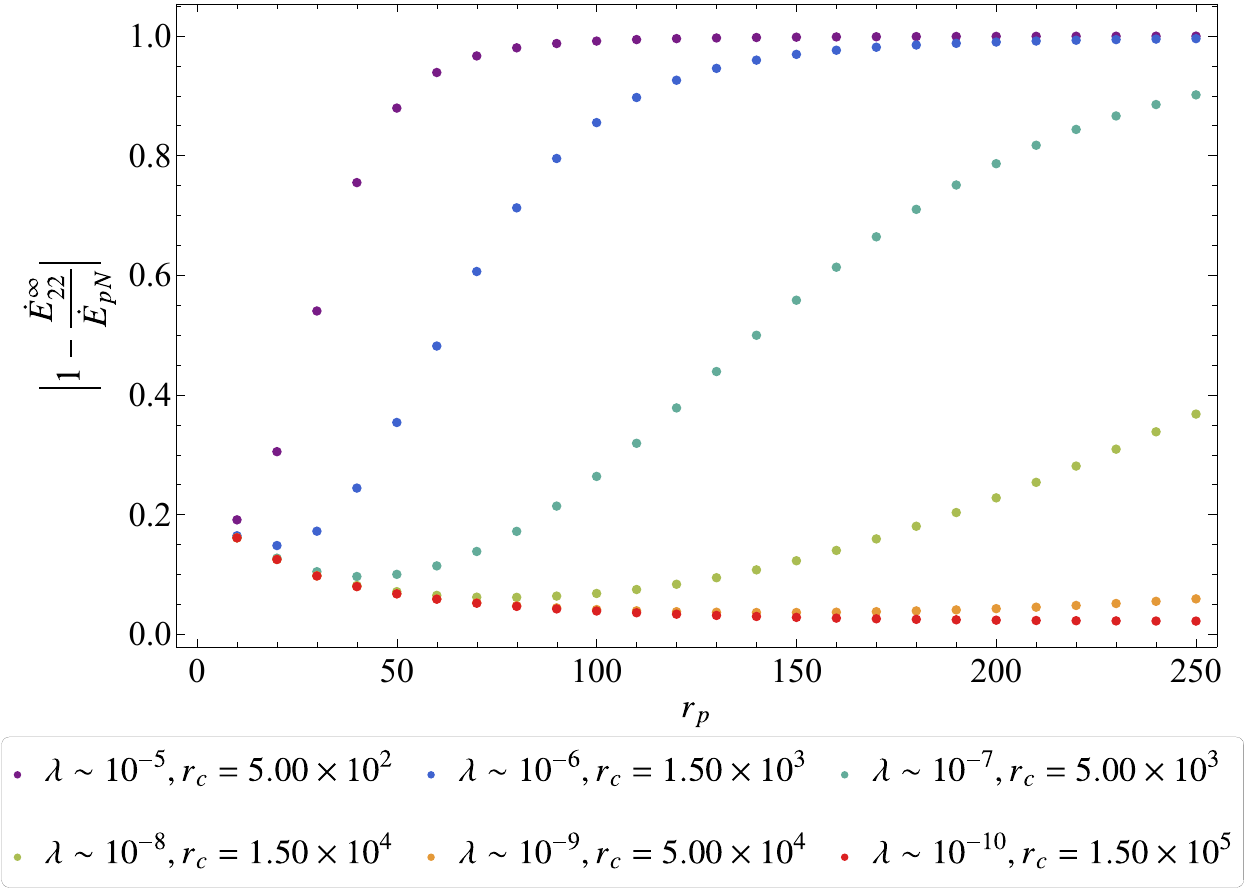}
	\caption{Deviation of the weak-field energy flux formula (Eq. \eqref{eq:Edot}) and the strong field energy flux ($l=2,m=2$ mode) from SdS perturbation theory (Eq. \eqref{eq:EdotSF}) for a circular orbit of radius $r_p$. The validity of the weak-field energy flux formula is stronger for larger $r_p$ and lower $\lambda$. For smaller $r_p$, the weak-field formula may still be valid if $\lambda$ is allowed to be of lower value.}
	\label{fig:e22vspnerror1}
\end{figure*}

In Fig. \ref{fig:e22vspnerror1}, we illustrate the deviation of the energy flux obtained from perturbation theory (Eq. \eqref{eq:EdotSF}) and that from a weak-field expansion (Eq.\eqref{eq:Edot}), $1-|\dot{E}_{22}^{\infty}/\dot{E}_{pN}|$, for increasing binary separation, $r_p$, and SdS parameter, $\lambda$. 
\begin{figure*}
	\centering
	\includegraphics[width=0.8\linewidth]{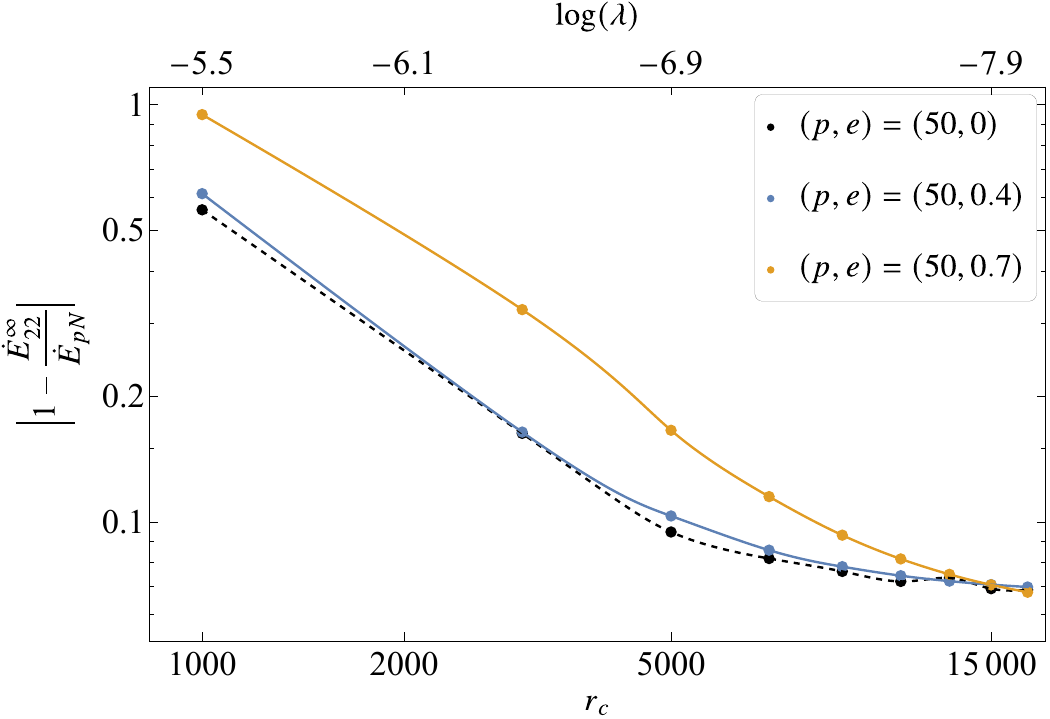}
	\caption{Deviation of the weak-field energy flux formula (Eq. \eqref{eq:Edot}) and the strong field energy flux ($l=2,m=2$ mode, $n_{\text{max}}=30$) from SdS perturbation theory (Eq. \eqref{eq:EdotSF}) for generic eccentric orbits in terms of the SdS parameter, $\lambda$, and cosmological horizon, $r_c$.}
	\label{fig:e22vspnerrorecc}
\end{figure*}
We also compare the energy flux of circular orbits to eccentric orbits, showing the deviations in Fig. \ref{fig:e22vspnerrorecc}. Our results indicate that for higher values of $\lambda$, significant deviations emerge at sufficiently large $r_p$, confirming that the weak-field quadrupolar energy flux (Eq. \eqref{eq:Edot}) remains valid only within the particle position satisfying the hierarchy (Eq.\eqref{eq:hierarchy}). 
Additionally, for smaller $\lambda$, the OSCO shifts farther from the black hole, leading to an increasingly larger region of validity for the weak-field fluxes. 
Overall, we find that higher $\lambda$ values lead to a scaling error in eccentric orbits relative to their circular counterparts. However, as expected, for smaller $\lambda$, both circular and eccentric orbits exhibit similar deviations in the fluxes.

\subsection{Orbits far away from the separatrix}
\label{sec:sec4.2}

It has been shown that in regions far from the separatrix, and thus the black hole, applying adiabatic approximation on orbital evolution results to no strict conditions for the mass ratio of the binary \cite{cutler1994gravitational}. We repeat the analysis here to determine whether the presence of a SdS parameter introduces new constraints. Similar to the derivation from previous subsection, the change in the semi-latus rectum in the large-$p$ limit yields the following constraint on the mass ratio:
\begin{equation}
	\frac{\mu}{M}\ll \alpha(e) p^{5/2}-\lambda(\beta(e)  p^{11/2}+\mathcal{O}(\lambda p^{7/2}))
	\label{eq:farmassratio}
\end{equation}
where $\alpha(e)$ and $\beta(e)$ are positive functions of eccentricity for all values of $e$. This result confirms that the constraint is automatically satisfied for large $p$, implying that the validity of the weak-field fluxes for adiabatic evolution remains independent of the mass ratio. Hence, at large distances, the evolution of the orbital parameters is necessarily adiabatic. We note however that the presence of a SdS parameter $\lambda$ strengthens the bound at which adiabaticity may be applied to eccentric orbits. This suggests that in the case of eccentric orbits, the strong-field region is effectively shifted towards much lower $p$. Thus, the weak-field fluxes will be sufficient for the balance formula governing the orbital trajectory and gravitational waveforms.  

We present below the gravitational waveforms, deriving the phase from the orbital mean motion and the strain magnitude from the binary mass quadrupole moment.  The leading order relationship of the time derivatives of the mass quadrupole moment to the waveform strain will be sufficient to show the effect of the SdS parameter up to order $\Lambda^{1/2}$ since the linearized radiation in asymptotically-de Sitter spacetime was shown to be \cite{date2016gravitational,he2018relationship}:
\begin{equation}
\chi_{ij}\sim\partial_t^2Q_{ij}+\mathcal{O}\Bigg(\frac{\sqrt{\Lambda}}{D}\Bigg)
\end{equation}
with $D$ as the wave zone distance.
 The wave zone radiation was derived from the 
perturbation of de Sitter spacetime (with the cosmological constant related to $H$ by $H^2=\Lambda/3$),
\begin{equation}
    g_{\mu\nu}=\gamma_{\mu\nu}+\epsilon h_{\mu\nu}
\end{equation}
with the background line element,
\begin{equation}
\gamma_{\mu\nu}dx^{\mu}dx^{\nu}=-dt^2+e^{2Ht}\delta_{ij}dx^{i}dx^{j}
\end{equation}
which should be valid far from the source. The resulting linearization of the field equations has the following form,
\begin{eqnarray}
    &&\eta^{\alpha \beta} \partial_\alpha \partial_\beta \chi_{\mu \nu}+\frac{2}{\eta^2} \partial_0 \chi_{\mu \nu}
    \nonumber\\
    &&-\frac{2}{\eta^2}\left(\delta_\mu^0 \delta_\nu^0 \eta^{\alpha \beta} \chi_{\alpha \beta}+\delta_\mu^0 \chi_{0 \nu}+\delta_\nu^0 \chi_{0 \mu}\right)=-16 \pi T_{\mu \nu}\nonumber\\
\end{eqnarray}
with $\eta_{\mu\nu}$ as the components of the Minkowski metric, and with the radiation field
\begin{equation}
    \chi_{\mu\nu}=\exp(-2Ht)\bar{h}_{\mu\nu}
\end{equation}
with $\bar{h}_{\mu\nu}$ as the usual trace-reversed perturbation metric. The radiation field can then be expressed in terms of the source multipole moments,
\begin{equation}
    \begin{aligned}
\chi_{i j}= & \frac{2}{R}\left(e^{-H t}+H R\right)\left(\ddot{Q}_{i j}^{(\rho)}-2 H \dot{Q}_{i j}^{(\rho)}+H \dot{Q}_{i j}^{(p)}\right) 
\\&-2 H\Big(\ddot{Q}_{i j}^{(\rho)}-3 H \dot{Q}_{i j}^{(\rho)}+H \dot{Q}_{i j}^{(p)}
\\
&\quad\quad+2 H^2 Q_{i j}^{(\rho)}-H^2 Q_{i j}^{(p)}\Big).
\end{aligned}
\end{equation}
with $R$ as the radial distance of the source, $R^2=x^2+y^2+z^2$. Higher order terms in $\Lambda$ would represent tail terms in the waveform due to backscattering \cite{ashtekar2015asymptotics2}, which we will ignore here since their contribution to the amplitude and phase is small enough with respect to the leading order. At spatial infinity, the polarization modes of the gravitational wave signal $h$ (in the transverse-traceless gauge), at a distance $D$ from the source, will then be related to the  mass quadrupole moment (we suppress the label $\rho$, $Q^{(\rho)}_{ij}=Q_{ij}$), as it is in an asymptotically flat spacetime (up to leading order in $\Lambda$) \cite{maggiore2008gravitational,he2018relationship} by the following expression:
\begin{eqnarray}
	h_{+}&=&\frac{1}{D}\left(\ddot{Q}_{11}-\ddot{Q}_{22}+2\sqrt{\frac{\Lambda}{3}}\left(\dot{Q}_{11}-\dot{Q}_{22}\right)\right)
    \nonumber\\
    &&+\frac{\Lambda}{3}\left(\dot{Q}_{11}-\dot{Q}_{22}+\frac{2\sqrt{\Lambda}}{3}\left(Q_{11}-Q_{22}\right)\right), \\
    h_{\times}&=&\frac{2}{D}\left(\ddot{Q}_{12}-2\sqrt{\frac{\Lambda}{3}}\dot{Q}_{12}\right)+\frac{\Lambda}{3}\left(\dot{Q}_{12}-2\sqrt{\frac{\Lambda}{3}}Q_{12}\right).\nonumber\\
	\label{eq:quadwaveform}	
\end{eqnarray}
We assume that the binary components lie in a single plane, say the $x-y$ plane, and the observer is aligned along the $z-$axis.  This orientation corresponds to a face-on view of the binary. We further set the orbital plane such that its origin coincides with one of the foci of the elliptic orbit, which would be the position of the central black hole. The mass quadrupole moment of the binary, with reduced mass $\mu$, is then expressed into:
\begin{equation}
	Q_{ij}=\mu\, r^2(\psi)\left(\begin{array}{ccc}
		\cos^2(\phi) &\quad \sin(\phi)\cos(\phi) &\quad 0\\\\ 
		\sin(\phi)\cos(\phi) &\quad \sin^2(\phi) &\quad 0\\\\ 
		0 &\quad 0 &\quad 0
	\end{array}\right).
	\label{quadmoment}
\end{equation}
Substituting in the solutions of the adiabatic osculating trajectory equations into the strain formulas \eqref{eq:quadwaveform}, allows us to generate adiabatic waveforms plots valid in the weak-field region of the parameter space. It would also be sufficient to plot the polarization strains \eqref{eq:quadwaveform} up to the leading order in $\Lambda$ (ignoring the second term) since we are already using the leading $\Lambda$-corrected solutions in the orbital parameters. We note here that under secular evolution \cite{pound2005limitations}, the waveform phase would gain additional contribution from the conservative gravitational self-force, which we do not include in this paper.

\begin{figure*}
	\centering
	\begin{tabular}{c}
		\includegraphics[width=0.8\linewidth]{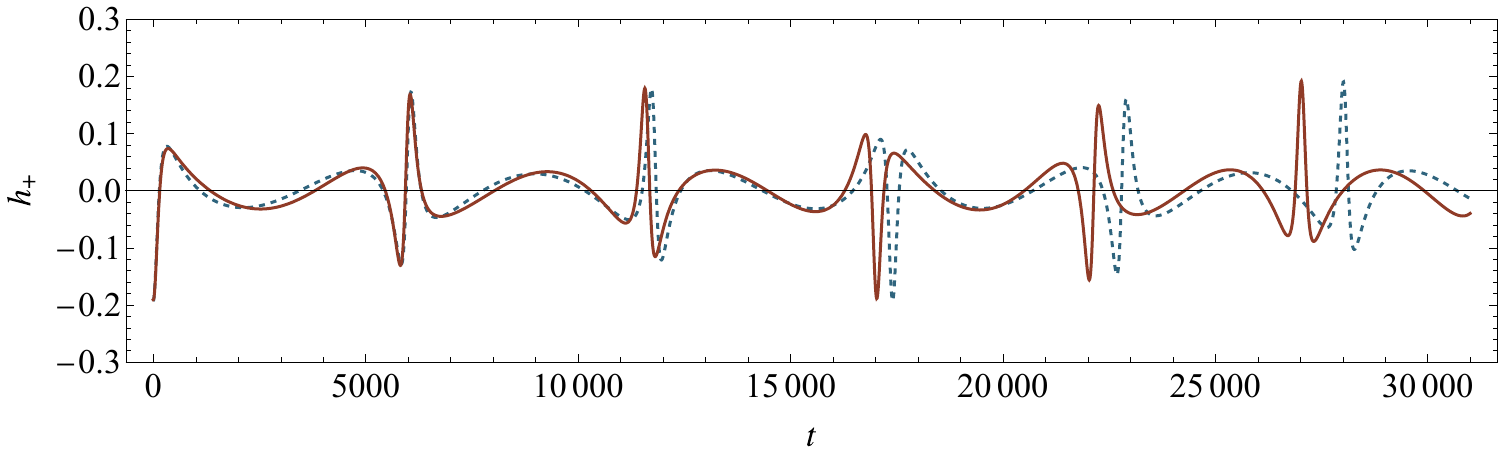}
		\\
		(a)
		\\
		\includegraphics[width=0.8\linewidth]{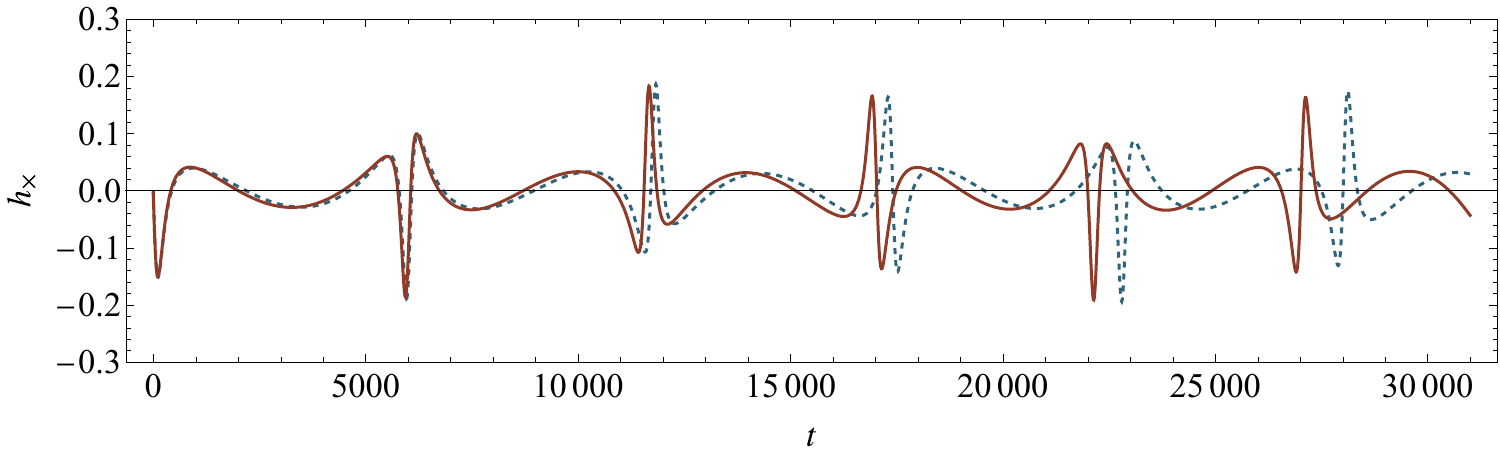}
		\\
		(b)
		\\
		\includegraphics[width=0.8\linewidth]{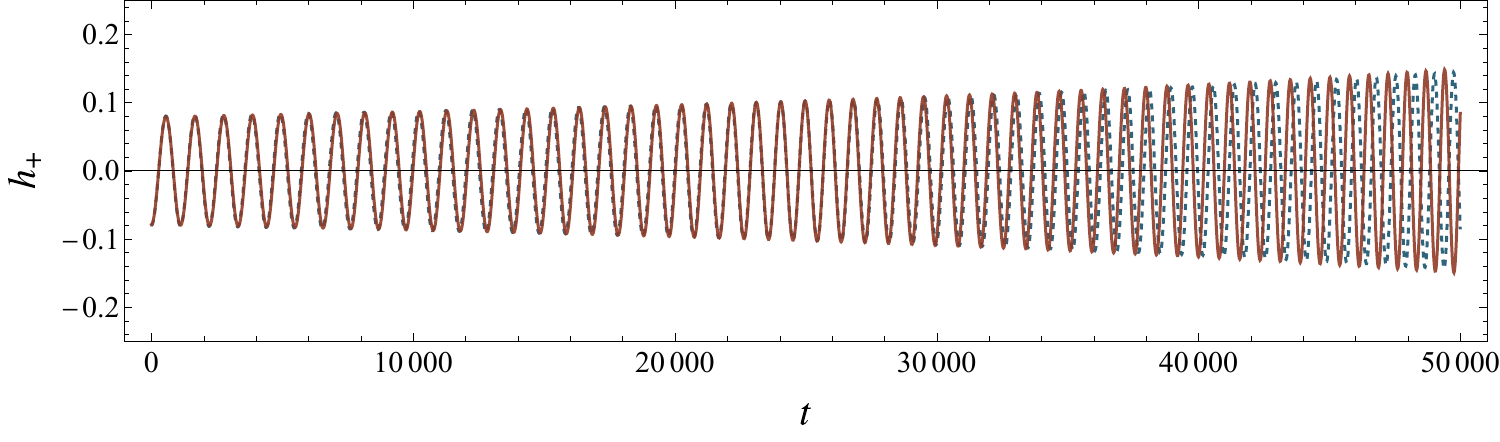}
		\\
		(c)
		\\
	\end{tabular}
	\caption{Adiabatic waveforms of $(p_0=50)$ for Schwarzschild with $\lambda=0$ (dashed, blue) and with SdS with $\lambda=10^{-7}$ (red). We show here the (a) plus polarization  as well as the (b) cross polarization  of the waveform in the adiabatic limit of an initially eccentric orbit with eccentricity $e_0=0.7$.  We show here the plus polarization (c) of the waveform in the adiabatic limit of an initially circular orbit. The main effect of the SdS parameter is the increase in amplitude and an initial delay in the wave phase.}
	\label{fig:waveforms}
\end{figure*}

In Fig. \ref{fig:waveforms}, we present a circular waveform under the assumption that the orbit begins with a small initial eccentricity and circularizes almost instantly, making $\dot{e}$ negligible. We observe an phase advance, which can be attributed to the de Sitter term in the mean motion. This early departure arises from modified angular frequency, with the energy and angular momentum being affected by the SdS parameter the most for nontrivial eccentricities. 

Lastly, we estimate the change both in the amplitude and the phase of the waveform due to the SdS gravitational flux in an EMRI orbit. We begin by considering the analytical expression for complex time-domain waveform \cite{maggiore2008gravitational}:
\begin{eqnarray}
	h(t)=\mathcal{A}(t)\;\exp(i\Psi(t))=h_{+}-i h_{\times}
	\label{eq:hcomplex}
\end{eqnarray}
where the SdS change in amplitude arises from an $\mathcal{O}(\lambda)$ deviation, $\delta\mathcal{A}_{SdS}$, in the amplitude

\begin{equation}
	\mathcal{A}=\mathcal{A}_{pN}(1+\delta\mathcal{A}_{SdS}).
\end{equation}
Similarly, the phase correction, $\delta\Psi_{SdS}$, appears in

\begin{equation}
	\Psi=\Psi_{pN}(1+\delta\Psi_{SdS}).
\end{equation}
 The post-Newtonian amplitude and phase are related to the Keplerian frequency $\omega_N$ as $\mathcal{A}_{pN}=4 \omega_N^{2/3}/D$ and $\Psi_{pN}=2 \omega_N t$. Expressing the semi-latus rectum $p$ in terms of $\omega_N$ via $\omega_{N}=p^{-3/2}$ and using the evolution equations (Eq. \eqref{eq:pdotsds} and \eqref{eq:edotsds}) we obtain first-order SdS corrections:
\begin{eqnarray}
	\delta\mathcal{A}_{SdS}&=&- \lambda\; \omega_N ^{-2} +\mathcal{O}(\lambda^2) \\
	\label{eq:ampshiftSDScirc}
	\delta\Psi_{SdS}&=&- \lambda\; \frac{\omega_N ^{-2}}{2}+\mathcal{O}(\lambda^2).
\label{eq:phaseshiftSDScirc}
\end{eqnarray}
These expressions indicate that the amplitude and phase shift are initially small but grow as the frequency increases. This suggests that both the amplitude and phase of the SdS waveform will eventually deviate significantly from their pN counterparts at an earlier stage in the inspiral.
The Keplerian mean motion, $\omega_N$ may be estimated as:
\begin{equation}
	\omega_N\sim1.22\times10^{-4}\mathrm{yr}^{-1}\left(\frac{1 \;\mathrm{pc}}{p}\right)^{3/2}\left(\frac{M}{10^{6}M_{\odot}}\right)^{1/2}.
	\label{eq:ampSDSdim}
\end{equation}
%\blue{Since both amplitude and phase shifts scale similarly with $\omega_N$, we then obtain an estimate for $\delta\mathcal{A}_{SdS}$ and $\delta\Psi_{SdS}$ in a circular EMRI: 
%\begin{eqnarray}
%\delta\mathcal{A}_{SdS}\sim\delta\Psi_{SdS}&\approx&10^{-24}\left(\frac{p}{1 \; \mathrm{pc}}\right)^{3}\left(\frac{M}{10^{6}M_{\odot}}\right)
%\nonumber\\
%&&\times\left(\frac{\Lambda}{10^{-46}\mathrm{  km}^{-2}}\right).
%	\label{eq:ampSDSestimate}
%\end{eqnarray}
%Using the previous inspiral rate estimate (Eq. \eqref{eq:dimpdot}), a circular inspiral over one year leads to an extremely small reduction in the amplitude/phase. However, as the frequency increases, the amplitude and phase shift eventually surpass their pN counterparts at critical frequency, $\omega_{crit}\sim \lambda^{-1/2}$. Before reaching $\omega_{crit}$, the SdS parameter primarily reduces the amplitude of the waveform and delays the phase evolution.}
As the frequency increases, the amplitude and phase shift eventually surpass their pN counterparts at critical frequency, $\omega_{crit}\sim \lambda^{-1/2}$. Before reaching $\omega_{crit}$, the SdS parameter primarily reduces the amplitude of the waveform and delays the phase evolution.
For initially eccentric orbits, the phase shift remains similar to the circular case, since we consider the waveform frequency to be related to the mean motion averaged over a full orbit. However, the amplitude shift is modified by the eccentricity:
\begin{equation}
	\mathcal{A}(e) D=\frac{2 (e+2) \omega ^{2/3}}{(e+1)^3}-\lambda\frac{4}{\omega ^{4/3} (e+1)^2}+\mathcal{O}(\lambda^2)+\mathcal{O}(\omega^{4/3}).
	\label{eq:ampshiftSDSecc}
\end{equation}
We see in this expression that  more eccentric orbits produce a lower overall amplitude shift due to the SdS parameter compared to circular orbits. The eccentricity affects $\omega_{crit}$ at most by a factor of order $e$, which increases for higher eccentricities. As a consequence, waveforms from highly eccentric SdS orbits take longer to surpass the amplitude and phase of their circular waveform counterparts.

\section{Timescales for gravitational wave detection}
\label{sec:sec5}

In this section, we apply the radiation reaction equations (Eqs. \eqref{eq:pdotsds} and \eqref{eq:edotsds}) to estimate the inspiral timescales for gravitational wave detection. Detectors will be operating for only a limited amount of time and therefore can only detect transient gravitational wave signals from sources for a given range of lengthscale or timescale. For the space-based detector LISA, the operation time is only up to 4 - 10 years before shutting down
%for improvements 
\cite{amaro2007intermediate,colpi2024lisa}. Signals can only enter the detector frequency range and can be detected at the ``lifetime" of the detector, which can translate to sources up to a range of length or timescale. 
%Sources that plunge less than the detector lifetime as well as those that cannot reach the frequency of the detector bandwidth will be irrelevant sources for detection.   
It will then be important to predict how long would relevant sources in a specific range of orbital parameters will stay in the LISA band for detection. Specifically, we obtain an analytical inspiral timescale formula in terms of orbital parameters that can describe the decay of a binary, similar to what was derived by  \cite{peters1964gravitational} and extract what effect the SdS parameter may bring. 

\begin{figure}
	\centering
	\includegraphics[width=\linewidth]{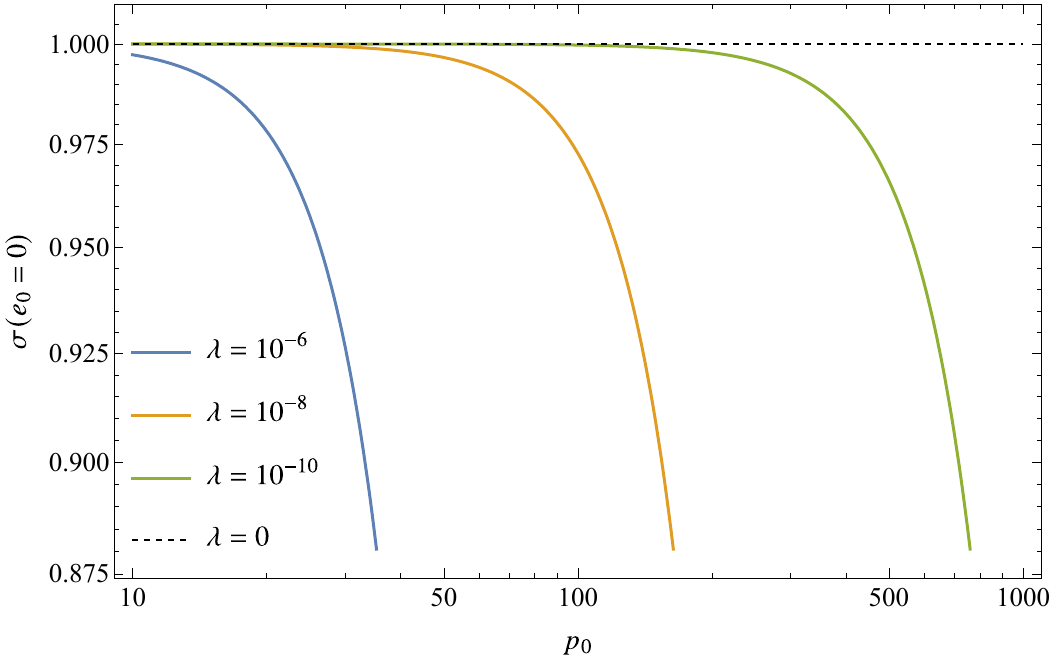} 
	\\ 
	\caption{Ratio, $\sigma(p_0,\lambda)$, of the Peters-Mathews inspiral timescale and the full SdS inspiral timescale  for various SdS parameter values with its $p$ dependence and (b) its $e$ dependence. We see here that larger orbits gain an increased effect by the SdS parameter as the orbit will decay faster compared to the Peters' decay time ($\lambda=0$). 
	}
	\label{fig:timescale-e0-ratio}
\end{figure}

For circular orbits, an exact analytic decay timescale with the effect of the SdS parameter can be derived similarly in \cite{peters1964gravitational}. Peters' argument regarding the minimal dependence of $e(p)$ can also be used to extend the suitability of the timescale to eccentric orbits. We start by deriving the inspiral timescale as the integral of the inverse of the $\dot{p}$ equation. Integrating up to near-coalescence, the timescale, $\tau$, becomes a function of the initial orbital parameters. Up to $\mathcal{O}(\lambda)$, the timescale with an SdS parameter is derived as a factor multiplied by Peters' timescale $\tau_P$, such that
\begin{equation}
	\tau\sim\tau_P\left(1+\lambda \; \tau_{\lambda}\right)
	\label{eq:correcteddecaytime}
\end{equation}
with Peters timescale in $(p,e)$ space equal to 
\begin{eqnarray}
	\tau_P&=&\frac{5 c^5 p_0^4}{256 G^3 M^3 q f(e_0)}\\
	&=&1.84\times10^{24}\text{yr}\left(\frac{p_0}{1\; \text{pc} }\right)^4\left(\frac{10^6 \,M_{\odot}}{M}\right)^3\left(\frac{10^{-5} }{q}\right)\frac{1}{\;f(e_0)}\nonumber
	\label{eq:Petersdecaytime}
\end{eqnarray}
wherein we brought back the physical units and scalings for increased readability. The semi-latus rectum $p$ here has units of length unlike the previous usage of the nondimensional $p$. The function $f(e_0)$ is the enhancement function for including initially eccentric orbits, and is equal to: 
\begin{equation}
	f(e)=\left(1-e^2\right)^{3/2} \left(1+7e^2/8\right).
	\label{eq:Peterseccfactor}
\end{equation}
The timescale function linear to $\lambda$ is given by:
\begin{equation}
	\tau_{\lambda}=-\frac{19 c^2 p_0^3 }{4 G M  }\frac{ F(e_0)}{f(e_0)}
	\label{eq:sdstimescale}
\end{equation}
which is related to a new eccentricity enhancement function:
\begin{eqnarray}
	F(e)&=&\frac{1}{228 \left(1-e^2\right)^{3/4}}\Big(60 \left(1-e^2\right)^{5/4}-26 \left(1-e^2\right)
    \nonumber\\
    &&-231 \sqrt{1-e^2}+425\Big)
	\label{eq:sdseccfactor}
\end{eqnarray}
that evaluates to 1 for $e\rightarrow0$. For circular orbits, we have the timescale ratio $\sigma(p,e=0,\lambda)=\tau_P/\tau$ has the form:
\begin{eqnarray}
	\sigma(p_0,e=0,\lambda)&=&1-\lambda\frac{19 c^2 p_0^3 }{4 G M}+\mathcal{O}(\lambda^2)\\
	&=&1-\lambda\frac{19   }{4 }\left(\frac{p_0}{1.5 \text{ km}}\right)^3\left(\frac{1 		M_\odot}{M}\right)^{3}.\nonumber
	\label{eq:sdstimefactor}
\end{eqnarray}
The main effect of the factor is to reduce the timescale due to the negative sign. Larger initial orbits will have an increased deviation to the Peters' timescale due to the $p_0^3$ scaling. As the orbiter starts initially near the BH, it will approximately follow the Peters' timescale but if it starts initially farther, the timescale correction goes larger as shown in Fig. \ref{fig:timescale-e0-ratio}.
For a stellar mass BH, with mass $M=\mathcal{O }(10 )M_\odot$, circular orbits will decay in a timescale reduced by a factor of $\sigma\approx1-10^{-3}\lambda \;p_0^3$, which will only become considerably large if the starting orbit has a size $p_0\approx10\lambda^{-1/3} \text{ km}$. Near the maximum SdS parameter, say $\lambda=10^{-6}$, the starting orbit must be around $p_0\sim\mathcal{O}(10^3)\text{ km}$ to have maximum deviation. Beyond this estimate, the ratio will be negative and is thus unphysical, which means that orbits are unbound at these regime. For a SMBH, the correction is very small with an initial orbit having $p_0\sim\mathcal{O}(10^{8})\text{ km}$ before being considerably large. This may be of astrophysical significance for scenarios of large-distance capture of inspiralling objects like EMRIs by SMBH \cite{amaro2007intermediate,amaro2018relativistic} with gravitational wave induced evolution estimated to be the dominant dynamics starting at a relatively large distance.
%We note here that a $\lambda=10^{-6}$ means either a relatively small $\Lambda$ ($=10^{-18}$) compared to the measured $\Lambda\;(\sim10^{-50})$ \cite{ade2016planck} or hypothetically large black hole (of order $M\approx10^{20}\,M_{\odot}$) inside a small universe (of order $\mathcal{O}(10^{9})$ km).

For the effect of the eccentricity, let us include the enhancement functions from Eqs. \eqref{eq:Peterseccfactor} and \eqref{eq:sdseccfactor}. To obtain this, we only need to multiply the earlier factors to the $\lambda$ dependent term, 
\begin{equation}
	\sigma(p_0,e_0,\lambda)\sim1-\lambda\frac{19 c^2 p_0^3 }{4 G M}\frac{F(e)}{f(e)}.
\end{equation}

\begin{figure}
	\centering
	\includegraphics[width=\linewidth]{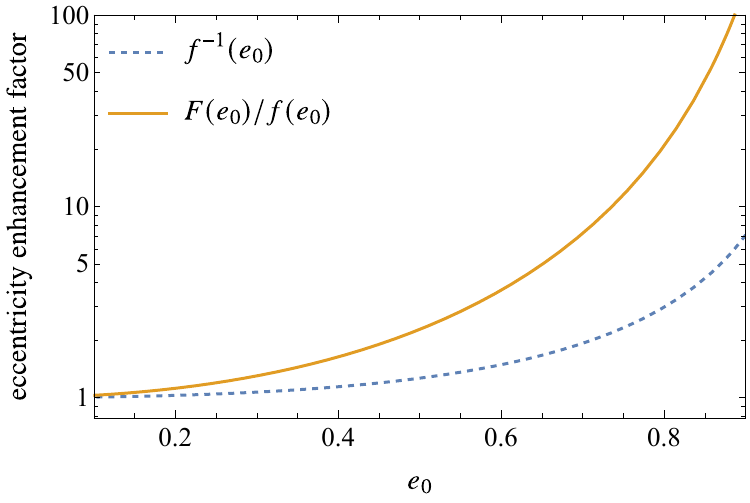}
	\caption{Eccentricity factors for the circular inspiral time. We see here that initially eccentric orbits are more affected by the SdS parameter which can result to a larger deviation in the decay time with Peters formula. Up to a factor of $\sim$100 is to be included for $e_0\sim0.7$ orbit decay time. However, this will still scale linearly with the value of $\lambda$ .}
	\label{fig:timescale-eccentricity-factor}
\end{figure}

The deviation brought by the eccentricity can also be visualized by noticing that the correction factor to the timescale is just the eccentricity functions:
\begin{eqnarray}
	\frac{\tau_P(e)}{\tau_P(e=0)}&=&f(e_0)^{-1},\\
	\frac{\tau_\lambda(e)}{\tau_\lambda(e=0)}&=&F(e_0)/f(e_0).
	\label{eq:eccfactors}
\end{eqnarray}
We show in Fig. \ref{fig:timescale-eccentricity-factor} the comparison of $f(e_0)^{-1}$ and $F(e_0)/f(e_0)$. The function $F(e_0)/f(e_0)$ increases faster than $f(e_0)$ for more eccentric orbits which means that the effect of eccentricity in modifying the timescale is strengthened by the presence of the SdS parameter. We see in Fig. \ref{fig:timescale-eccentricity-factor} that for an initially eccentric orbit, with $e_0\sim0.7$, the deviation in the inspiral timescale is amplified by a factor of $\mathcal{O}(100)$. This implies that as the eccentricity continues to decrease, its value affects the SdS inspirals faster as compared than the pN inspirals. 

\begin{figure*}
	\centering
	\begin{minipage}{0.45\textwidth}
		\includegraphics[width=1.05\linewidth]{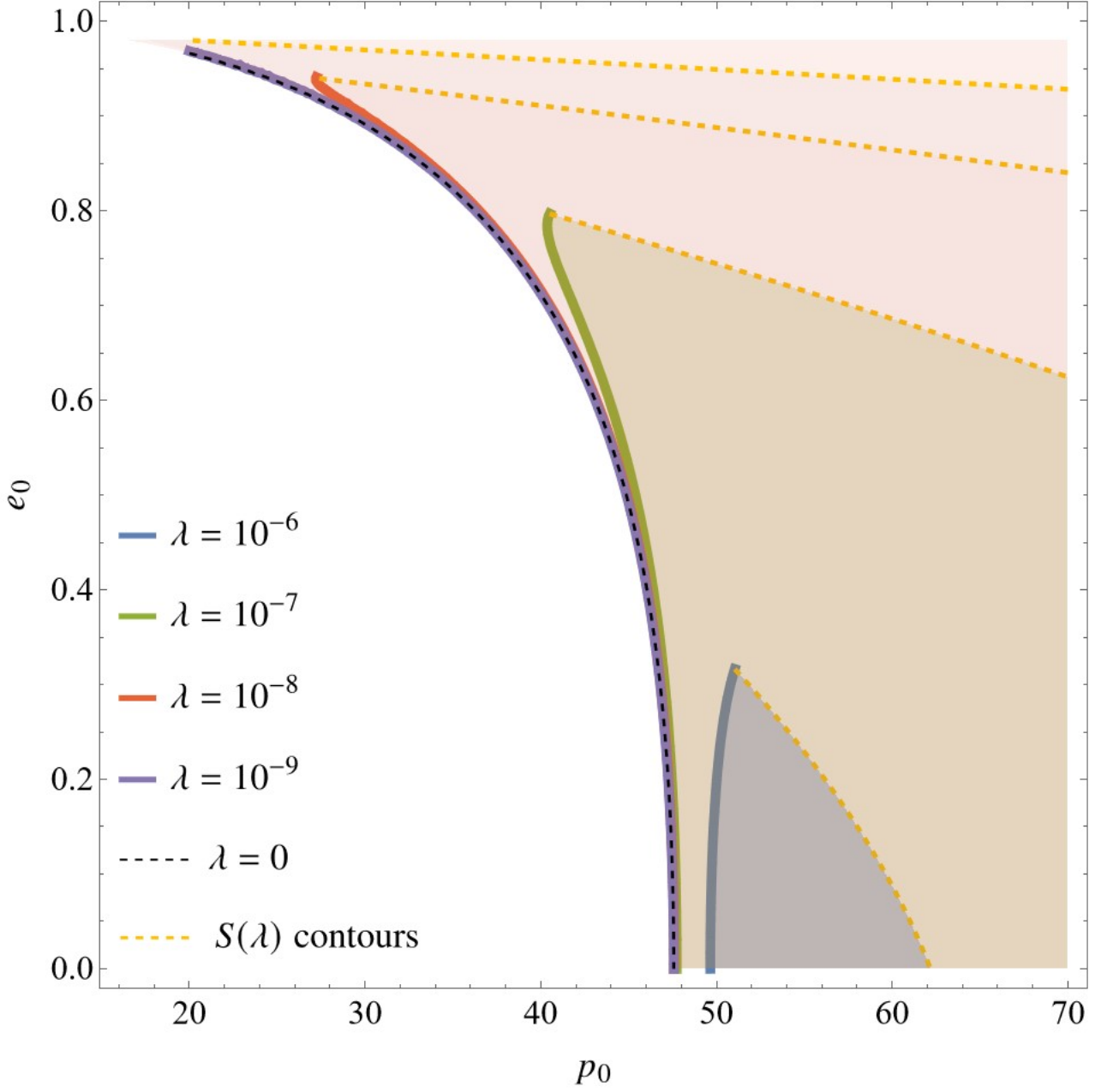}
		\text{(a) Contours of equal $\lambda$ for a fixed inspiral time.}
	\end{minipage}
	\hspace{1cm}
	\begin{minipage}{0.45\textwidth}
		\includegraphics[width=\linewidth]{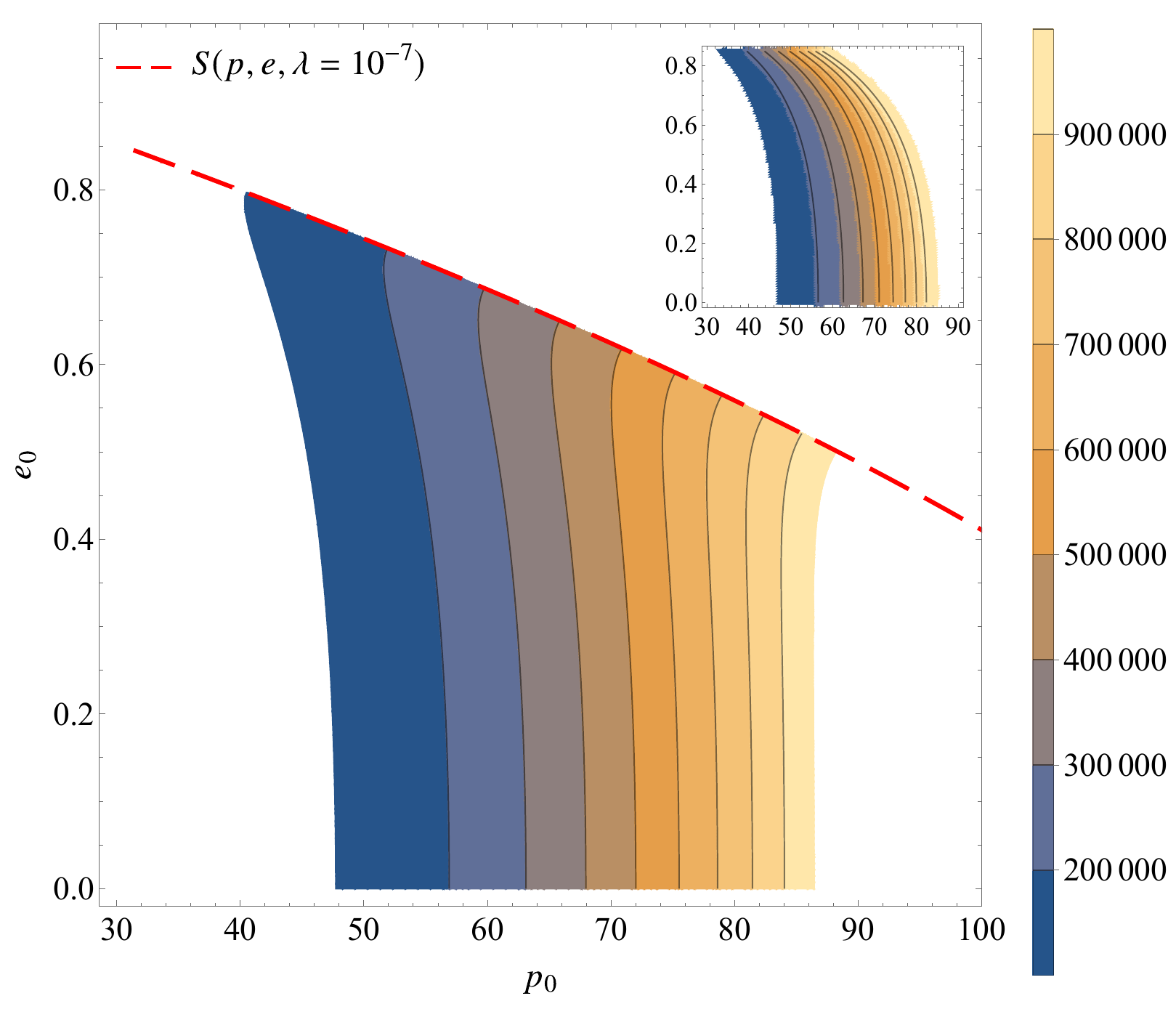}
		\text{(b) Isochrones for a fixed $\lambda$}
	\end{minipage}
	\caption{Analytical contours for a fixed $\tau$ and fixed $\lambda$. In (a), we have an increasing $\lambda$ for a fixed $\tau=10^5$ with the filled region as the source population with higher decay timescale ($\tau>10^5$). The fixed timescale contours show that by increasing the SdS parameter, the total area covered by a particular timescale contour decreases which could mean that the source population will decrease, most notably in $e$, for a given SdS parameter and time. However in (b), we compare decreasing $\tau$ for a fixed $\lambda$. By comparing timescale contours for a specific SdS parameter, we observe that the source population increases in $p$ as compared to the $\lambda=0$ case.}
	\label{fig:timescale-contours-iso-lambda}
\end{figure*}

We show in Fig. \ref{fig:timescale-contours-iso-lambda} the $\lambda$ contours and isochrones in the phase space of orbital parameters $(p,e)$. If we relate the area in the phase space to the number of potential sources, Fig. \ref{fig:timescale-contours-iso-lambda}(a) shows that higher values of the SdS parameter increases the population of detectable sources at a fixed inspiral time. In Fig. \ref{fig:timescale-contours-iso-lambda}(b) we see that the SdS inspiral time shows a different curve from the $\lambda=0$ inspiral time implying a shift in the orbital parameters of potential sources. This means that, for a given observation time, we can detect much larger orbits due to the SdS parameter. The general observation here is that the population of sources will be limited to their initial eccentricity and size at capture and will increase due to higher eccentricity with larger orbits decaying faster from the effect of $\lambda$.

\begin{figure*}
	\begin{tabular}{cc}
		\includegraphics[width=0.47\linewidth]{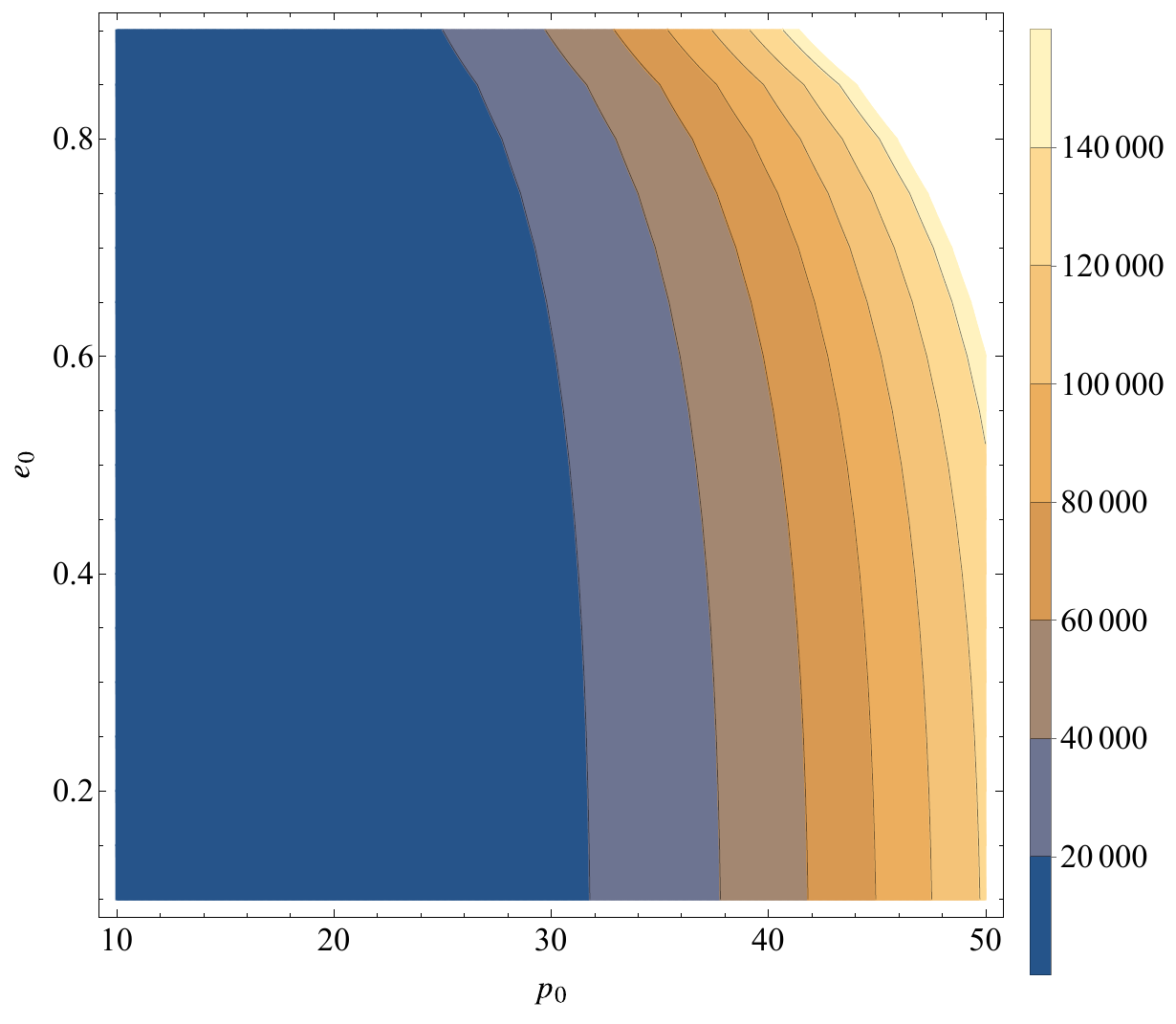}& \includegraphics[width=0.47\linewidth]{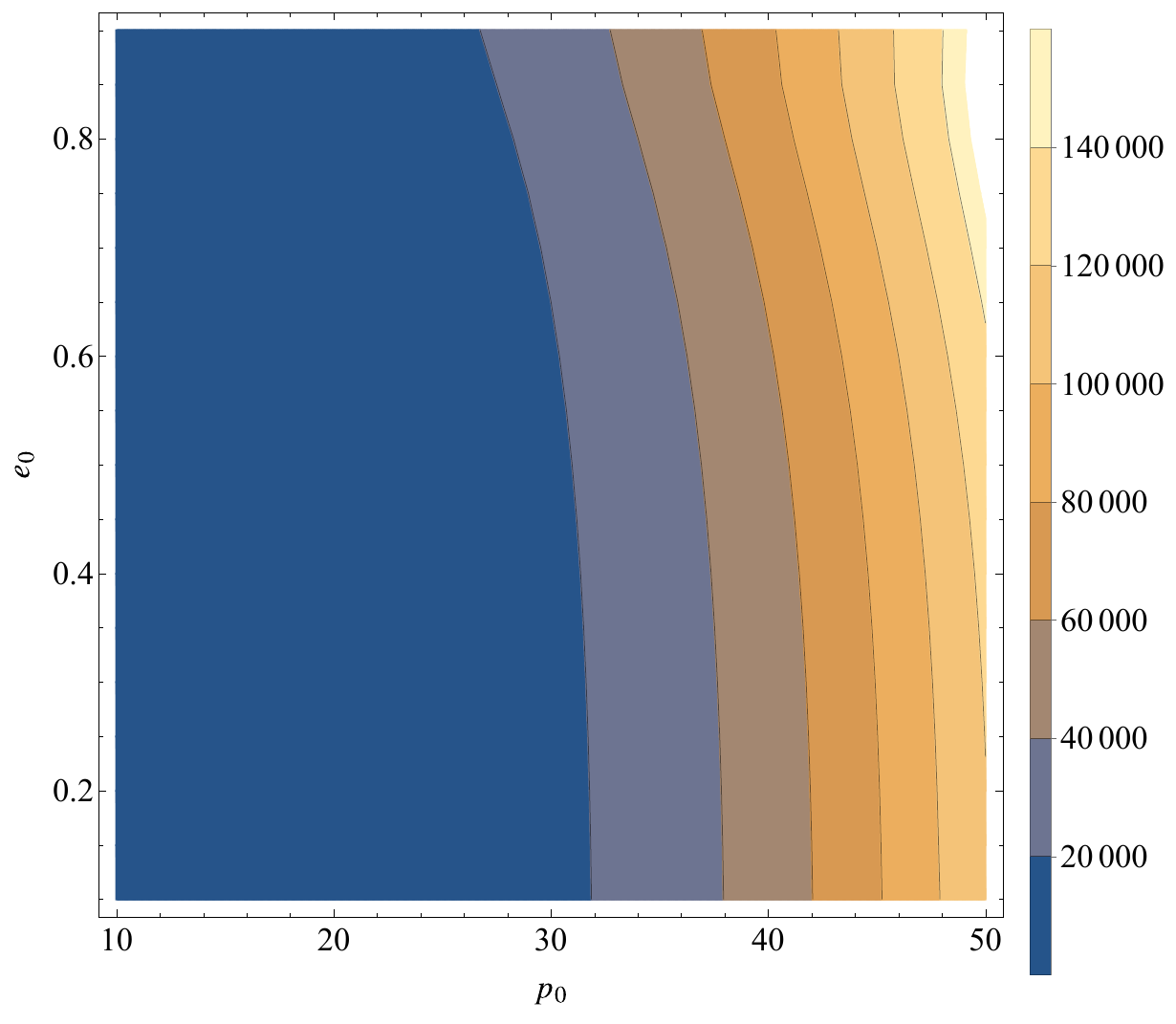}\\ 
	\end{tabular} 
	\caption{Numerical contour plots of the plunge times in a range of initial orbital parameters for Schwarzschild $(\lambda=0)$ (left) and SdS $(\lambda=10^{-7})$ (right) spacetime. We note here that incorporating a self-consistent evolution in $(p,e)$ decreases plunge times as the CO starts to inspiral further away from the BH. Given a fixed timescale, say at $\tau=10^{5}$, a larger region is covered in the initial parameter which can be expected as an increase in the population of potential gravitational waves sources. There is also a notable decrease in the population for highly eccentric orbits as also seen from the analytical contours previously, which means that highly eccentric orbits are expected to decay more slowly for higher values of SdS parameter. }
	\label{fig:numericaltimescales}
\end{figure*}

We can now numerically integrate the orbital evolution equations to obtain a more accurate inspiral time measurement across the parameter space. We start by solving the plunge times for a range of initial parameters, then we deduce if those parameters are detectable within a given time. We numerically integrate the radiation reaction equations (Eqs. \eqref{eq:pdotsds} and \eqref{eq:edotsds}), considering all possible initial parameters at a given range. Then we stop the integration if the parameters near the separatrix and we then collect the time per initial parameter for the integration to end. We consider the range $p_0=(10,50),\; e_0=(0,1)$ with $\lambda=10^{-7}$.  The isochrones with the evolving eccentricity is obtained numerically in Fig. \ref{fig:numericaltimescales}. We show in Fig. \ref{fig:numericaltimescales} that for a given range of time, the range of initial parameters is expanded due to the presence of the SdS parameter. Orbits of the same $p$ will indeed decay faster with the help of the $\lambda$, but with higher eccentricity, the factor of reduction in the timescale is decreased.

\section{Conclusions}

The SdS parameter introduces both conservative and dissipative effects on the dynamics of a binary system. Key findings of this paper include an accelerated rate of orbital decay, with shorter plunge and circularization times due to the $\lambda$. Additionally, we demonstrate that eccentric orbits are more strongly influenced by $\lambda$, with this effect being independent of the binary’s separation.

For the conservative effects, we analyzed changes in the effective potential, where the SdS parameter introduces additional turning points for a given orbital parameter. Depending on its value, the SdS parameter alters the nature of bound and unbound orbits, with a critical threshold causing the complete absence of bound orbits. This leads to an analysis of the $\lambda$’s influence on the orbital parameters governing the existence of bound orbits. We derive the separatrix, represented as a 2D surface, a polynomial function of $(p,e,\lambda)$, and identify the critical value of the SdS parameter that allows for a single stable bound orbit. Although small in magnitude, the SdS parameter significantly affects the parameter space. We also find the presence of an outer separatrix, which further constrains the region of bound orbits; outside this boundary, the $\lambda$ term dominates, rendering orbits unstable.

For the dissipative effects, we use the prescription by Hoque and Aggarwal to compute the post-Newtonian gravitational fluxes for a binary system influenced by the SdS parameter. We show that the SdS parameter increases the rate of decay of the orbital parameters when translated into radiation reaction equations. Due to the increase in Keplerian binding energy and angular momentum, the SdS parameter introduces new dynamical effects in the radiation reaction. While the SdS parameter generally reduces the rate of orbital decay compared to standard post-Newtonian radiation reaction in flat spacetime, we find a nontrivial increase in eccentricity evolution at initially low eccentricities. We attribute this to the non-trivial coupling of the SdS parameter with eccentricity, where binaries with nonzero eccentricity are more sensitive to its effects.

We have investigated the effect of the SdS parameter on physical quantities such as circularization, plunge times, and orbital trajectories. Deviations from post-Newtonian orbital shapes are observed for eccentric binaries. We also generate waveforms produced by binaries under radiation reaction, showing an increase in amplitude and phase advance due to the SdS parameter. The phase advance is linked to a decrease in orbital frequency, also caused by the SdS parameter. Finally, we calculate the detection timescales for these waveforms, demonstrating that the SdS parameter increases the number of detectable sources within a given time.

Although the contribution from $\lambda$ is typically subleading compared to the dominant post-Newtonian terms, its relative importance grows in certain regions  of the parameter space (where the environmental contribution might dominate over the radiation reaction dynamics) potentially becoming comparable in magnitude. Again, our results may be interpreted more generally as describing a quadratic term in the effective gravitational potential, $\delta V \sim \lambda r^2$, with $\lambda$ serving as an effective parameter for environmental perturbations exhibiting the same scaling. The calculations performed here therefore extend beyond the purely cosmological context, providing a unified framework for analyzing systems subject to $r^2$ corrections to the effective potential and clarifying their combined conservative and radiation-reaction–driven evolution. Future work related to this claim may involve a deeper examination of the mapping of $\lambda$ to other astrophysically relevant parameters.

Lastly, we emphasize that adopting the SdS parameter $\lambda$ as an environmental proxy constitutes an idealized framework rather than an exact metric for a specific astrophysical system. By capturing the leading-order $\mathcal{O}(r^2)$ scaling characteristic of external energy densities, such as the tidal quadrupole potential in a binary system or the spacetime curvature induced by a uniform magnetic field, $\lambda$ allows us to analytically isolate how an isotropic, secular outward perturbation influences dissipative EMRI dynamics. We note, however, that this isotropic approximation fundamentally breaks down in realistic astrophysical environments. 
%Furthermore, our formalism isolates the static potential while explicitly neglecting velocity-dependent interactions—such as the Lorentz force, the Coriolis force, and relativistic frame-dragging, which become highly relevant in strong-gravity regimes. Finally, this proxy assumes a uniform background curvature, neglecting the spatial decay of realistic localized environments, such as the drop-off of accretion disk magnetic fields.} 
Results presented here are best interpreted as a qualitative exploration of perturbed conservative and dissipative dynamics. This formalism serves as a mathematically clean baseline to guide future investigations before transitioning to computationally expensive, fully three-dimensional general relativistic magnetohydrodynamic (GRMHD) or N-body simulations.

%Lastly, we are currently extending our investigation of the cosmological constant's effects on radiation reaction to the strong-field regime. Additionally, much work is needed to incorporate other environmental effects from realistic astrophysical scenarios. Future studies may explore cases with non-constant cosmic acceleration, such as the McVittie black hole spacetime \cite{nolan2014particle}, which we leave for subsequent research.

\begin{acknowledgments}
The authors would like to thank Niels Warburton, Reggie Bernardo, and Barry Wardell for helpful correspondence on the draft and also the organizers of the 27th Capra Meeting on Radiation Reaction in General Relativity, held at the National University of Singapore where this project was presented. We would also like to thank Jezreel Castillo, Sean Fortuna, Jerome Mecca, Delo Procurato, and Jane BDM Garcia for help with the numerics and visualizations in the paper. This work was made possible by the support of the University of the Philippines Diliman OVPAA through Grant No. OVPAA-BPhD-2016-13 and the University of the Philippines Diliman OVCRD through Project No. 191937 ORG. JAN Villanueva would like to thank DOST ASTHRDP Scholarship for funding their PhD program in the University of the Philippines Diliman.
\end{acknowledgments}

\appendix

%\section{Appendixes}

%\section{A little more on appendixes}

%\subsection{\label{app:subsec}A subsection in an appendix}

\bibliography{sdsrefs}% Produces the bibliography via BibTeX.

\end{document}